\documentclass[10pt,letterpaper]{article}
\usepackage{blindtext,titlefoot}
\usepackage{authblk}
\usepackage{marvosym}
\usepackage[utf8]{inputenc}
\usepackage[english]{babel}
\usepackage{amsmath}
\usepackage{amsfonts}
\usepackage{amssymb}
\usepackage{makeidx}
\usepackage{graphicx}
\usepackage{lmodern}
\usepackage{kpfonts}
\usepackage{dsfont}
\usepackage{cancel}
\usepackage{amsthm} 
\usepackage[left=2cm,right=2cm,top=2cm,bottom=2cm]{geometry}
\usepackage{subcaption}
\usepackage{multirow}
\usepackage{float}
\usepackage{url}
\usepackage{breakurl}
\usepackage[breaklinks]{hyperref}
\usepackage{multicol}
\usepackage{mathtools, cuted}
\usepackage{lipsum}
\usepackage{booktabs}
\usepackage{comment}
\usepackage{cite}

\usepackage{longtable}

\DeclareMathOperator{\re}{Re}
\DeclareMathOperator{\im}{Im}
\DeclareMathOperator{\card}{card}
\DeclareMathOperator{\Gen}{Gen}
\providecommand{\nset}[1]{\mathbb{#1}}
\providecommand{\com}[1]{``#1"}
\newcommand{\set}[1]{\left\{#1\right\}}
\newcommand{\norm}[1]{\left\lVert #1 \right\rVert}
\newtheorem{corollary}{Corollary}

\usepackage{enumitem} 
\setlist[itemize]{noitemsep} 

\usepackage{titlesec} 
\titleformat{\section}[block]{\large\bfseries\scshape\centering}{\thesection.}{1em}{} 
\titleformat{\subsection}[block]{\large\bfseries\scshape\centering}{\thesubsection.}{1em}{}
\titleformat{\subsubsection}[block]{\large\bfseries\scshape\centering}{\thesubsubsection.}{1em}{} 

\title{\Huge\bfseries Simulation of the behavior of the \textit{Leopardus guigna} using random walkers}

\author[,a]{A. Torres-Hernandez  \footnote{Email: anthony.torres@ciencias.unam.mx; ORCID: 0000-0001-6496-9505}}

\author[,b]{Byron C. Guzmán \footnote{Email: b.guzman.marin@outlook.com; ORCID: 0000-0001-7051-1154}}

\author[,b]{Melanie Kaiser \footnote{Email: melanie.kaiser.pv@gmail.com; ORCID: 0000-0002-8592-926X}}

\author[,c]{Julio C. Hernández \footnote{Email: biol.julio@gmail.com; ORCID: 0000-0002-1286-2404}}

\affil[b]{Foundation for Wild Cat Coordination, Santiago, Chile}

\affil[c]{Pronatura Veracruz A. C. México}

\affil[a]{Department of Physics, Faculty of Science - UNAM, Mexico}

\date{}

\begin{document}

\maketitle


\begin{abstract}

Considering that there is very little information on the behavior habits of guigna cats, as well as investigations in which small populations are captured to place radiocollars on them and then release them in the place where they were captured, which is done with the intention of collecting data on their positions in a territory and thus make estimates of the mean distances they usually travel. Under the hypothesis that guignas maintain a sedentary behavior in a specific area of a given territory, this paper shows one way to simulate a distribution of points in a territory using random walkers to emulate the distribution of the data that would be obtained by placing radiocollars in a population of guignas, with which it is possible to make estimates of the mean distances that move away from a certain fixed position, and the interactions they can have with points in the territory that represent a high probability of lethality, such as farms, packs of dogs, roads, urban areas, etc. It is necessary to mention that by estimating the possible interactions that a guignas population may have with possible predators in a territory with the help of a satellite image, it is possible to evaluate the points of a territory that represent a potentially lethal risk for the guignas, and thus generate relocation strategies that help preserve them.

\end{abstract}

\section{Introduction}

\textit{Leopardus guigna} is not only the smallest wild cat in the Neotropics, but also one of the felids with the most narrow distribution range. The guigna cat is endemic to Central and Southern Chile and a thin strip of Western Argentina. In detail, northernmost observations of this species have been obtained in the Chilean Coquimbo Region, at about $31^\circ$ South, while the southern limit of its distribution range is in the Aysén Region of Chile, at $48^\circ$ South \cite{napolitano2020new}. For a long time, this area has been characterized by vast native forests that provided ample space to these territorial cats. Unfortunately though, progressive urbanization and changes of land use have led to the loss of large parts of the guigna cats habitat. Especially in southern Chile, increasing deforestation and parcelling of previously intact areas is putting pressure on the population of guigna cats \cite{fleschutz2016response}. Parcelling goes hand in hand with the presence of people, dogs, and vehicles – which constitute the main threats for \textit{L. guigna}. As of today, the species is classified as vulnerable both by the International Union for Conservation of Nature and national species classification regulations \cite{napolitano2015leopardus}.

Because the reliable prediction of individual guigna cats' movements may support the successful design of mitigation strategies, data on sighting records, home ranges, and movement behavior of this species should be taken into account during the decision-making on urban planning, roadbuilding, and the designation of protected areas. There are several reasons why this is not currently done satisfactorily, not the least of which is the lack of population and behavioral data.

One of the most recent studies on \textit{L. guigna} home ranges was carried out in the fragmented landscape of the Chilean Araucanía Region and found that these cats had mean $95\%$ usage areas of $6.23 \pm 4.00 \ km^2$ \cite{schuttler2017habitat}. The authors of that study related their findings to a prior publication, where smaller home ranges were reported for guigna cats living in two large, pristine national parks (mean $90\%$ kernel home range = $1.19 \ km^2$; \cite{dunstone2002uso}). Schüttler et al. hypothesized that cats may increase their home range to compensate for lesser forest coverage \cite{schuttler2017habitat}. Notwithstanding, Sanderson and colleagues reported home ranges of $0.8 - 3.7 \ km^2$ for guigna cats living in highly fragmented areas and observed two males occupying $1.8 \ km^2$ and $22.4 \ km^2$, respectively, in contiguous forest with less human influence \cite{sanderson2002natural}.

Because there exits very little information on the behavior habits of the guigna cats, there are investigations in which small populations are captured to place radiocollars on them and then release them in the place where they were captured, which is done with the intention of collecting data on their positions in a territory and thus make estimates of the mean distances that they usually travel or of the hours in which the guignas tend to be more active. An example of these investigations may be found in the reference \cite{sanderson2002natural}, in which the authors obtained the following result through the data obtained from radiocollars placed on guignas:

\begin{eqnarray}
\begin{minipage}{0.8\textwidth}
\com{
$\cdots$ habitat fragmentation and loss may affect male and female guignas differently. Females were sedentary and rarely explored areas beyond their established home ranges. In contrast, males constantly moved about their territories, presumably marking their boundaries and visiting females.  $\cdots$ . Males were
therefore more likely to come into contact
with humans, their pets, and domestic fowl.}
\end{minipage}
\end{eqnarray}

The result that females have acquired a sedentary behavior due to the fact that they rarely explore beyond a delimited area in their territory, could be interpreted as an evolutionary response to the harassment that guignas suffer when exploring territories close to human populations by free-roaming dogs and cats, as well as the constant shots they receive from farmers when they approach their poultry. So, due to deforestation and the increase in urban areas, it would not be strange that as an evolutionary mechanism the males adopt a sedentary behavior similar to that of the females to increase their chances of survival. Therefore, under the hypothesis that guignas maintain a sedentary behavior in a specific area of a given territory, this paper shows one way to simulate a distribution of points in a territory using random walkers to emulate the distribution of the data that would be obtained by placing radiocollars in a population of guignas, with which it is possible to make estimates of the mean distances that move away from a certain fixed position, and the interactions they can have with points in the territory that represent a high probability of lethality, such as farms, packs of dogs, roads, urban areas, etc.

\section{Random Walkers}

Denoting by $\set{\hat{e}_k}_{k=1}^n$ the canonical basis of $\nset{R}^n$. So, if $x\in \nset{C}^n$, it may be written using Einstein notation as follows

\begin{eqnarray*}
x=\hat{e}_k[x]_k=\hat{e}_k\re\left( [x]_k\right)+i \hat{e}_k\im\left([x]_k \right) ,
\end{eqnarray*}

where $[x]_k$ denotes the $k$-th component of vector $x$. On the other hand, considering a set $\Omega_z\subset \nset{Z}$, defined as follows

\begin{eqnarray*}
 \Omega_z:=\set{-z,-(z-1),\cdots,z-1,z},
\end{eqnarray*}

it is possible to define the following set of vectors

\begin{eqnarray}
\mathcal{S}\left(\nset{C}^n, \Omega_z \right):=\set{s\in \nset{C}^n \ : \ \re\left([s]_k\right), \im\left([s]_k\right) \in \Omega_z \ \forall k\geq 1}.
\end{eqnarray}

So, denoting by $\card\left( \cdot \right)$ the cardinality of a set, it is possible to obtain the following results:

\begin{eqnarray}
\mbox{If }\card\left( \Omega_z\right)=k \ \Rightarrow \ \card\left(\mathcal{S}\left(\nset{C}^n, \Omega_z \right) \right)=2k^n \ \mbox{ and } \ \card\left(\mathcal{S}\left(\nset{R}^n, \Omega_z \right) \right)=k^n,
\end{eqnarray}

\begin{eqnarray}
\mbox{If }\card\left( \Omega_z\right)=k \ \Rightarrow \ \exists M>0 \ \mbox{ such that } \ \norm{s}\leq M \ \forall s\in \mathcal{S}\left(\nset{C}^n,\Omega_z \right).
\end{eqnarray}

Therefore, considering a value $r>0$ and some set $ \mathcal{S}\left(\nset{C}^n, \Omega_z \right)$, it is possible to define a function $\Phi_r:\nset{C}^n \to \nset{C}^n$ as follows

\begin{eqnarray}
\Phi_r(x):=x+rs,
\end{eqnarray}

with which it is possible to define the following iterative method

\begin{eqnarray}\label{eq:1}
x_{i+1}:=\Phi_r(x_i)=x_i+rs_i, & i=0,1,2,\cdots.
\end{eqnarray}

\begin{figure}[!ht]
        \begin{subfigure}[c]{0.4\textwidth}
        \centering
 \includegraphics[width=\textwidth, height=0.57\textwidth]{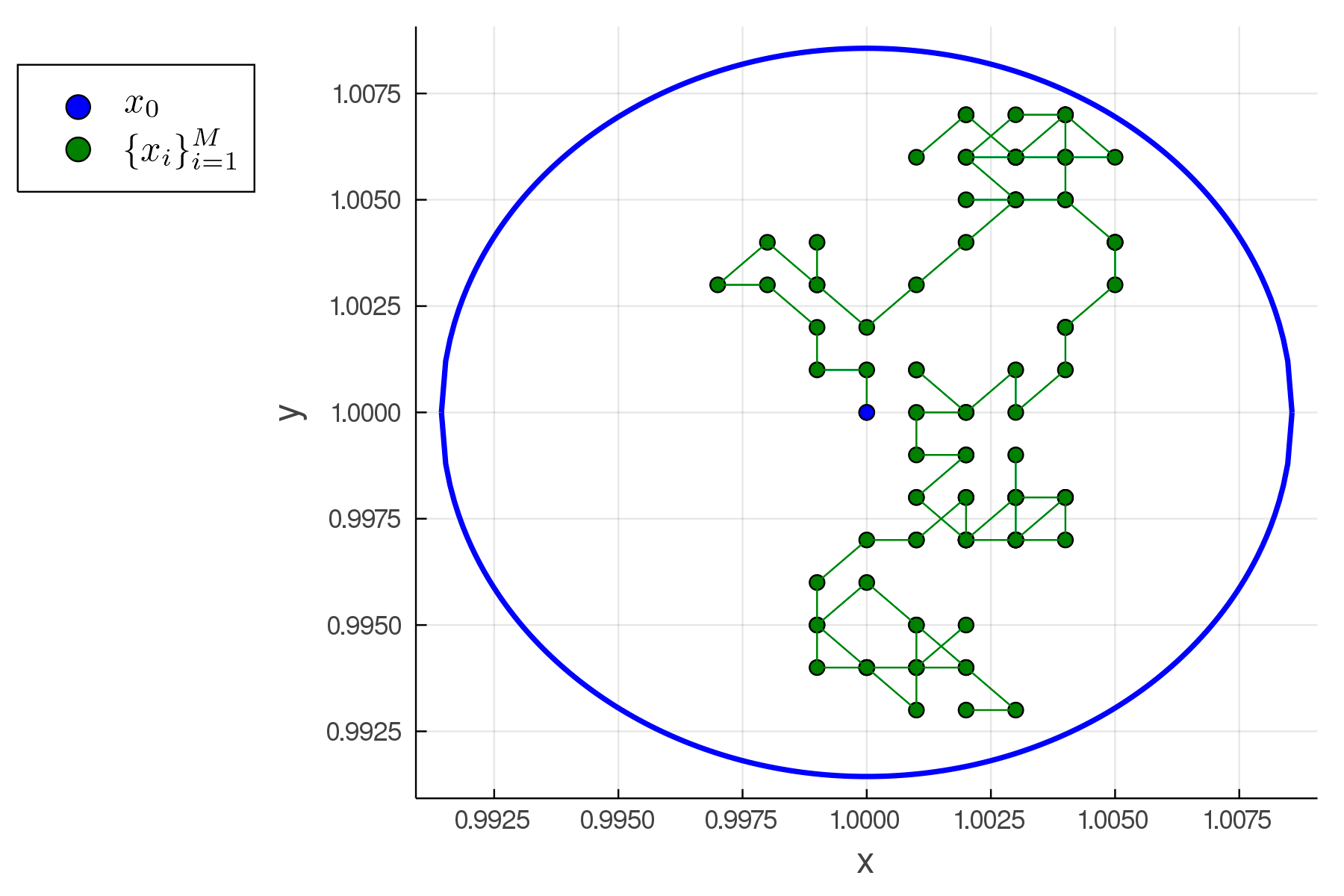}      
    \caption*{a) $r=0.001$}
    \end{subfigure}
        \begin{subfigure}[c]{0.4\textwidth}
        \centering
 \includegraphics[width=\textwidth, height=0.57\textwidth]{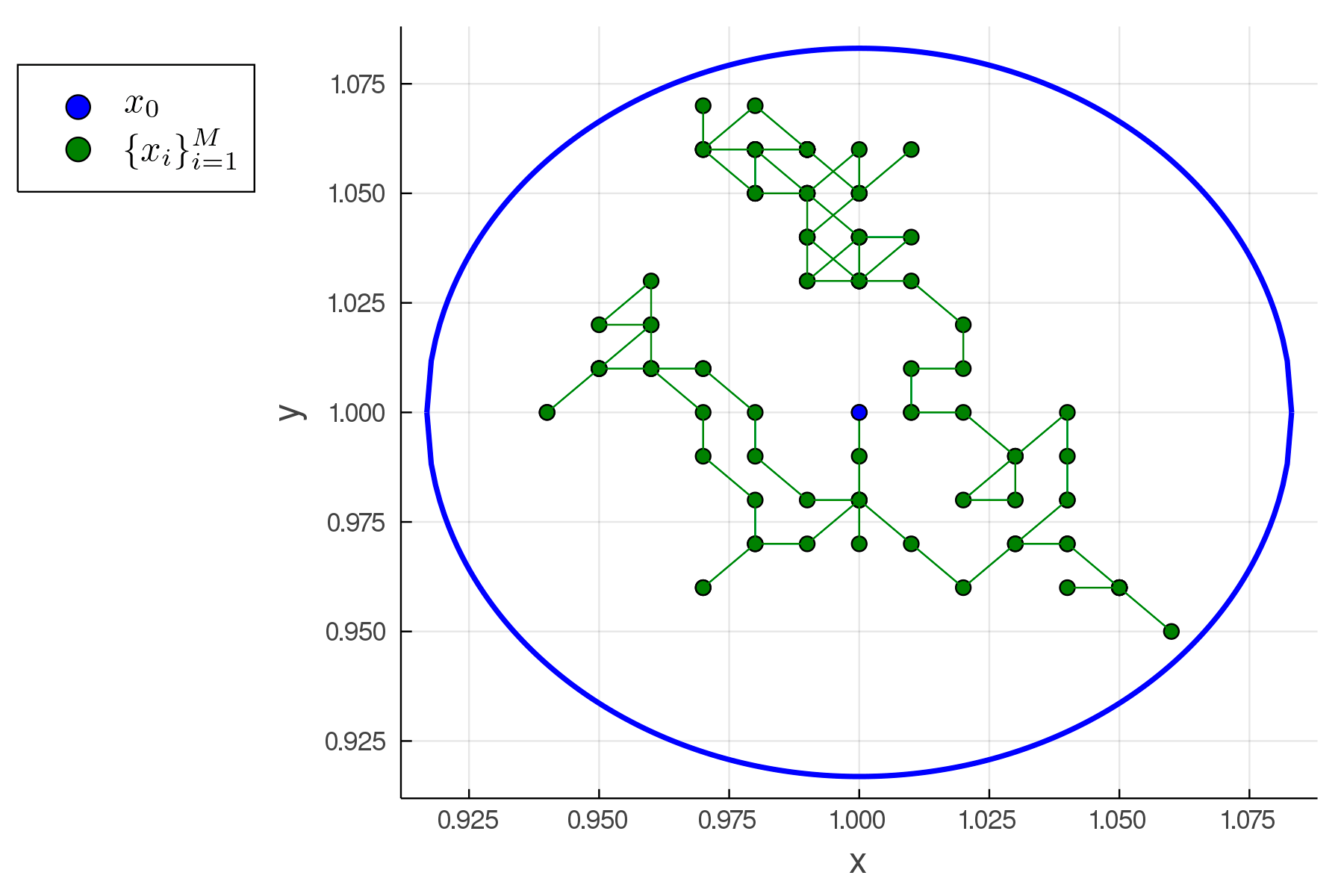}     
    \caption*{b) $r=0.01$}
    \end{subfigure} 
    \centering
    \begin{subfigure}[c]{0.4\textwidth}
    \centering
 \includegraphics[width=\textwidth, height=0.57\textwidth]{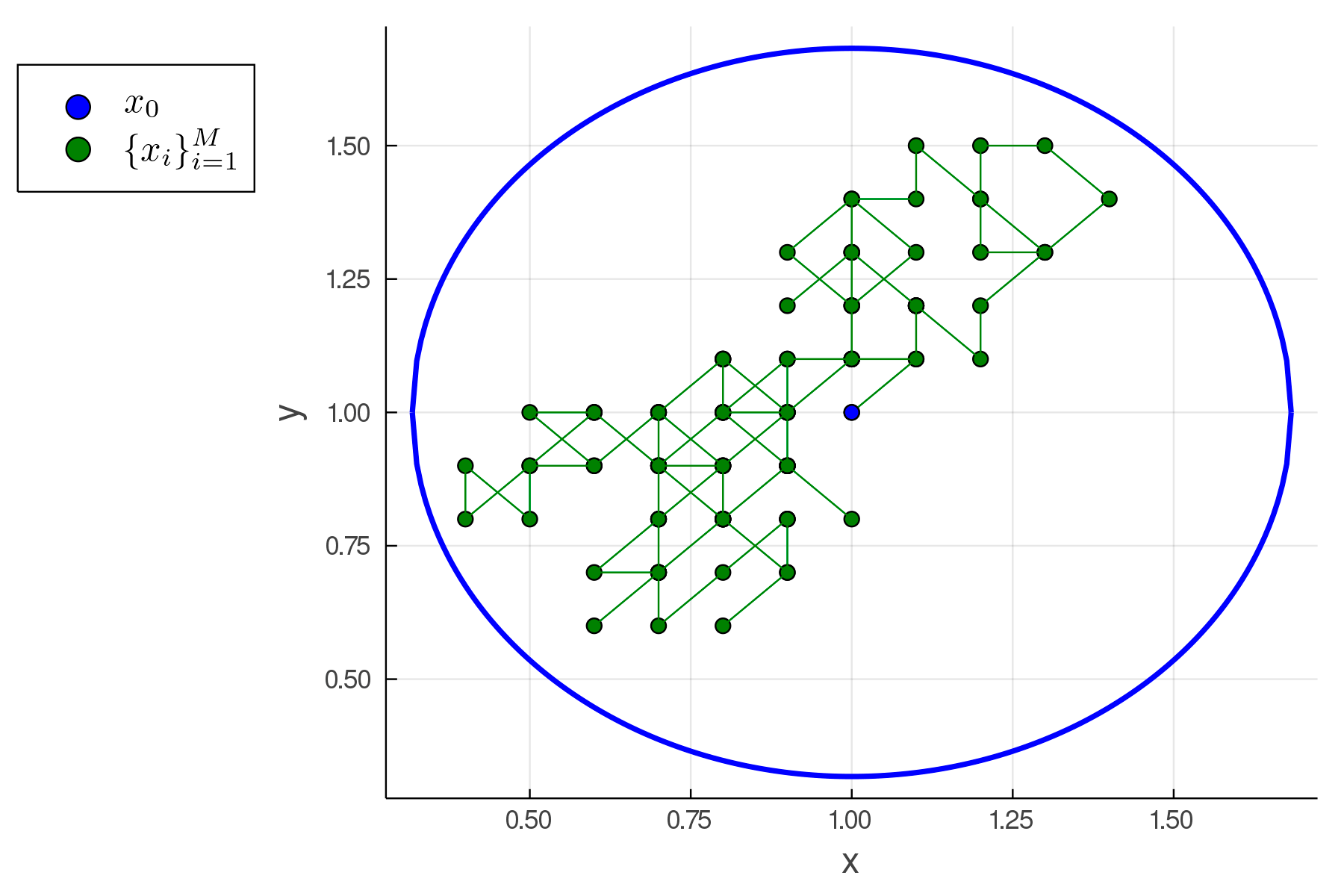}      
    \caption*{c) $r=0.1$}
    \end{subfigure}       
        \caption{Illustrations of some trajectories of random walkers in $\nset{R}^2$ for the same initial condition $x_0$ but with different step sizes $r$ using the set $\mathcal{S}\left(\nset{R}^2, \Omega_1 \right)$.}\label{fig:01}
\end{figure}

So, considering the following set

\begin{eqnarray*}
B(x_0;R):=\set{x \ : \ \norm{x_0-x}<R},
\end{eqnarray*}

and allowing the vectors $s_i$ to be chosen randomly from the set $ \mathcal{S}\left(\nset{C}^n, \Omega_z \right)$, it is possible to obtain the following result (see Figure \ref{fig:01}):

\begin{eqnarray}\label{eq:2}
 \forall x_0 \in \nset{C}^n \ \exists R=R(r)>0 \ \mbox{ such that } \ \set{x_i}_{i=1}^M\in B(x_0;R),
\end{eqnarray}

where $\set{x_i}_{i=1}^M$ denotes a finite sequence of random variables generated through the function $\Phi_r$, which may be named as an iteration function of a random walker. It is necessary to mention that the sequence $\set{x_i}_{i=1}^M$ defines a trajectory that may be termed as the trajectory of a random walker, while the value $r$ may be termed as the step size of a random walker.

\begin{figure}[!ht]
        \begin{subfigure}[c]{0.4\textwidth}
        \centering
 \includegraphics[width=\textwidth, height=0.57\textwidth]{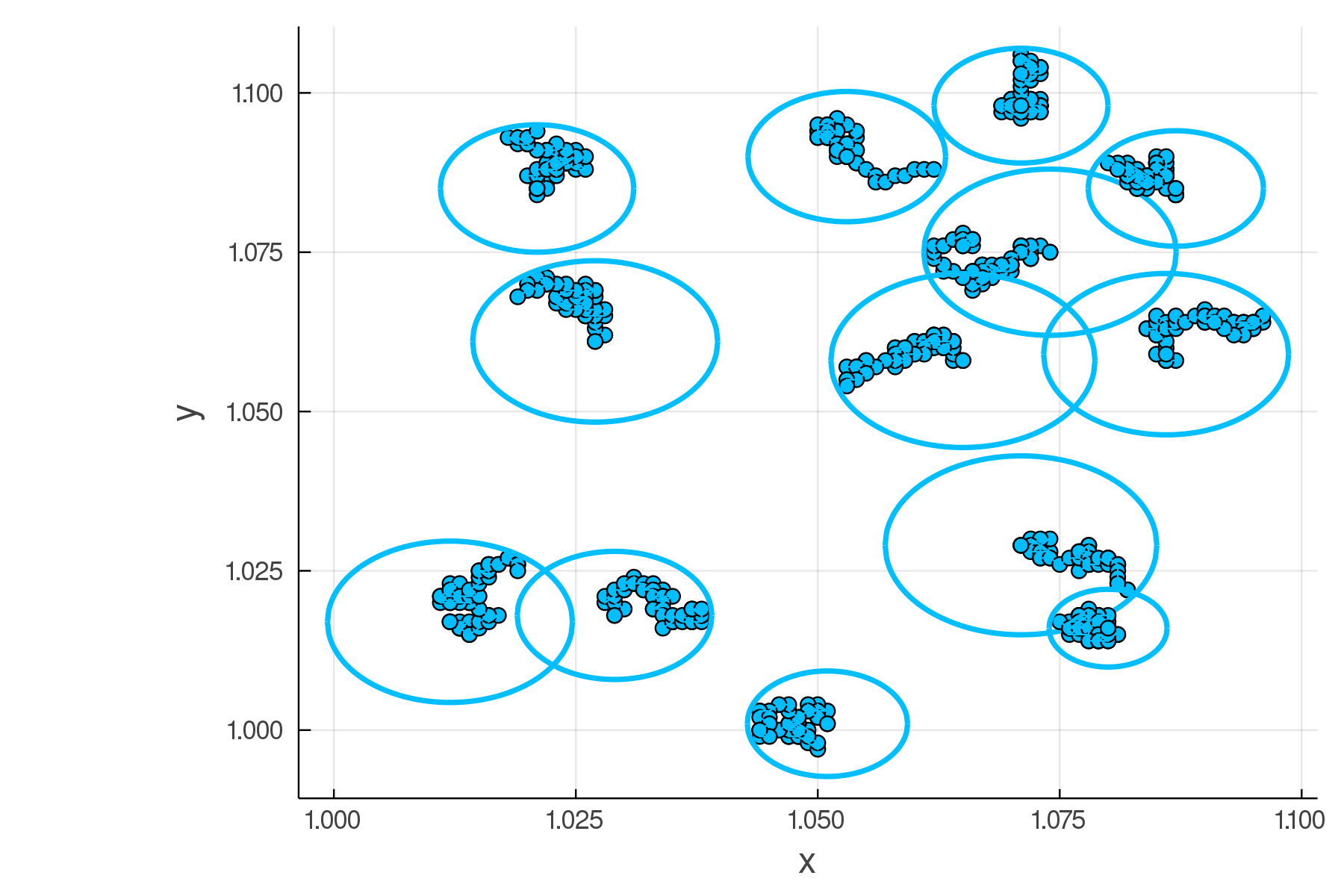}      
    \caption*{a) $r=0.001$}
    \end{subfigure}
        \begin{subfigure}[c]{0.4\textwidth}
        \centering
 \includegraphics[width=\textwidth, height=0.57\textwidth]{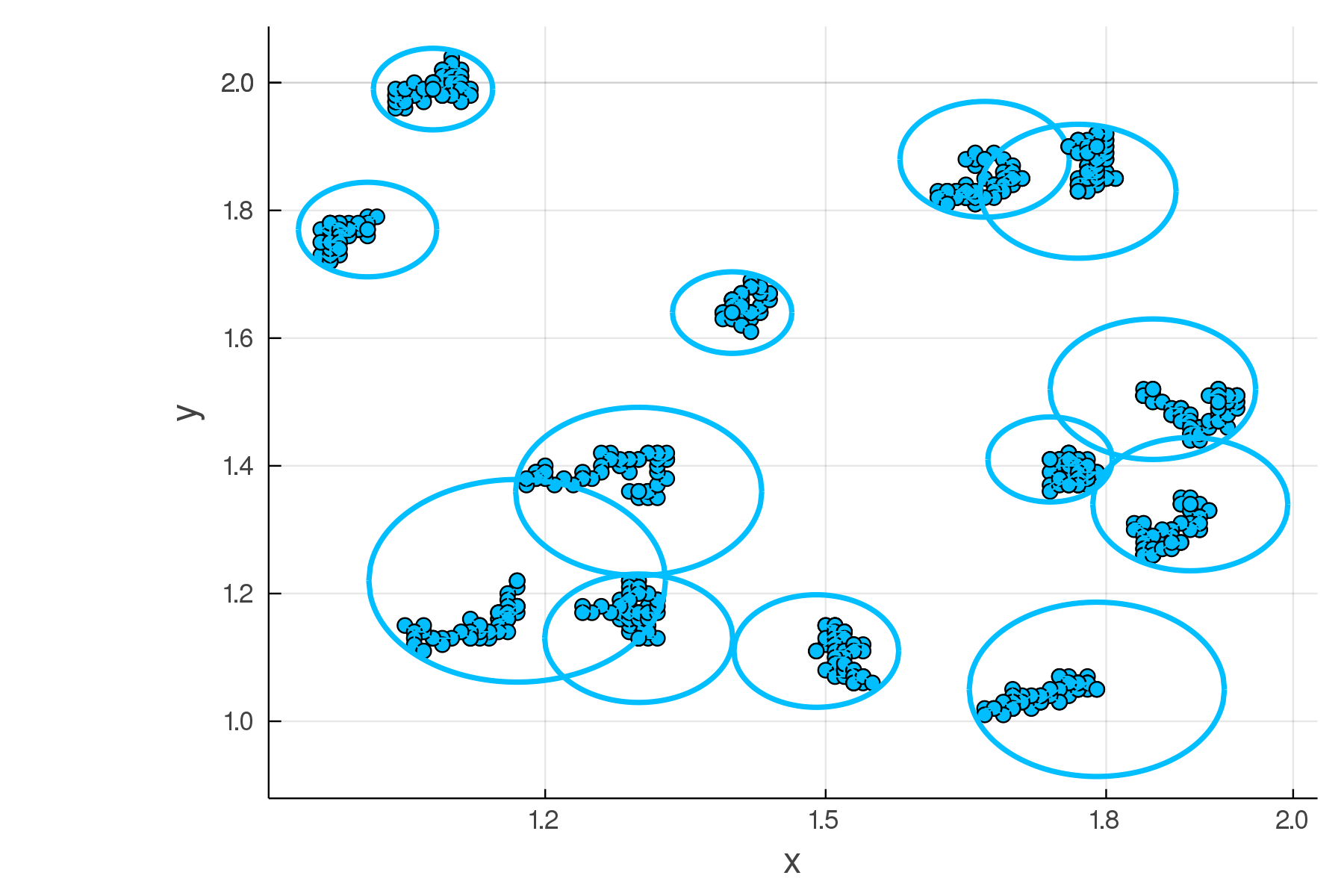}     
    \caption*{b) $r=0.01$}
    \end{subfigure} 
    \centering
    \begin{subfigure}[c]{0.4\textwidth}
    \centering
 \includegraphics[width=\textwidth, height=0.57\textwidth]{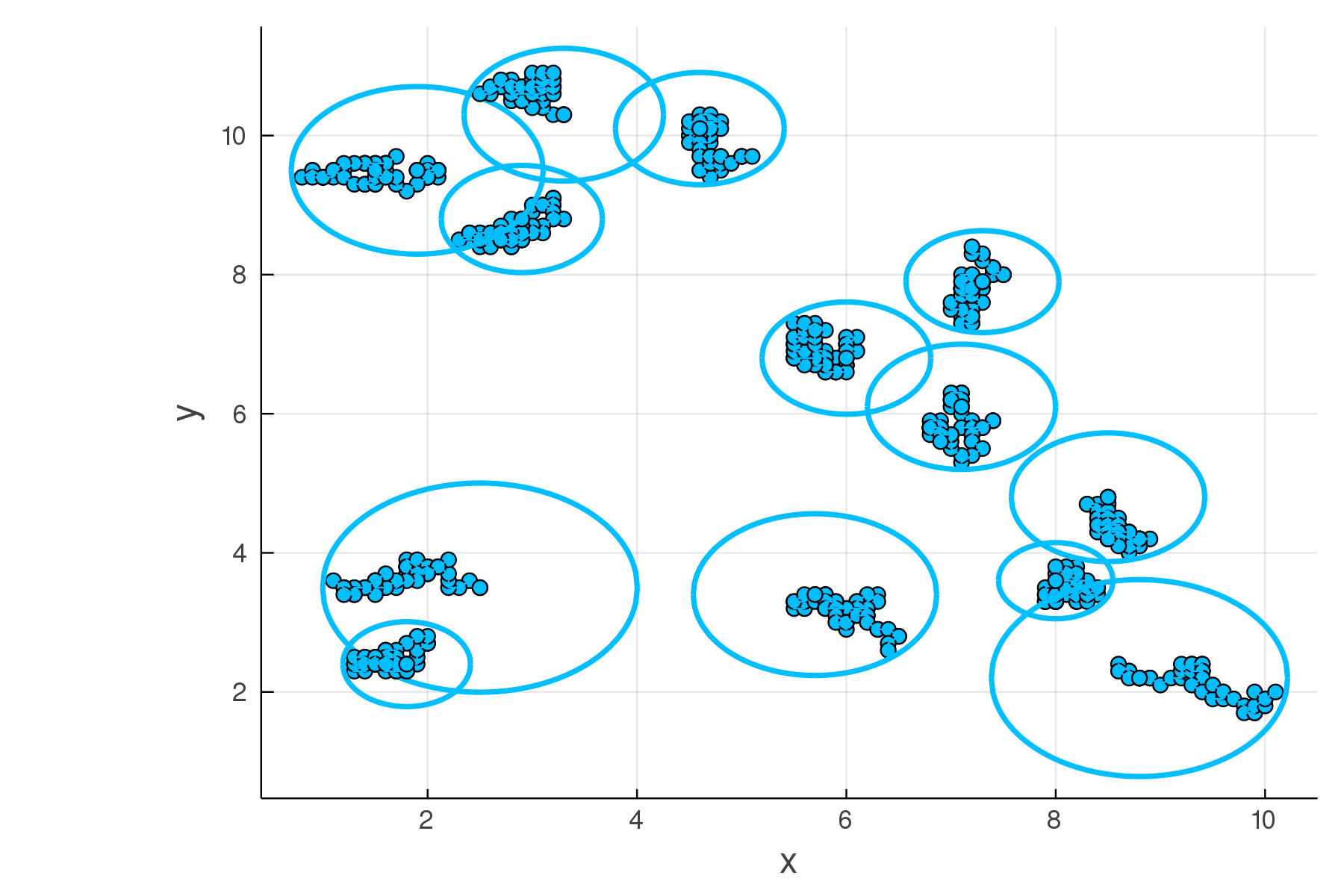}      
    \caption*{c) $r=0.1$}
    \end{subfigure}       
        \caption{Illustrations of some trajectories of random walkers in $\nset{R}^2$ for different initial conditions $\set{x_i}_{i=1}^{13}$ and different step sizes $r$ using the set $\mathcal{S}\left(\nset{R}^2, \Omega_1 \right)$.}\label{fig:02}
\end{figure}

Defining the following set

\begin{eqnarray}
\Gen_M(x_0):=\set{\Phi_r \ : \   \set{\Phi_r(x_i)}_{i=0}^{M-1} \in B(x_0,R) },
\end{eqnarray}

which may be interpreted as the set of iteration functions of random walkers defined in some set $\mathcal{S}\left(\nset{C}^n,\Omega_z \right)$, which using the initial value  $x_0$ generate a finite sequence of random variables $\set{x_i}_{i=1}^M\in B(x_0;R)$. So, considering the result of the equation \eqref{eq:2}, it is possible to obtain the following corollary (see Figure \ref{fig:02}):

\begin{corollary}
Let $\Phi_r: \nset{C}^n \to \nset{C}^n$ be an iteration function of a random walker defined in some set $\mathcal{S}\left(\nset{C}^n, \Omega_z\right)$. So, for any sequence $\set{x_i}_{i\geq 1}\in \nset{C}^n$, it is fulfilled that

\begin{eqnarray*}
 \Phi_r\in \Gen_M(x_j)  \ \forall x_j\in \set{x_i}_{i\geq 1},
\end{eqnarray*}

and therefore 

\begin{eqnarray}
\forall x_j\in \set{x_i}_{i\geq 1} \ \exists B(x_j;R_j).
\end{eqnarray}

\end{corollary}

\section{Proposed Model}

From the previous section it is possible to obtain the following observations:

\begin{itemize}

\item[$i)$] Any random walker may be considered as an iterative method through an iteration function $\Phi_r$, defined on a set $\mathcal{S}\left(\ \cdot\ , \Omega_z \right)$ with a fixed step size $r$.

\item[$ii)$] Any random walker may be characterized by its initial position $x_0$ and its iteration function $\Phi_r$, which by generating a finite trajectory, generates a finite sequence of random variables $\set{x_i}_{i=1}^M$.

\item[$iii)$] For any random walker that generates a finite sequence of random variables$\set{x_i}_{i=1}^M\in B(x_0;R)$, there exists an associated radius $R_M<R$, such that

\begin{eqnarray}
\norm{x_0-x_j}\leq R_M \hspace{0.1cm} \forall x_j\in \set{x_i}_{i=1}^M, 
\end{eqnarray}

where for an arbitrary positive constant $k$, it is fulfilled that

\begin{eqnarray}
R_M=kr.
\end{eqnarray}

\item[$iv)$] Any random walker that generates a finite sequence of random variables $\set{x_i}_{i=1}^M$, may be characterized by its initial position $x_0$ and its associated radius $R_M$.

\item[$v)$] The behavior of a random walker, without considering external forces, is invariant of the scale.

\end{itemize}

\newpage

\begin{figure}[!ht]
\centering
        \begin{subfigure}[c]{0.4\textwidth}
        \centering
 \includegraphics[width=\textwidth, height=0.65\textwidth]{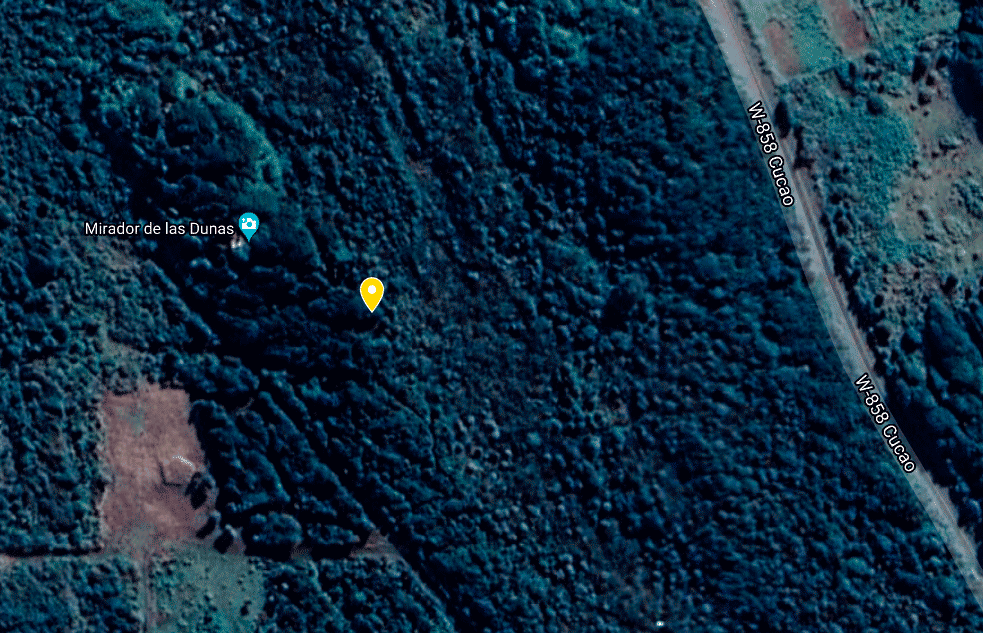}      
    \end{subfigure}
 \begin{subfigure}[c]{0.4\textwidth}
        \centering
 \includegraphics[width=\textwidth, height=0.65\textwidth]{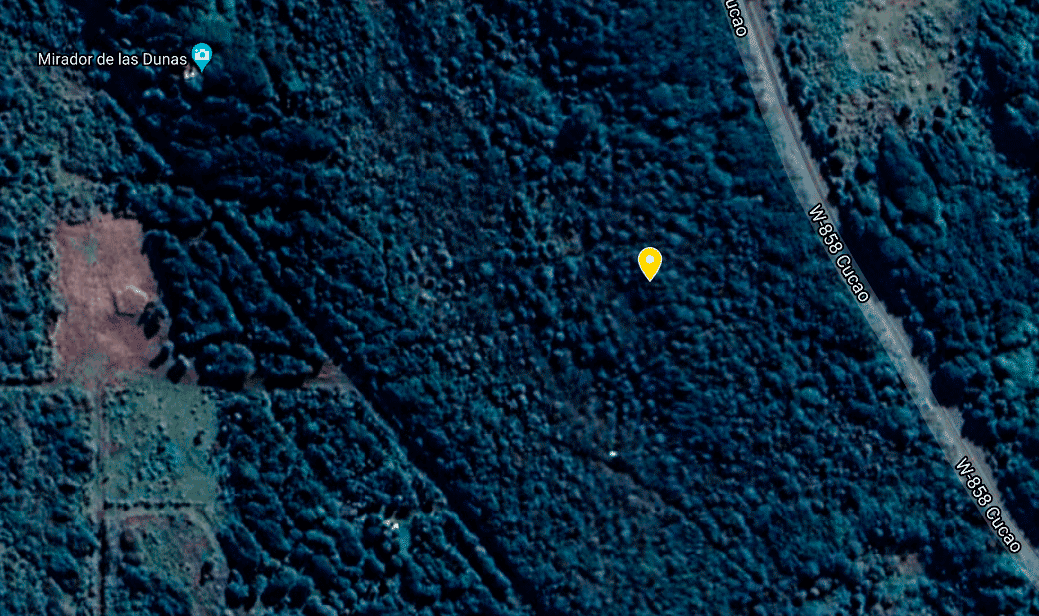}     
    \end{subfigure}  
        \caption*{a) Search for regions with possible presence of guignas}
\centering
        \begin{subfigure}[c]{0.4\textwidth}
        \centering
 \includegraphics[width=\textwidth, height=0.65\textwidth]{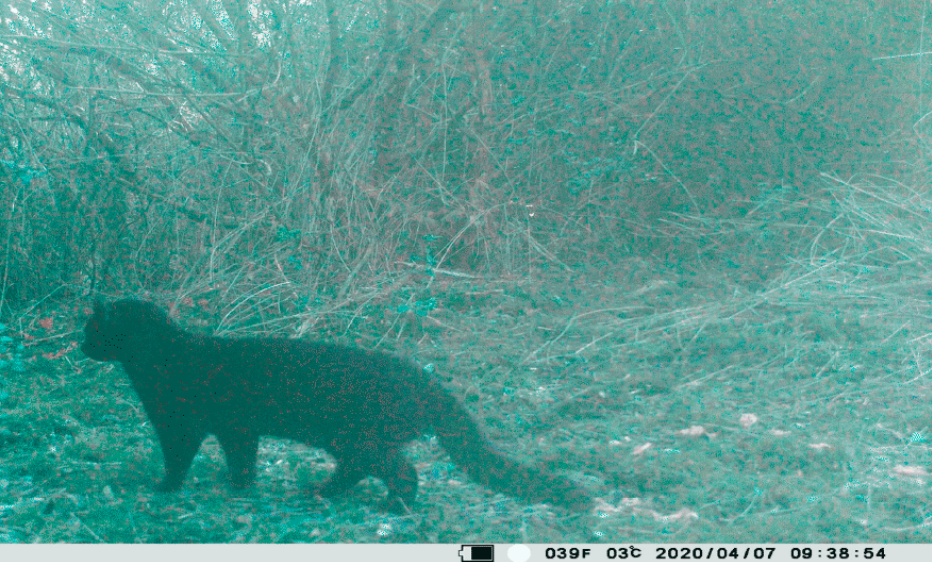}      
    \end{subfigure}
 \begin{subfigure}[c]{0.4\textwidth}
        \centering
 \includegraphics[width=\textwidth, height=0.65\textwidth]{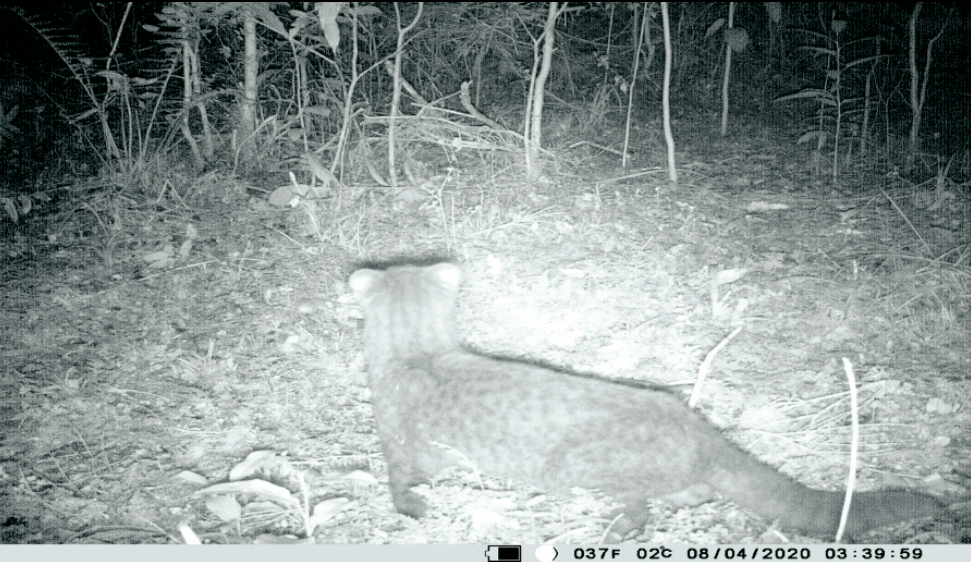}     
    \end{subfigure}  
        \caption*{b) Confirmation of the presence of guignas using trap cameras}
    \centering
        \begin{subfigure}[c]{0.4\textwidth}
        \centering
 \includegraphics[width=\textwidth, height=0.65\textwidth]{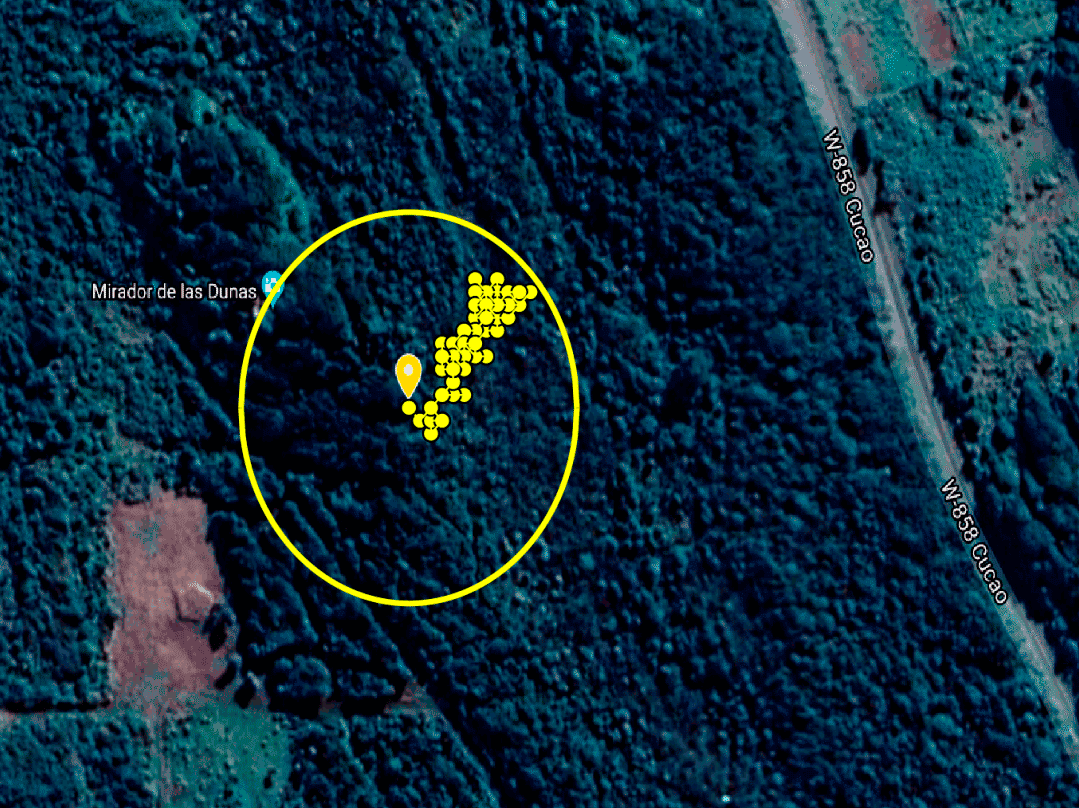}      
    \end{subfigure}
 \begin{subfigure}[c]{0.4\textwidth}
        \centering
 \includegraphics[width=\textwidth, height=0.65\textwidth]{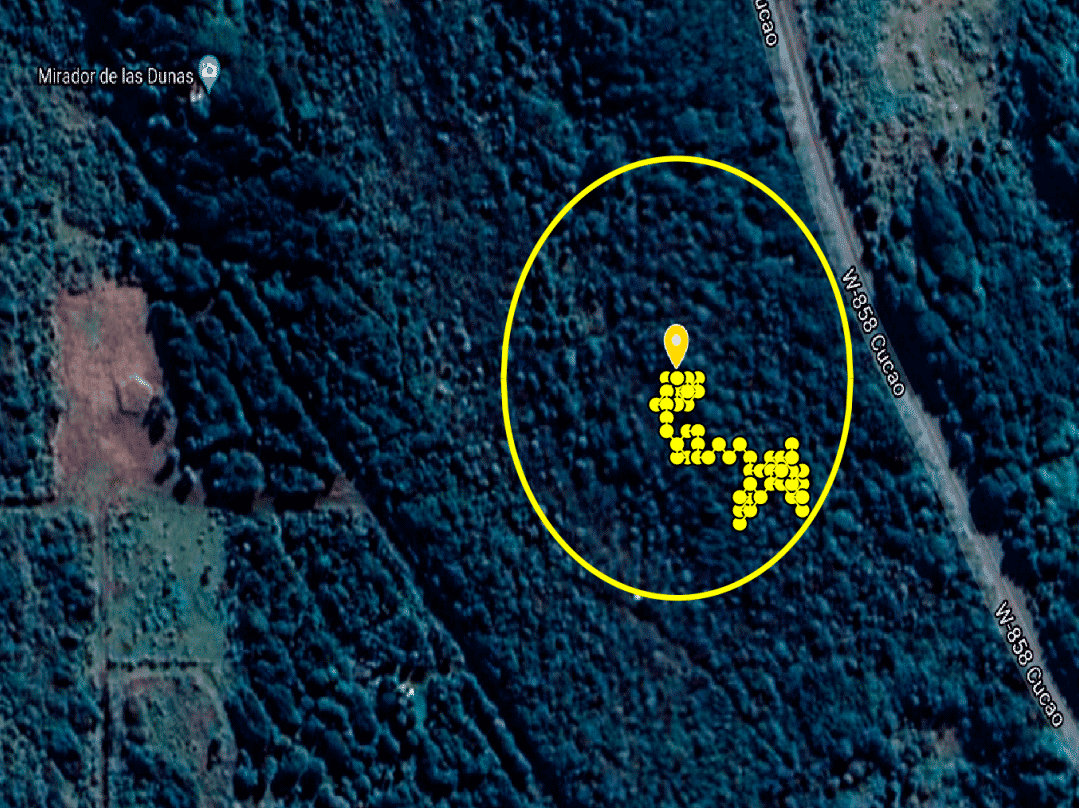}     
    \end{subfigure}  
        \caption*{c) Simulation of the behavior of some guignas using random walkers}       
        \caption{Summary of the steps necessary to simulate the behavior of guignas in specific areas using random walkers, it is necessary to mention that the images in part $c)$ are analogous to the data that would be obtained when using radiocollars in animals (see reference \cite{lopez2021free}).}\label{fig:03}
\end{figure}

So, considering the previous observations, it is possible to abstract the behavior of a sedentary animal as an iterative method that starts from a fixed initial condition (such as a burrow), and that generates a random trajectory 
(when searching for food in a specific area of a given territory) which is bounded by a circumference (see Figure \ref{fig:03}). The previous description coincides with a random walker, however, to generate data through random walkers that may be useful to simulate the behavior of guignas in a region that allow estimating the mean distances that they can travel in a suitable unit of measure, as well as the interactions that they may have with points that represent a high probability of lethality near their territory, it is necessary to have real initial positions in a region of interest where the presence of populations of guignas is confirmed. Confirmation of the presence of guignas may be done through camera traps or records of sightings of people living near the area of interest, however, to carry out the simulation initial positions may be generated considering regions with possible presence of guignas or through pseudorandom numbers, which may be considered as test data to analyze the interactions that they may have in the territory.

\subsection{Example}

Before continuing it is necessary to define the following set:

\begin{eqnarray}
Predators:=\set{x \ : \ x  \ \mbox{represents any external agent that is lethal to guignas}}, 
\end{eqnarray}

with which it is possible to generalize all the points of a territory that are potentially fatal for the guignas, such as urban areas, roads, packs of dogs, etc.

\begin{figure}[!ht]
        \begin{subfigure}[c]{0.43\textwidth}
        \centering
 \includegraphics[width=\textwidth, height=0.65\textwidth]{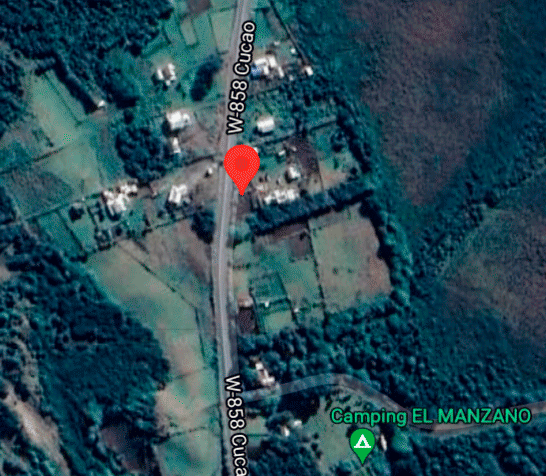}      
    \caption*{a) Urban areas}
    \end{subfigure}
        \begin{subfigure}[c]{0.43\textwidth}
        \centering
 \includegraphics[width=\textwidth, height=0.65\textwidth]{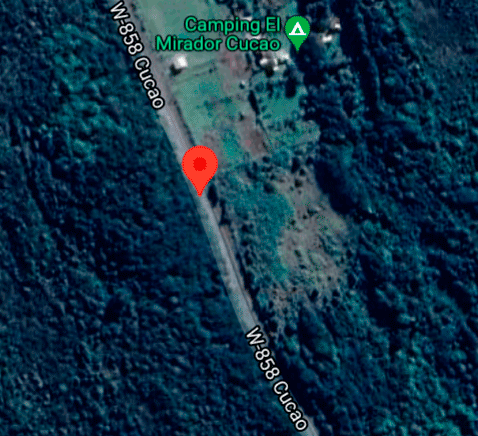}     
    \caption*{b) Roads}
    \end{subfigure} 
    \centering
    \begin{subfigure}[c]{0.43\textwidth}
    \centering
 \includegraphics[width=\textwidth, height=0.65\textwidth]{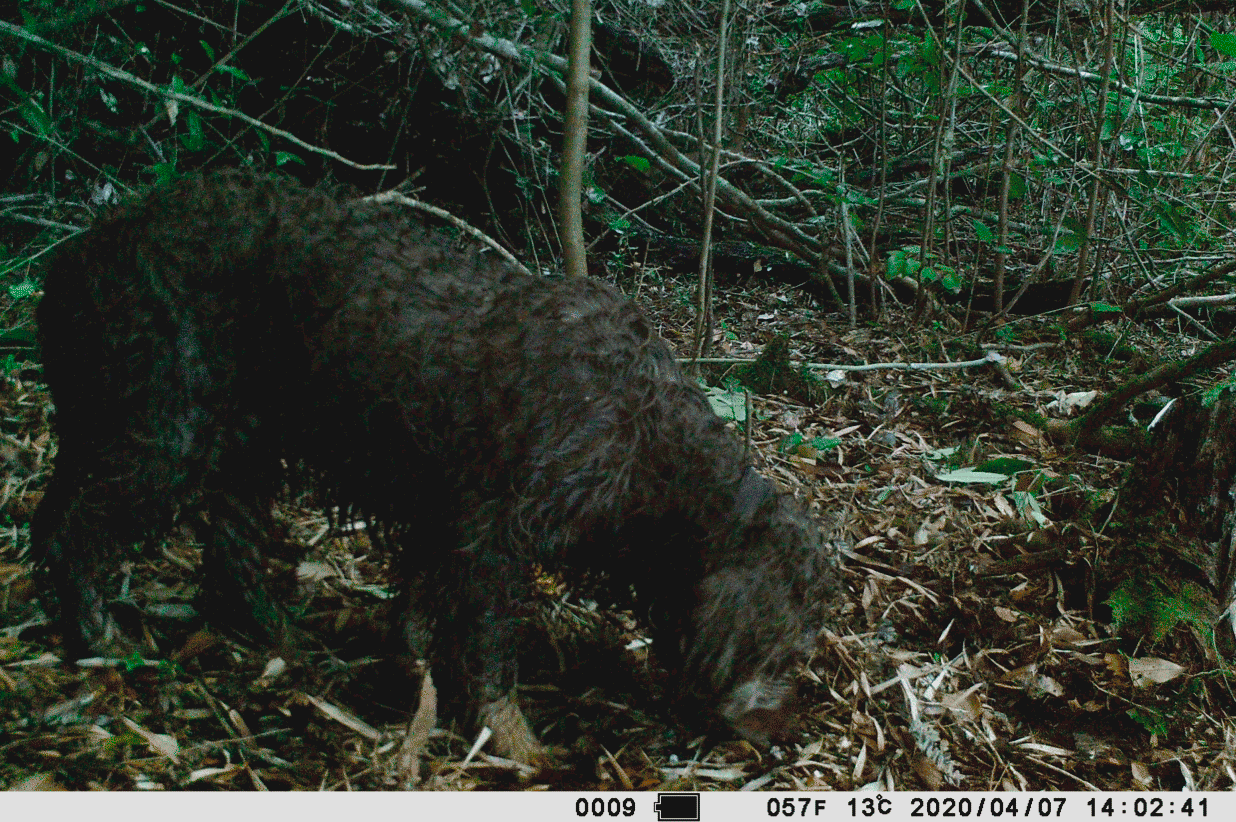}      
    \end{subfigure}
    \begin{subfigure}[c]{0.43\textwidth}
    \centering
 \includegraphics[width=\textwidth, height=0.65\textwidth]{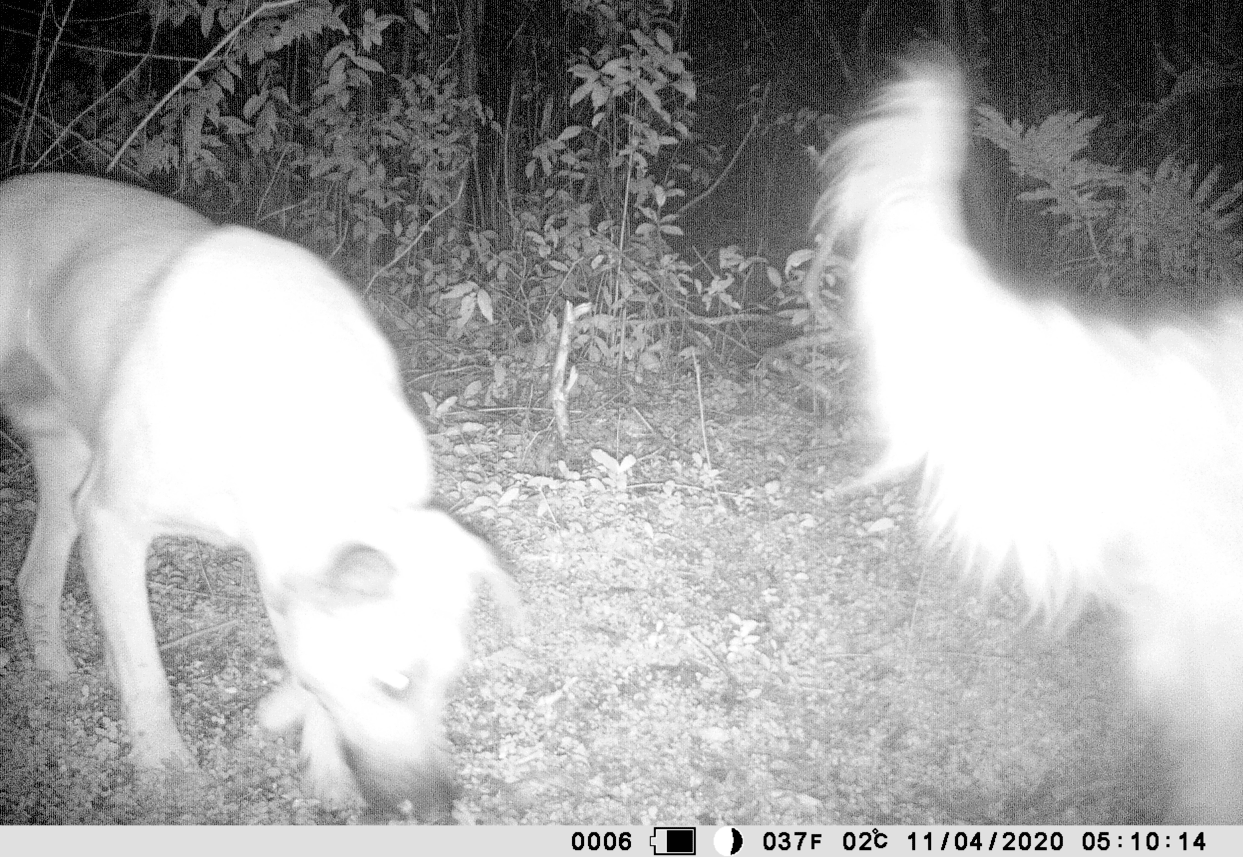}      
    \end{subfigure}   
        \caption*{c) Packs of dogs}     
        \caption{Images of some elements belonging to the set Predators.}
\end{figure}

To carry out the simulation, it is necessary to generate a record of real positions of points belonging to the set Predators in a given territory with the presence of guignas (see Table \ref{tab:01}).

\begin{footnotesize}
\begin{longtable}{c|ccc}
\toprule
    Predators & Latitude & Longitude \\
\midrule
    Predator 1 & -42.6149730 & -74.1147730 \\
    Predator 2 & -42.6208730 & -74.1121220 \\
    Predator 3 & -42.6228700 & -74.1112320 \\
    Predator 4 & -42.6213250 & -74.1049990 \\
    Predator 5 & -42.6142930 & -74.1153090 \\
    Predator 6 & -42.6219000 & -74.1115460 \\
    Predator 7 & -42.6160470 & -74.1140780 \\
    Predator 8 & -42.6194878 & -74.1129435 \\
    Predator 9 & -42.6181298 & -74.1137374 \\
    Predator 10 & -42.6171034 & -74.1145528 \\
    Predator 11 & -42.6161402 & -74.1119564 \\
    Predator 12 & -42.6164560 & -74.1087378 \\
    Predator 13 & -42.6147822 & -74.1092313 \\
    Predator 14 & -42.6161244 & -74.1059483 \\ 
    \bottomrule
    \caption{Coordinates of the predators.}\label{tab:01}
\end{longtable}
\end{footnotesize}

\begin{figure}[!ht]
        \begin{subfigure}[c]{0.43\textwidth}
        \centering
 \includegraphics[width=\textwidth, height=0.65\textwidth]{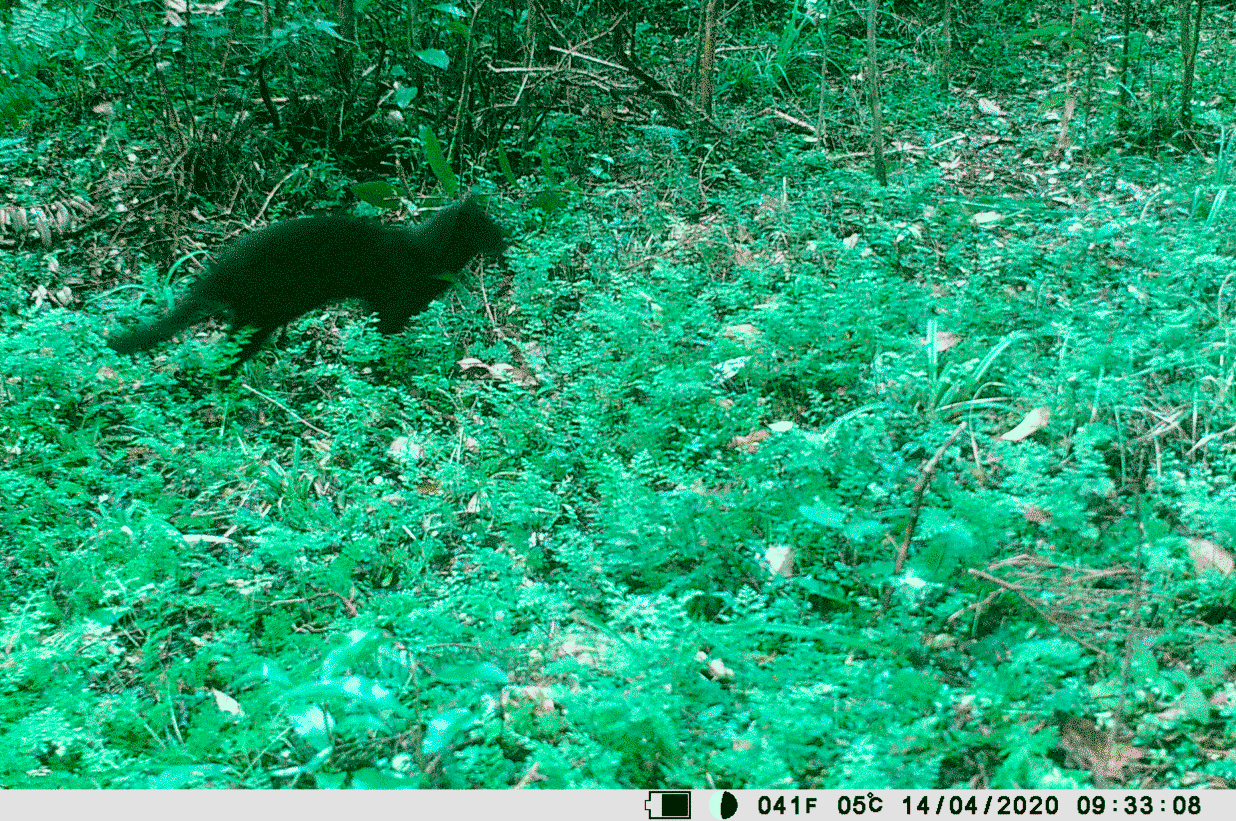}      
    \end{subfigure}
        \begin{subfigure}[c]{0.43\textwidth}
        \centering
 \includegraphics[width=\textwidth, height=0.65\textwidth]{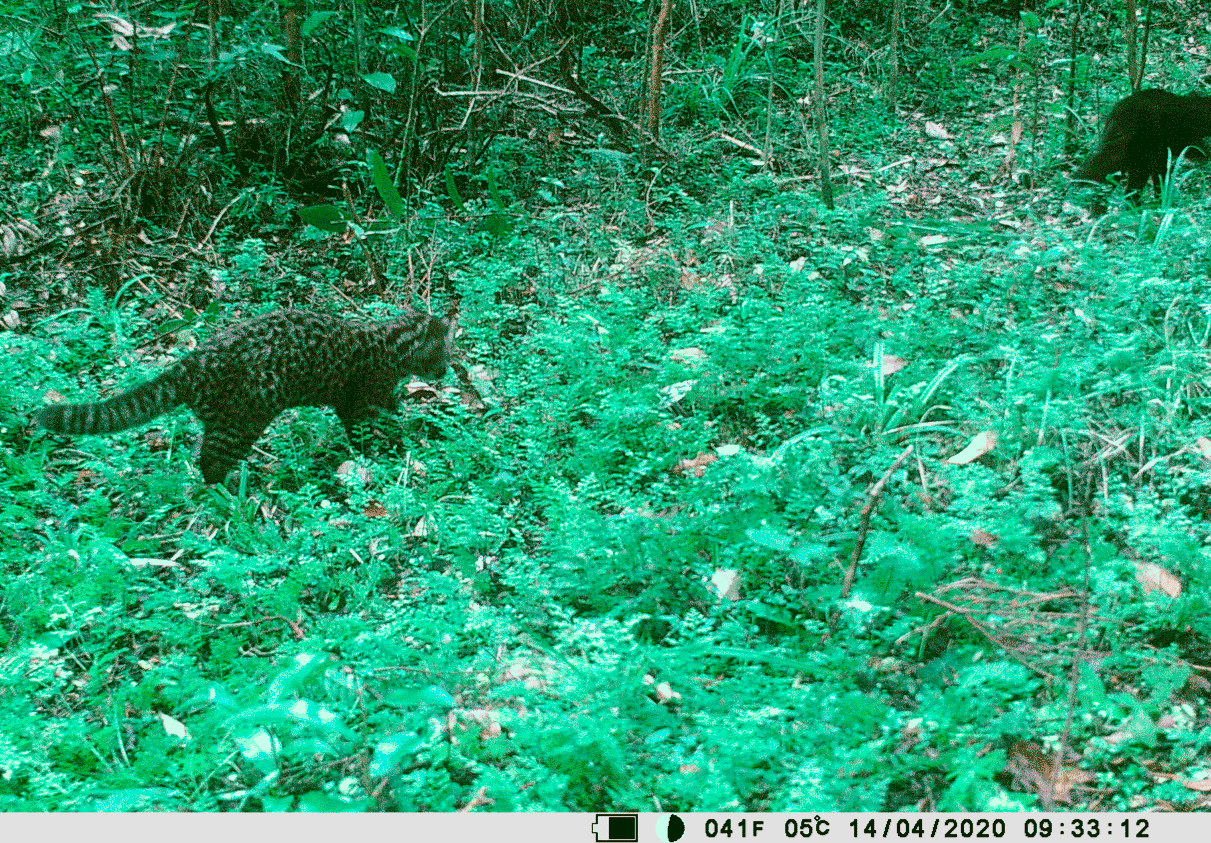}     
    \end{subfigure} 
    \centering
    \begin{subfigure}[c]{0.43\textwidth}
    \centering
 \includegraphics[width=\textwidth, height=0.65\textwidth]{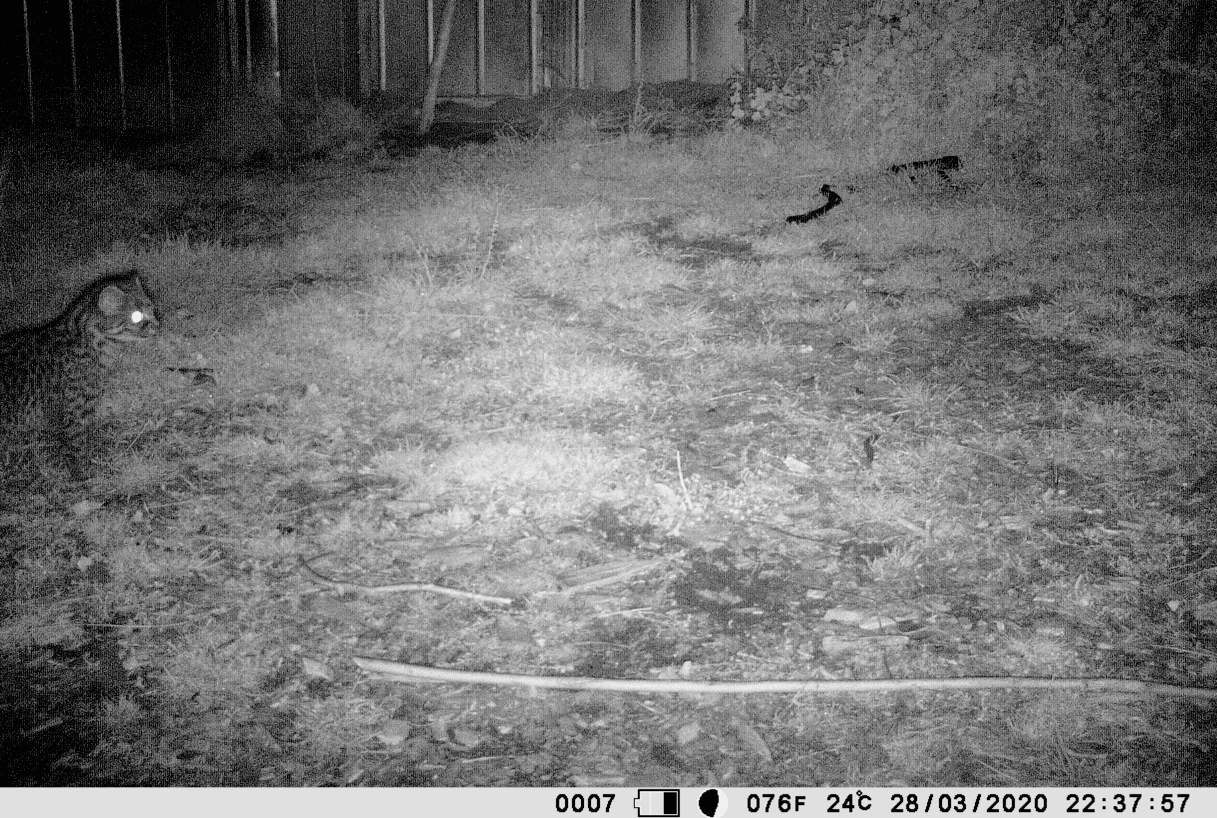}      
    \end{subfigure}
    \begin{subfigure}[c]{0.43\textwidth}
    \centering
 \includegraphics[width=\textwidth, height=0.65\textwidth]{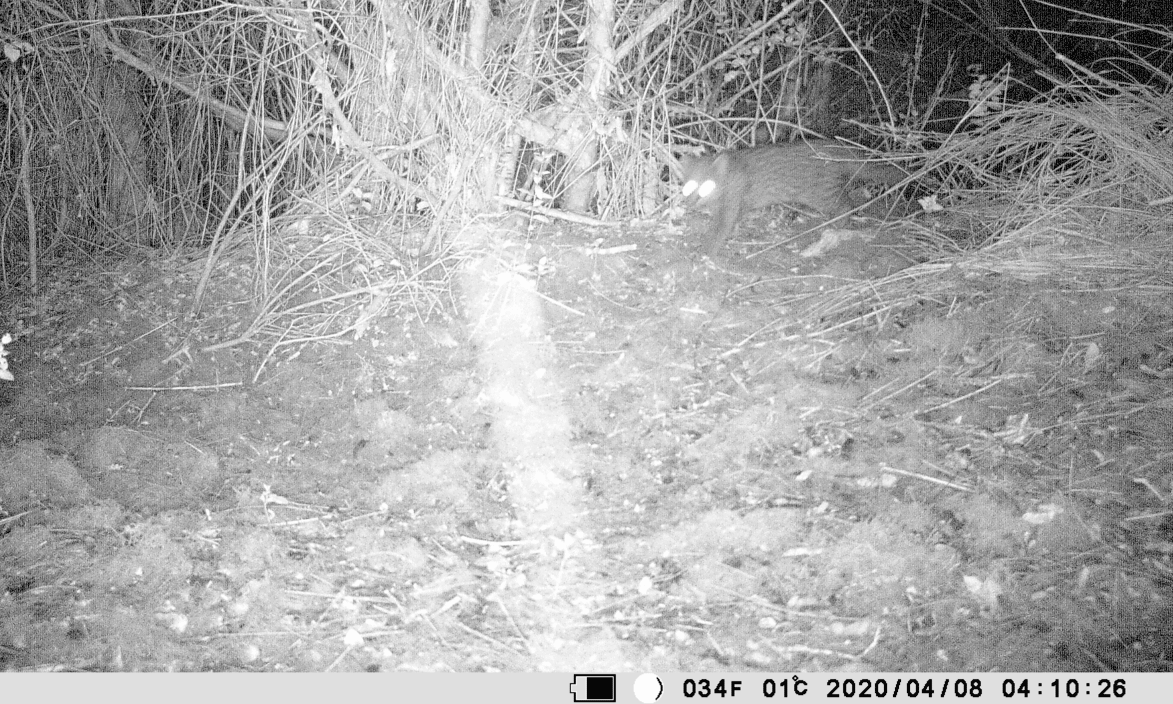}      
    \end{subfigure}    
        \caption{Images of some guignas in the territory of interest for the simulation.}\label{fig:04}
\end{figure}

Analogously, it is necessary to generate a record of real positions of points representing potential territories of guignas, which may be obtained through camera trap records (see Figure \ref{fig:04}). However, in the absence of records confirming the presence of guignas, test data generated through pseudorandom numbers may be used. Due to the small number of available records, a combination of real data and test data is used in the Table \ref{tab:02}.

\begin{footnotesize}
\begin{longtable}{c|ccc}
\toprule
    Guignas & Latitude & Longitude \\
\midrule
    Guigna 1 & -42.6215620 & -74.1139580 \\
    Guigna 2 & -42.6220470 & -74.1124180 \\
    Guigna 3 & -42.6185106 & -74.1069997 \\
    Guigna 4 & -42.6174691 & -74.1117204 \\
    Guigna 5 & -42.6211949 & -74.1090382 \\
    Guigna 6 & -42.6166947 & -74.1166342 \\
    Guigna 7 & -42.6145629 & -74.1123856 \\
    Guigna 8 & -42.6202633 & -74.1143812 \\
    Guigna 9 & -42.6190475 & -74.1099823 \\
    \bottomrule
    \caption{Coordinates of the guignas.}\label{tab:02}
\end{longtable}
\end{footnotesize}

\begin{figure}[!ht]
\centering
        \begin{subfigure}[c]{0.8\textwidth}
        \centering
 \includegraphics[width=\textwidth, height=0.55\textwidth]{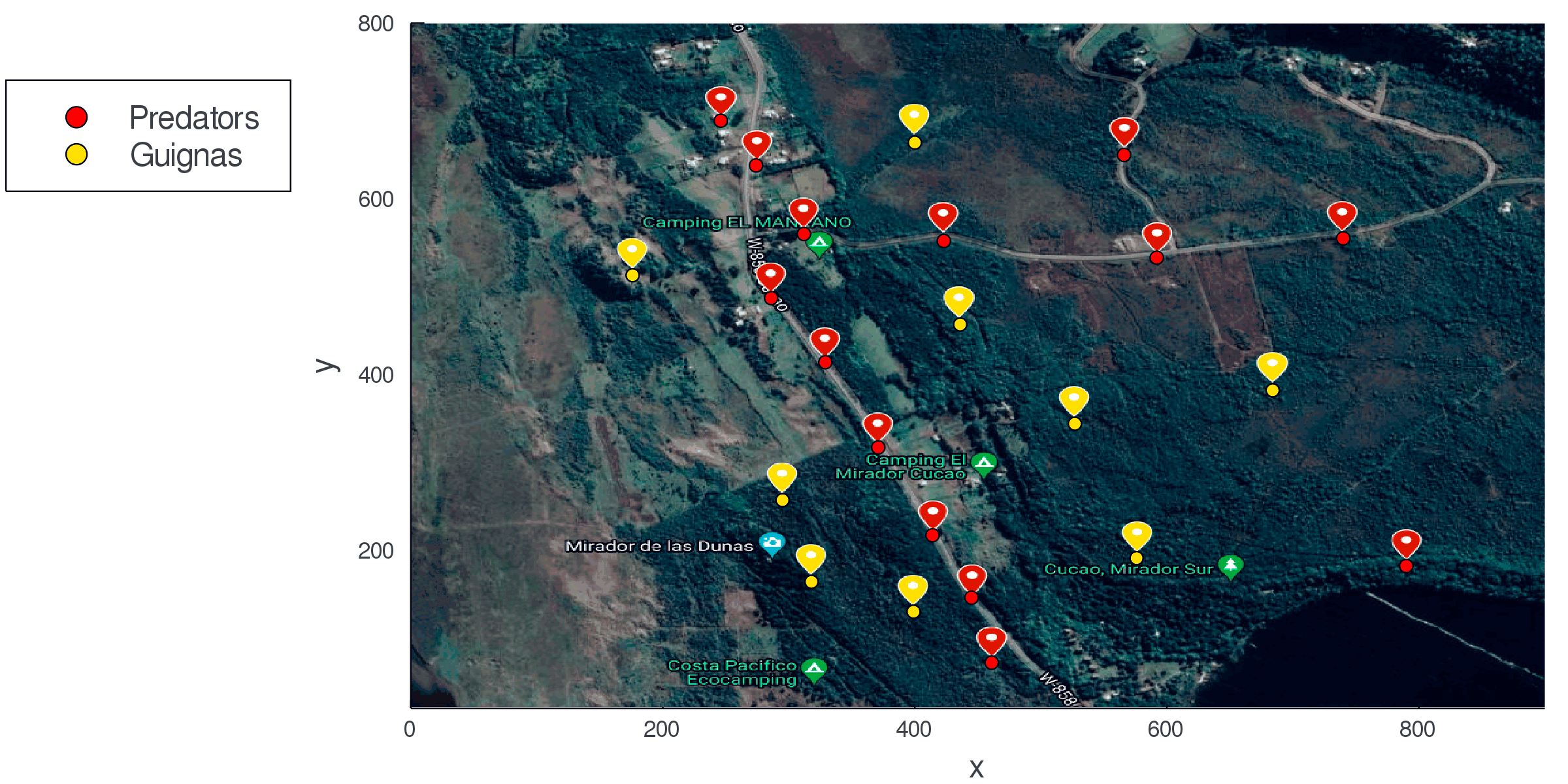}   
         \caption*{a) Generating a distribution of predators and of guignas in a region of interest}   
    \end{subfigure}
\centering
        \begin{subfigure}[c]{0.8\textwidth}
        \centering
 \includegraphics[width=\textwidth, height=0.55\textwidth]{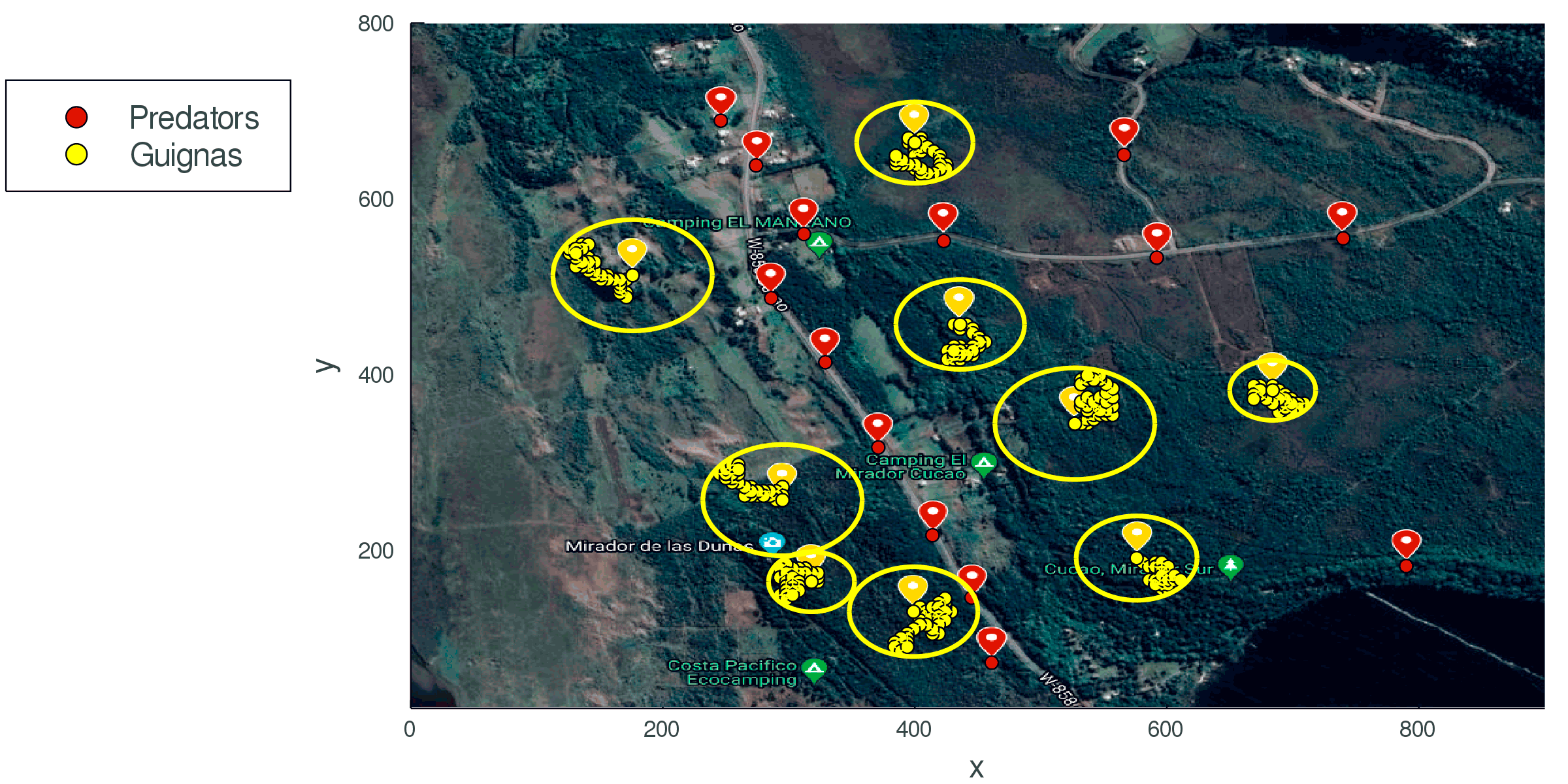}   
         \caption*{b) Simulation of the behavior of some guignas using random walkers}      
    \end{subfigure}   
        \caption{Summary of the steps to follow to generate a simulation of a population of guignas in a given region through random walkers.}\label{fig:05}
\end{figure}

Using the coordinates of the predators and of the guignas, the distribution of points that will be analyzed in the simulation is generated with the help of a satellite image (see Figure \ref{fig:05}). So, a numerical scale is assigned in the satellite image and the values that will be used to generate the iteration function of the random walkers are specified, it is necessary to mention that for this example the following values were chosen:

\begin{eqnarray}
r=5,  & z=2, & M=200,
\end{eqnarray}

with which the following iteration function is obtained to generate the simulation of the behavior of a population of guignas in a given region through random walkers:

\begin{eqnarray}
\Phi_5(x)=x+5s \hspace{0.1cm} \mbox{ with } \hspace{0.1cm} s\in \mathcal{S}\left(\nset{R}^2,\Omega_2 \right) \hspace{0.1cm} \mbox{ and } \hspace{0.1cm} \Phi_5\in \Gen_{200}(x_0),
\end{eqnarray}

for each initial position $x_0$ that represents a guigna.

\newpage

\begin{figure}[!ht]
\centering
        \begin{subfigure}[c]{0.33\textwidth}
        \centering
 \includegraphics[width=\textwidth, height=0.85\textwidth]{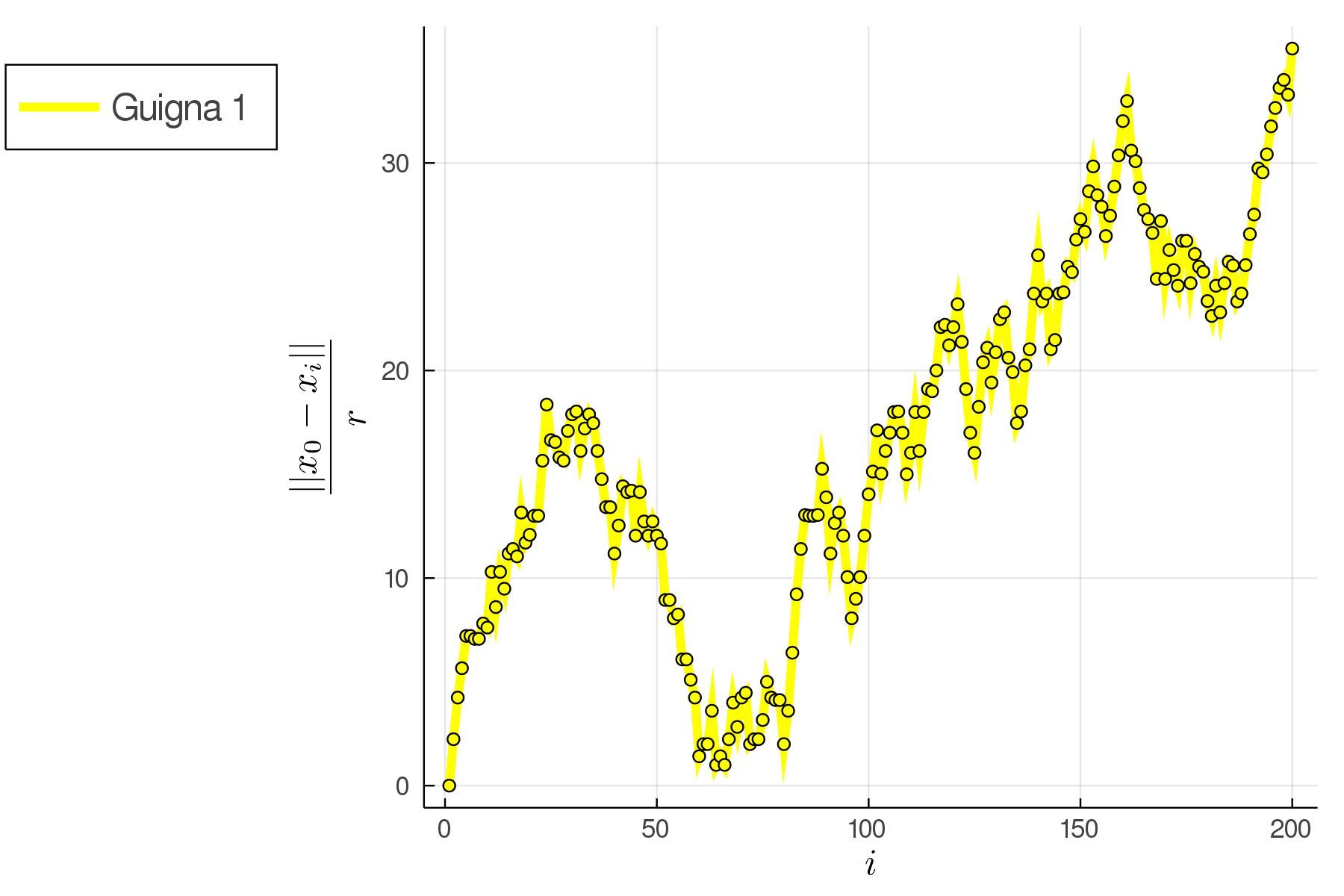}  
         \caption*{a) Sequence of distances of the guigna 1}    
    \end{subfigure}
        \begin{subfigure}[c]{0.33\textwidth}
        \centering
 \includegraphics[width=\textwidth, height=0.85\textwidth]{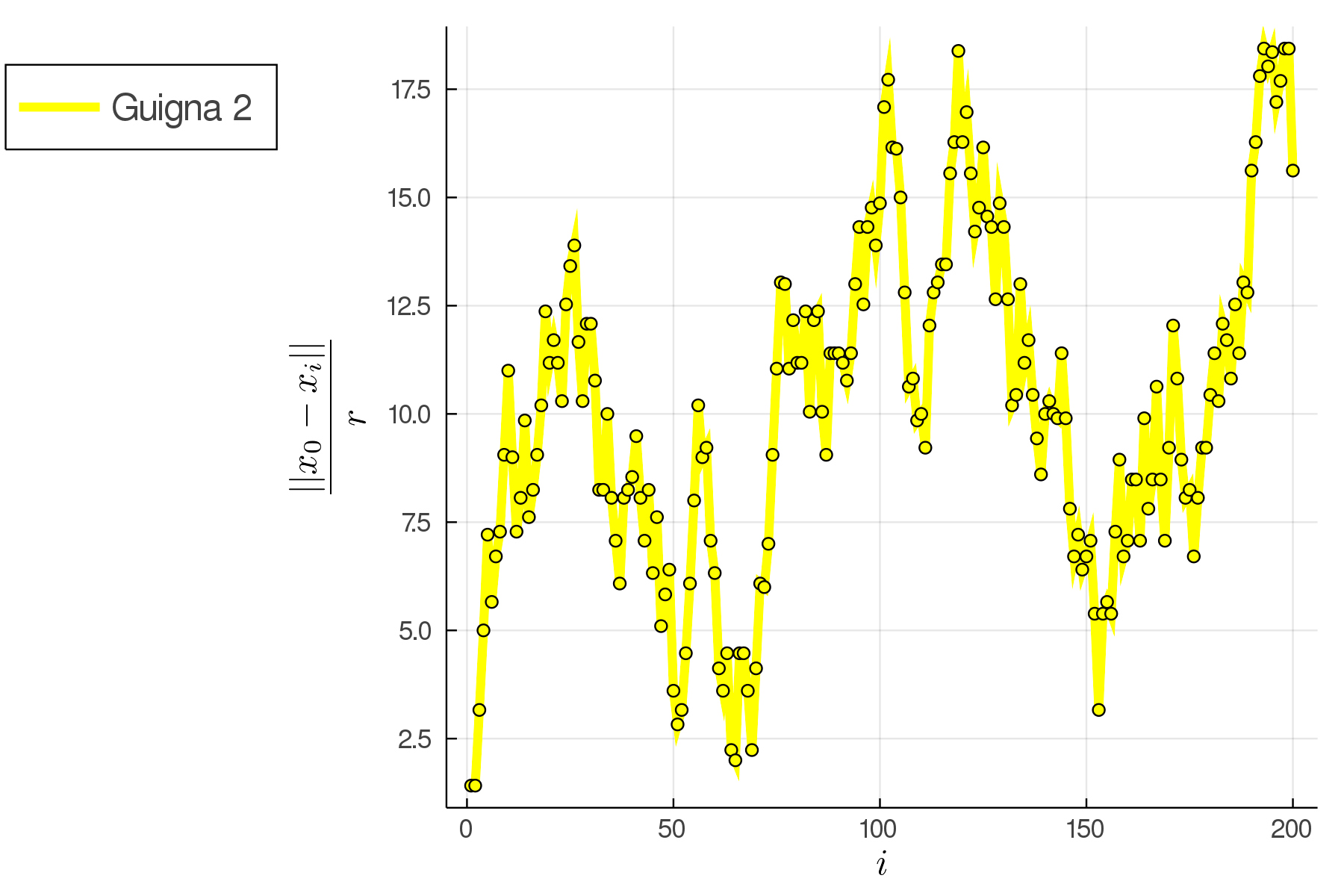}  
         \caption*{b) Sequence of distances of the guigna 2}      
    \end{subfigure} 
        \begin{subfigure}[c]{0.33\textwidth}
        \centering
 \includegraphics[width=\textwidth, height=0.85\textwidth]{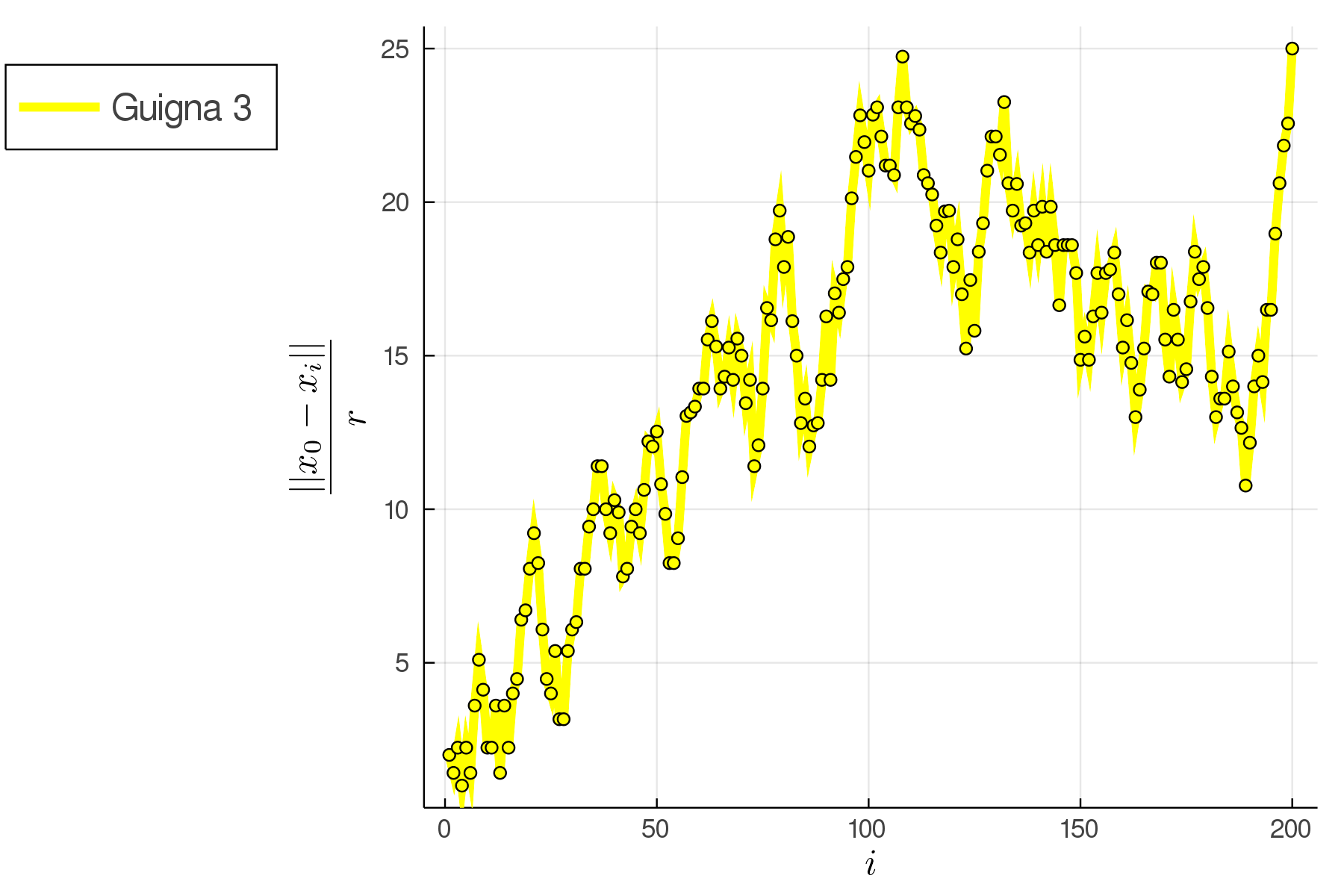}    
         \caption*{c) Sequence of distances of the guigna 3}       
    \end{subfigure} 
\centering
        \begin{subfigure}[c]{0.33\textwidth}
        \centering
 \includegraphics[width=\textwidth, height=0.85\textwidth]{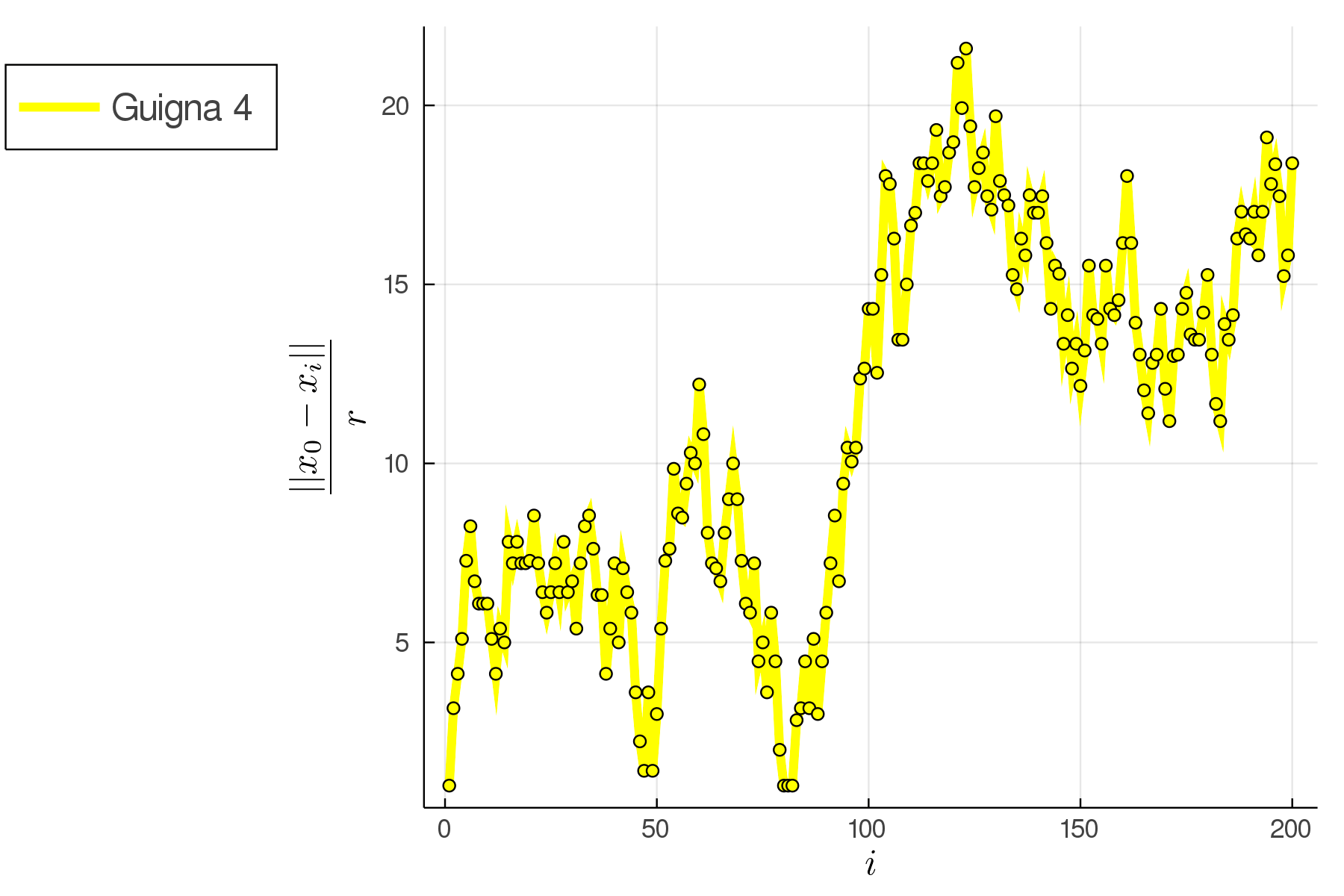}  
         \caption*{d) Sequence of distances of the guigna 4}    
    \end{subfigure}
        \begin{subfigure}[c]{0.33\textwidth}
        \centering
 \includegraphics[width=\textwidth, height=0.85\textwidth]{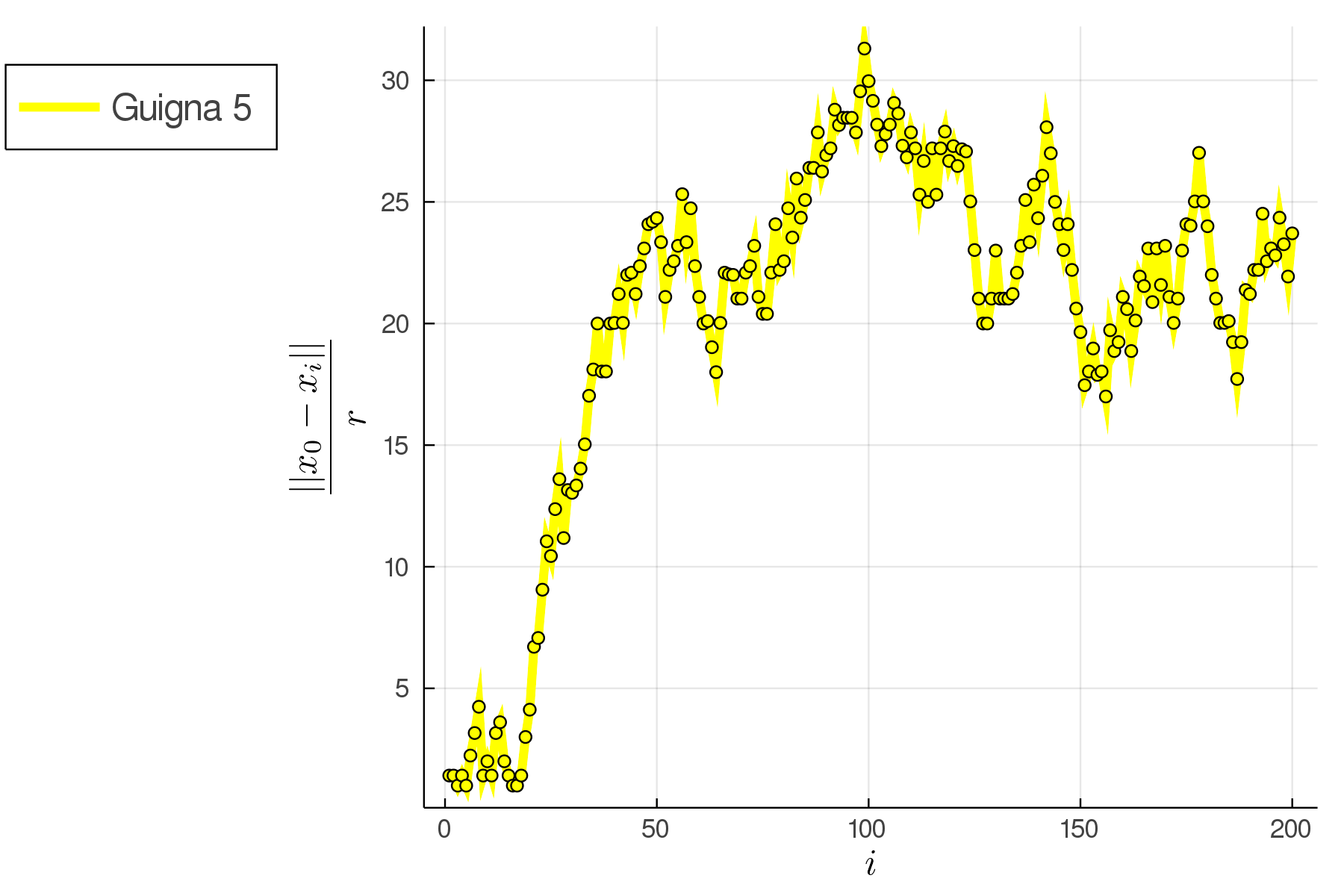}  
         \caption*{e) Sequence of distances of the guigna 5}      
    \end{subfigure} 
        \begin{subfigure}[c]{0.33\textwidth}
        \centering
 \includegraphics[width=\textwidth, height=0.85\textwidth]{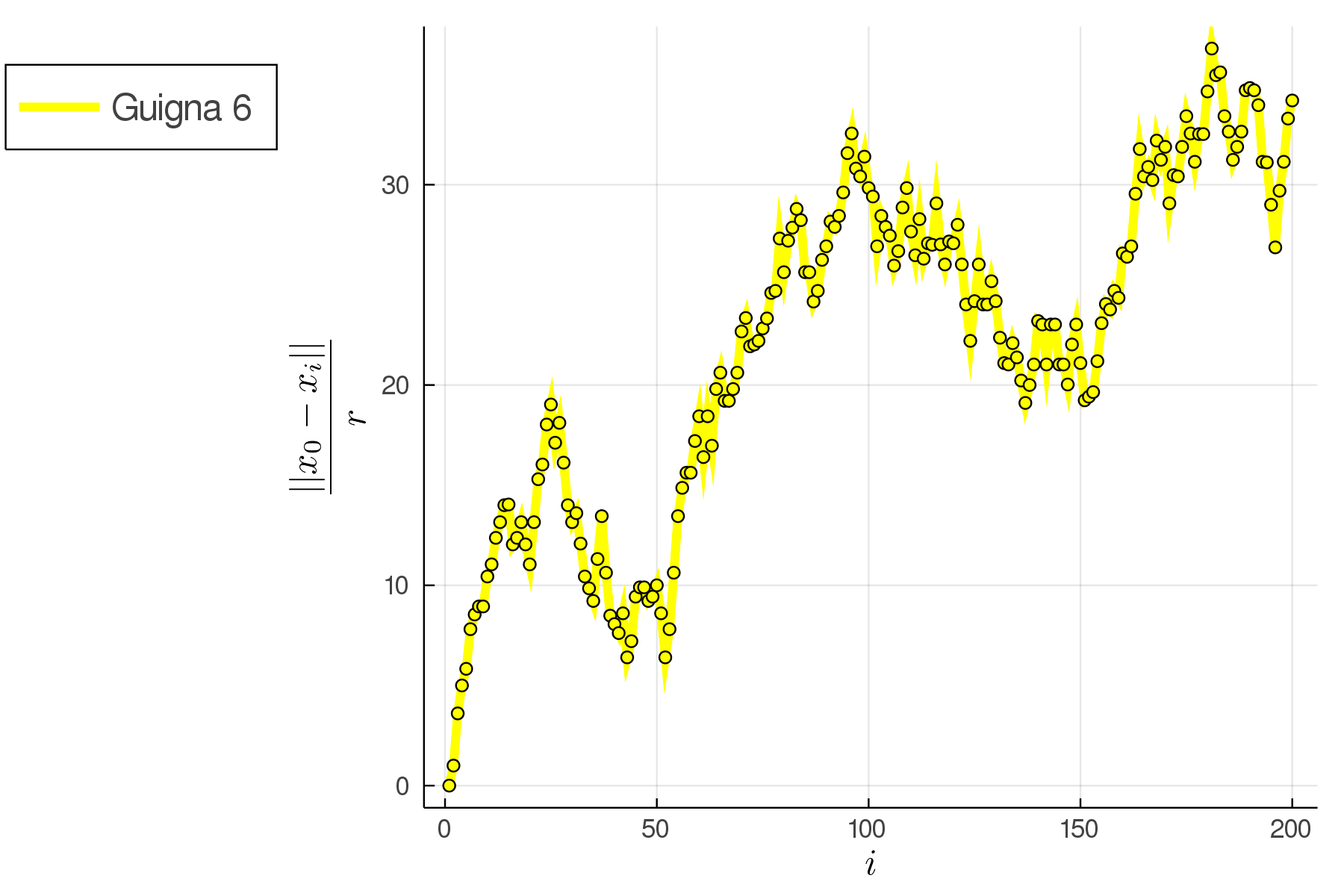}    
         \caption*{f) Sequence of distances of the guigna 6}       
    \end{subfigure} 
\centering
        \begin{subfigure}[c]{0.33\textwidth}
        \centering
 \includegraphics[width=\textwidth, height=0.85\textwidth]{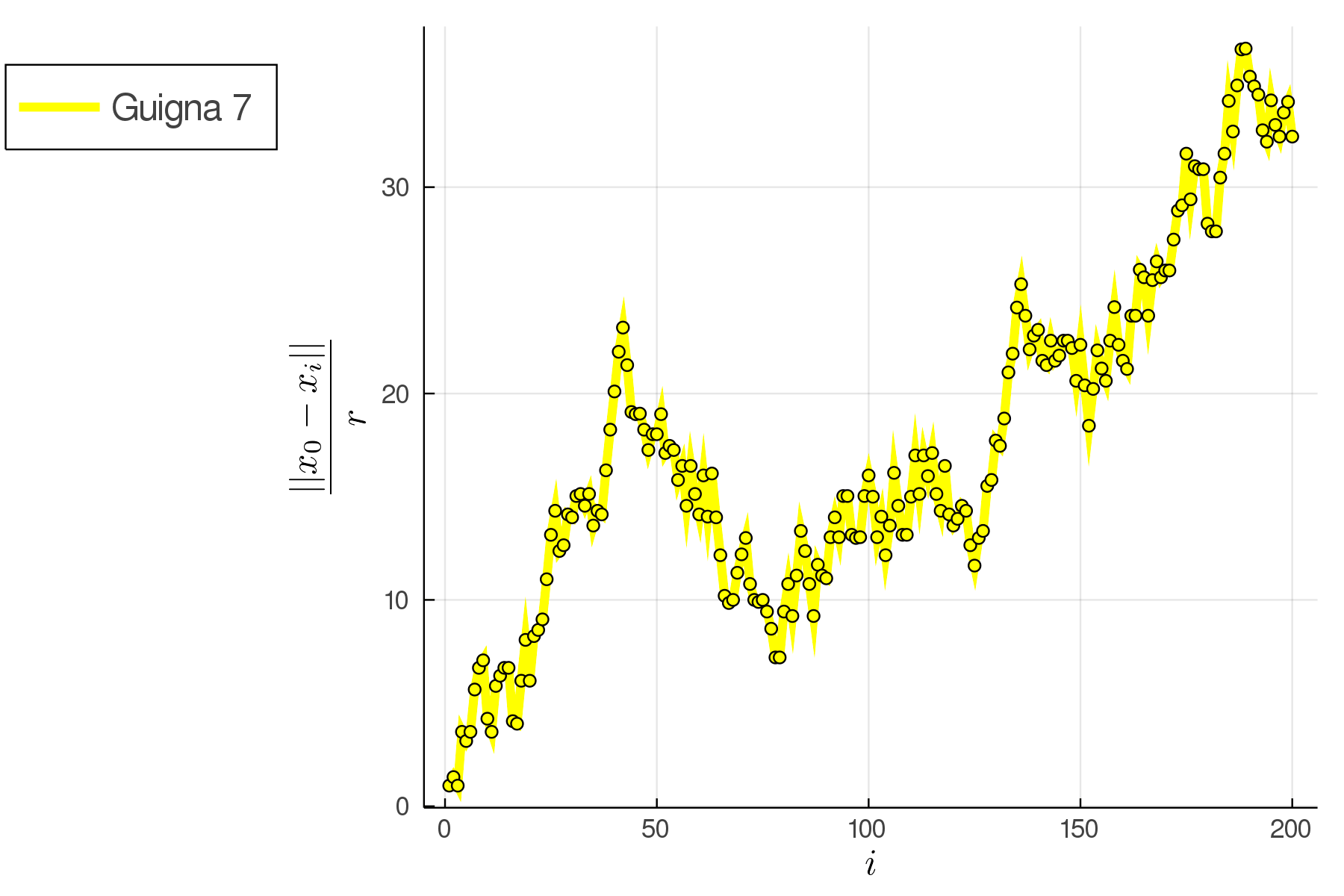}  
         \caption*{g) Sequence of distances of the guigna 7}    
    \end{subfigure}
        \begin{subfigure}[c]{0.33\textwidth}
        \centering
 \includegraphics[width=\textwidth, height=0.85\textwidth]{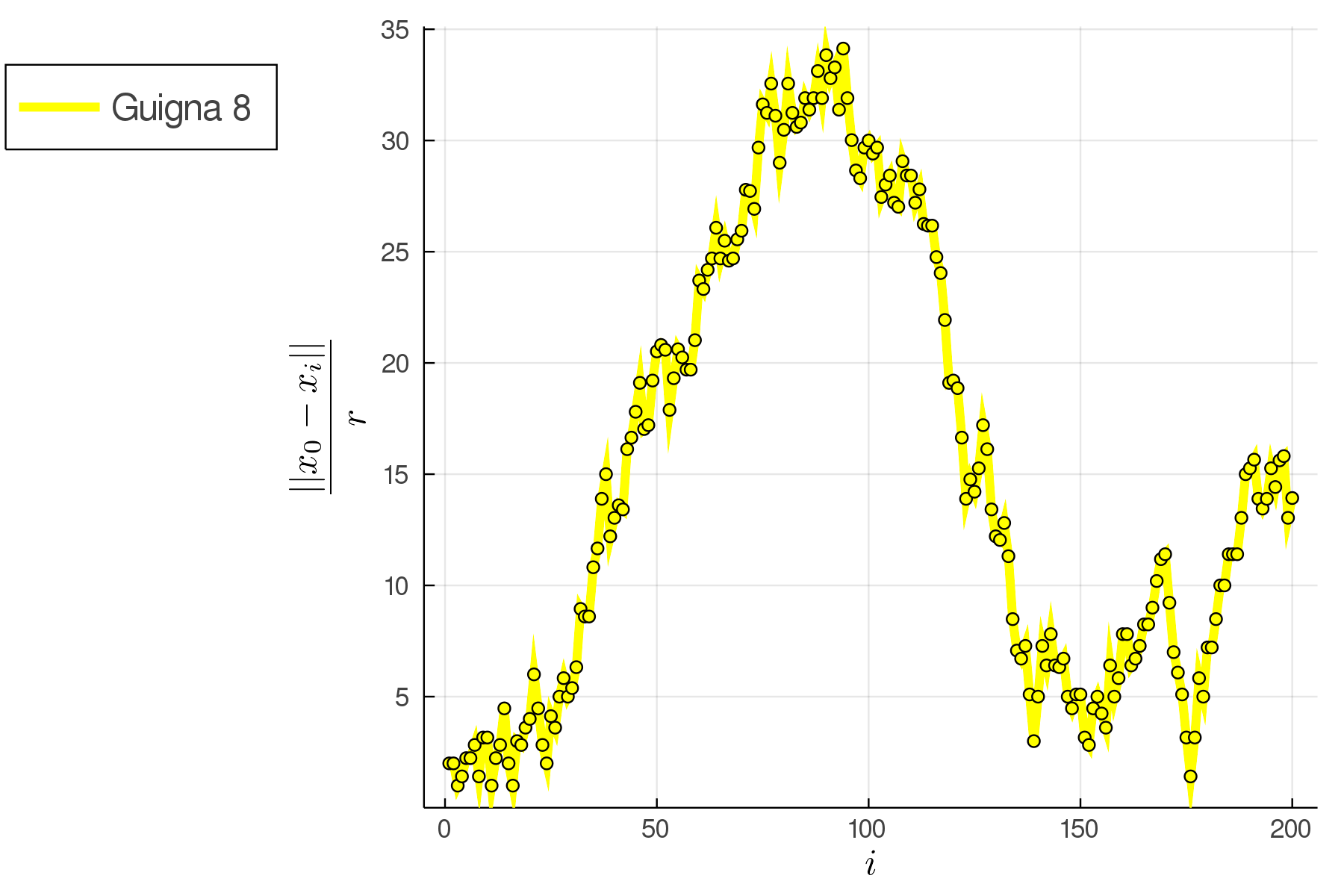}  
         \caption*{h) Sequence of distances of the guigna 8}      
    \end{subfigure} 
        \begin{subfigure}[c]{0.33\textwidth}
        \centering
 \includegraphics[width=\textwidth, height=0.85\textwidth]{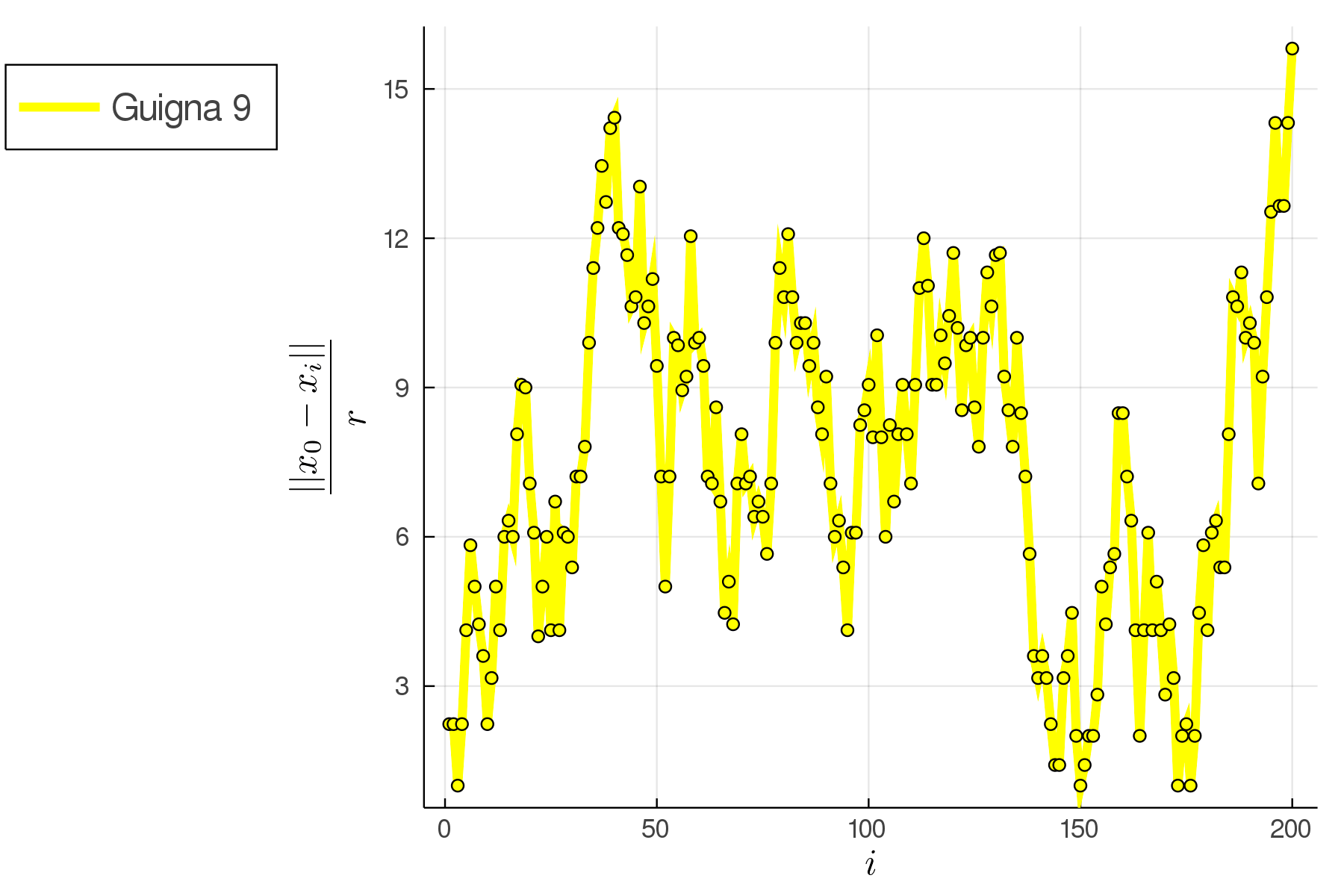}    
         \caption*{i) Sequence of distances of the guigna 9}       
    \end{subfigure}  
        \caption{Graphs of some sequences of distances $\set{\frac{\norm{x_0-x_i}}{r}}_{i=1}^M$ of the different guignas.}\label{fig:06}
\end{figure}

So, since $\Phi_r\in \Gen_{200}(x_0)$, for each initial condition $x_0$ it is possible to generate a finite trajectory of random variables $\set{x_i}_{i=1}^{200}$ that emulates the behavior of a guigna when looking for food in its territory. On the other hand, considering that the behavior of random walkers is invariant of the scale when external forces are not considered, it is more convenient to work with a scale of distances that depends on the step size $r$ chosen for the random walkers (see Figure \ref{fig:06}), which may be obtained through the following result:

\begin{eqnarray}
\mbox{If }\Phi_r\in \Gen_{M}(x_0) \ \Rightarrow \ \forall x_j\in \set{x_i}_{i=1}^{M} \hspace{0.2cm} \exists k_j=\dfrac{\norm{x_0-x_j}}{r},
\end{eqnarray}

where $k_j$ represents the number of steps $r$ that a random walker must travel in a straight line to get from the initial position $x_0$ to the position $x_j$.

\newpage

It is necessary to mention that the objective of the simulation is to generate a distribution of points over a given territory, with the objective of emulating the data received from radiocollars when they are placed on animals in a way analogous to that shown in the reference \cite{lopez2021free}. So, considering an iteration function of a random walker $\Phi_r$ that fulfills the following condition

\begin{eqnarray}
\Phi_r\in \Gen_m(x_0),
\end{eqnarray}

it is possible to generate multiple finite sequences of random variables $\set{x_i}_{i=1}^m$ from the same initial condition $x_0$, that is,

\begin{eqnarray}
\set{x_{i_1}}_{i_1=1}^m, \ \set{x_{i_2}}_{i_2=1}^m,\  \cdots, \ \set{x_{i_N}}_{i_N=1}^m,
\end{eqnarray}

which subsequently may be concatenated to generate one sequence of random variables as follows

\begin{eqnarray}\label{eq:3}
\set{\set{x_{i_1}}_{i_1=1}^m, \set{x_{i_2}}_{i_2=1}^m,\cdots , \set{x_{i_N}}_{i_N=1}^m}:=\set{x_i}_{i=1}^{M}.
\end{eqnarray}

\begin{figure}[!ht]
\centering
        \begin{subfigure}[c]{0.33\textwidth}
        \centering
 \includegraphics[width=\textwidth, height=0.85\textwidth]{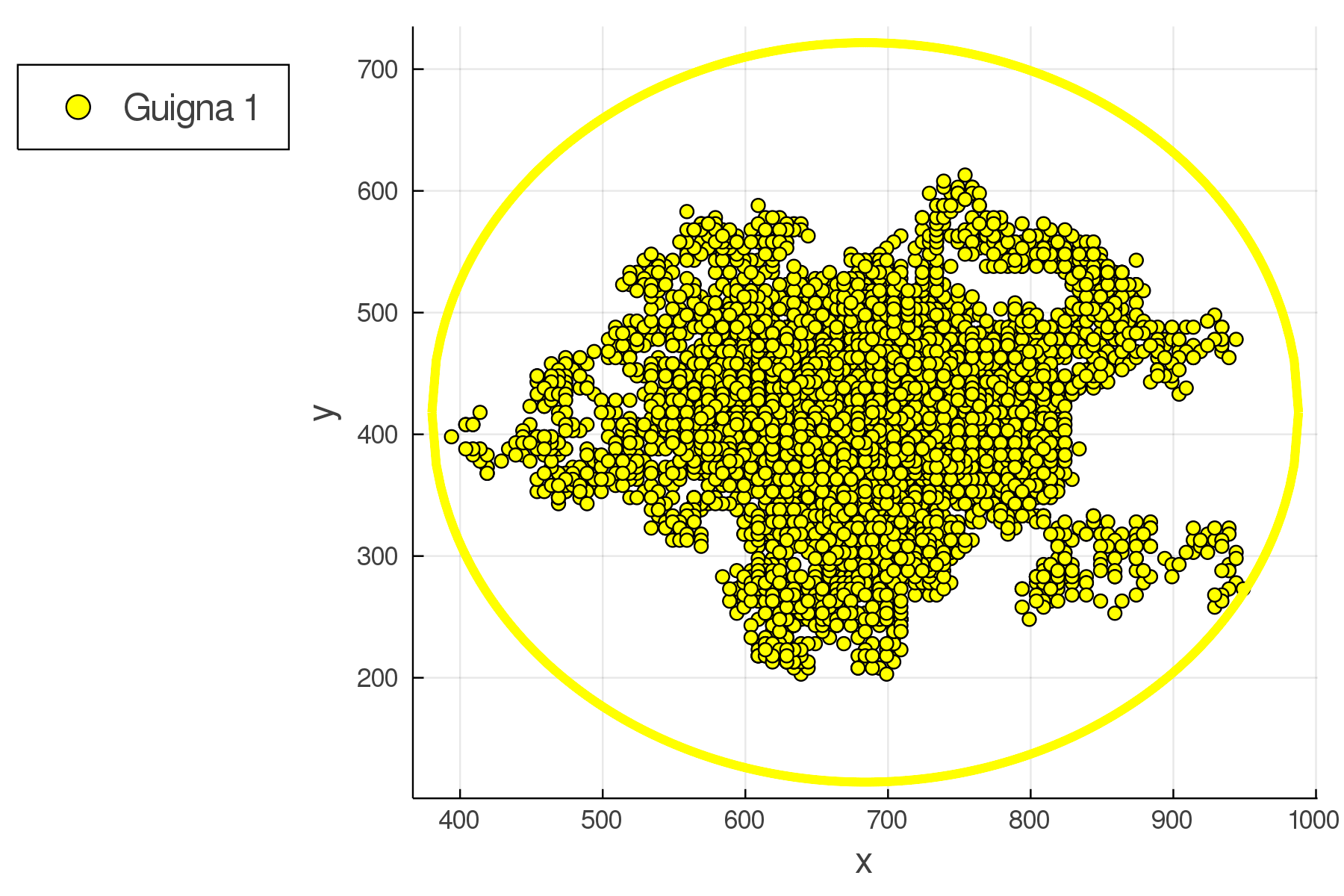}  
         \caption*{a) Positions of the guigna 1}    
    \end{subfigure}
        \begin{subfigure}[c]{0.33\textwidth}
        \centering
 \includegraphics[width=\textwidth, height=0.85\textwidth]{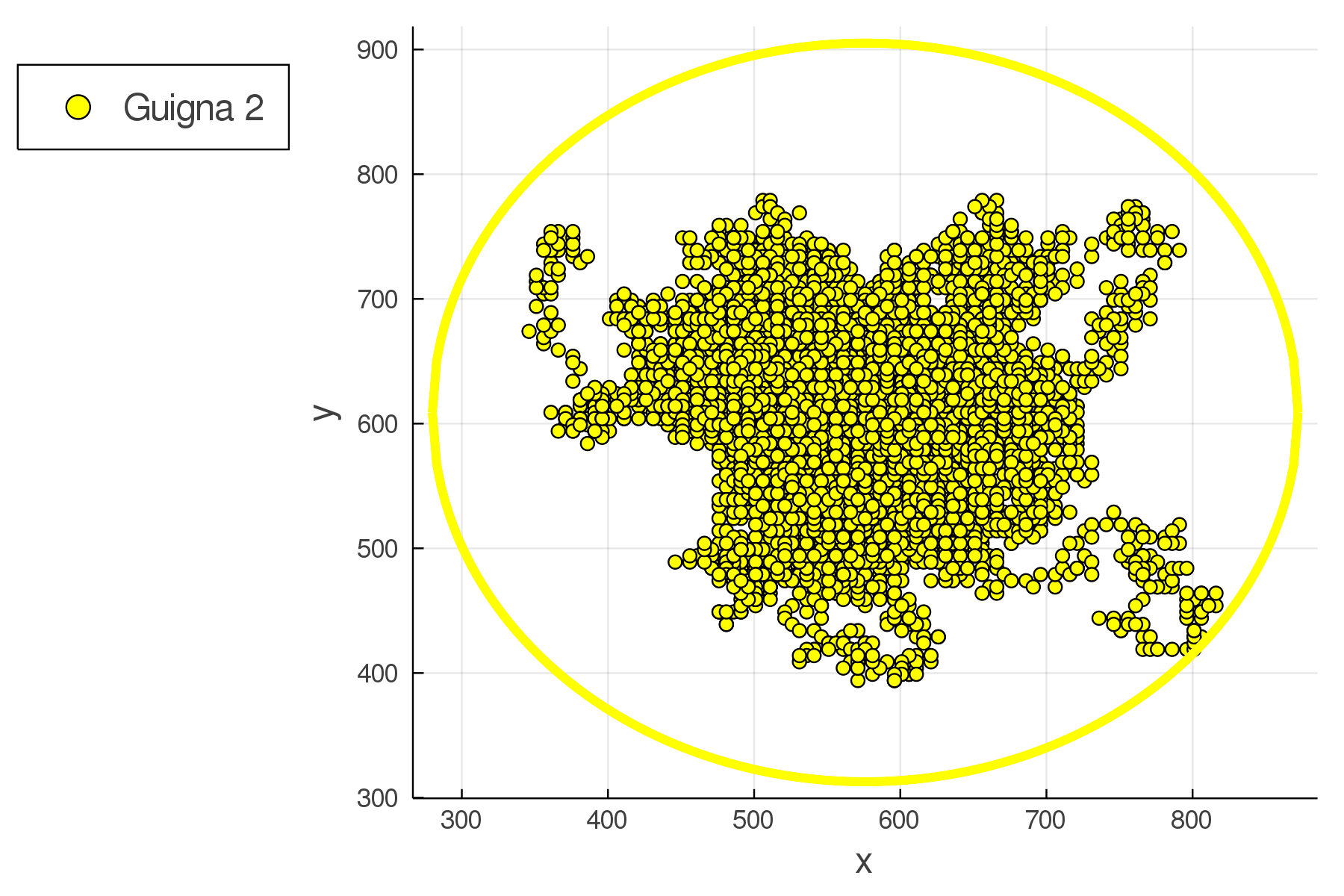}  
         \caption*{b) Positions of the guigna 2}      
    \end{subfigure} 
        \begin{subfigure}[c]{0.33\textwidth}
        \centering
 \includegraphics[width=\textwidth, height=0.85\textwidth]{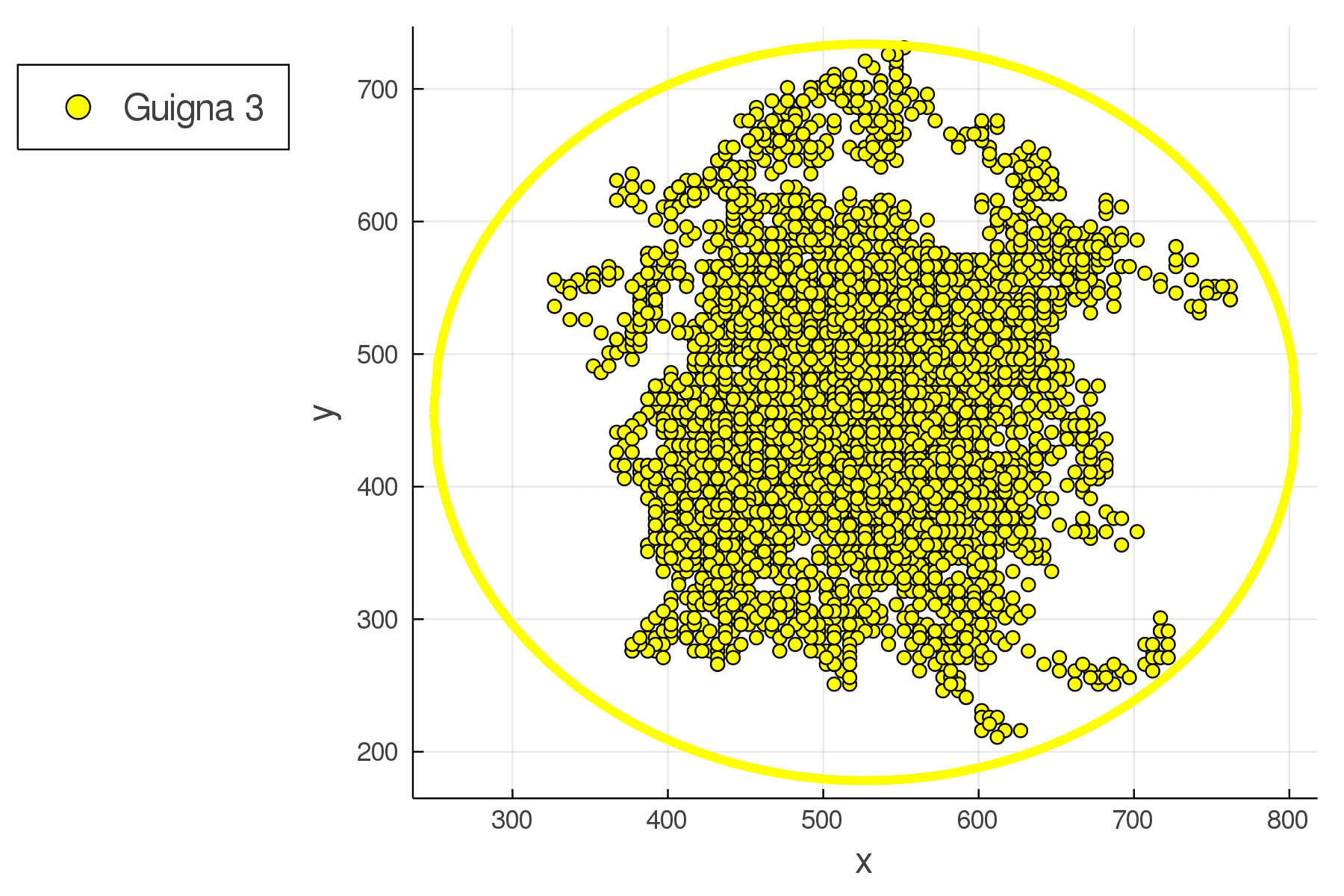}    
         \caption*{c) Positions of the guigna 3}       
    \end{subfigure} 
\centering
        \begin{subfigure}[c]{0.33\textwidth}
        \centering
 \includegraphics[width=\textwidth, height=0.85\textwidth]{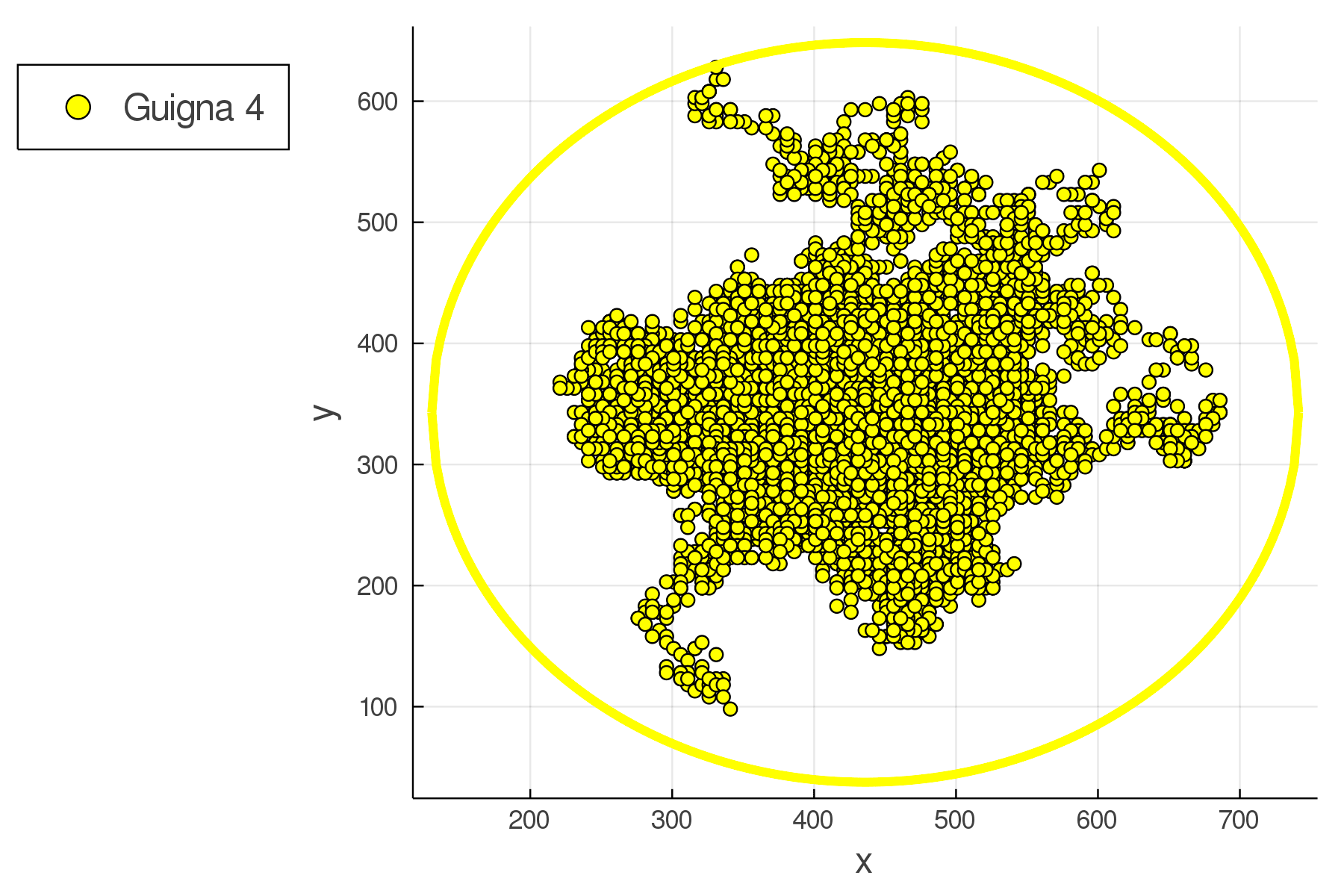}  
         \caption*{d) Positions of the guigna 4}    
    \end{subfigure}
        \begin{subfigure}[c]{0.33\textwidth}
        \centering
 \includegraphics[width=\textwidth, height=0.85\textwidth]{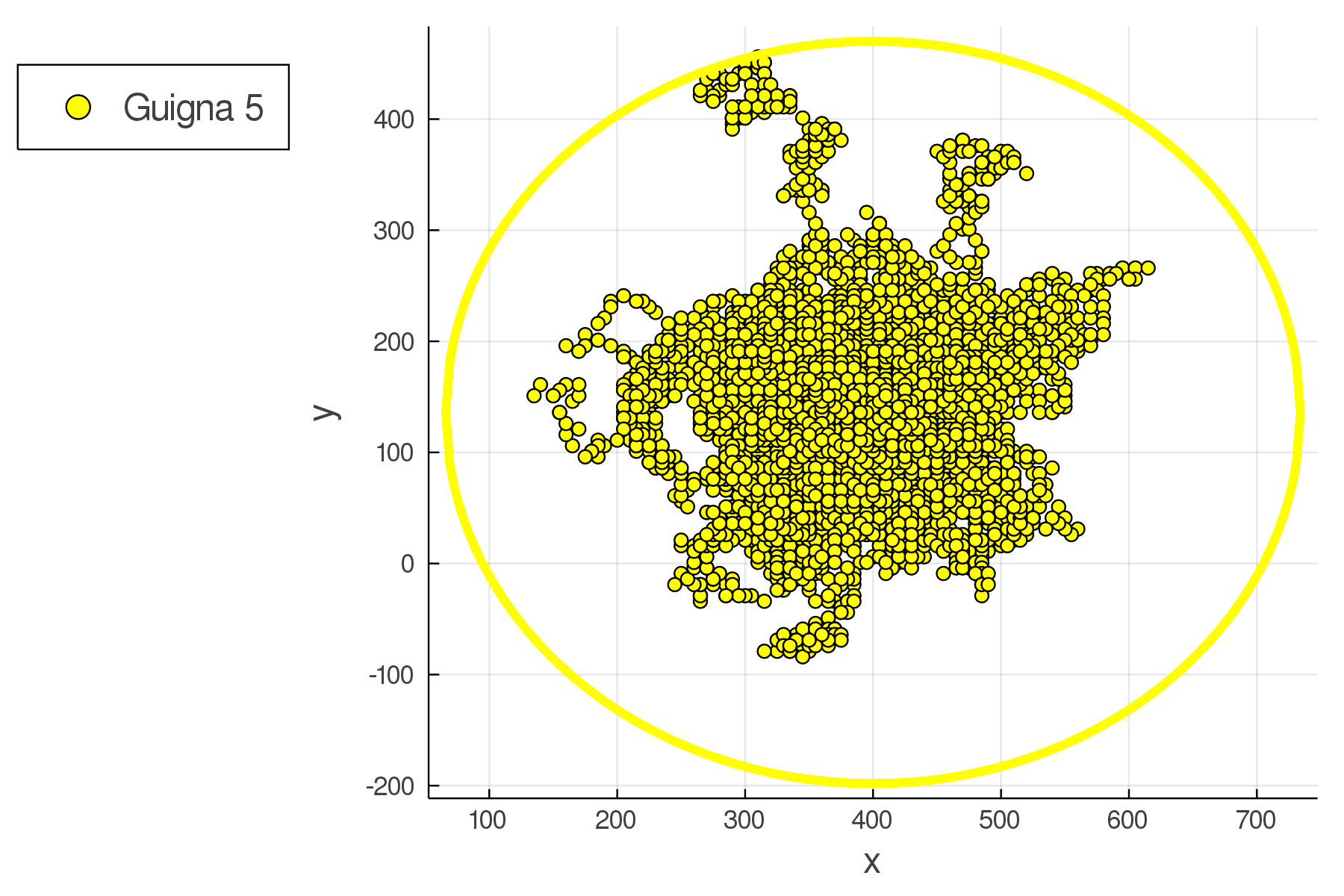}  
         \caption*{e) Positions of the guigna 5}      
    \end{subfigure} 
        \begin{subfigure}[c]{0.33\textwidth}
        \centering
 \includegraphics[width=\textwidth, height=0.85\textwidth]{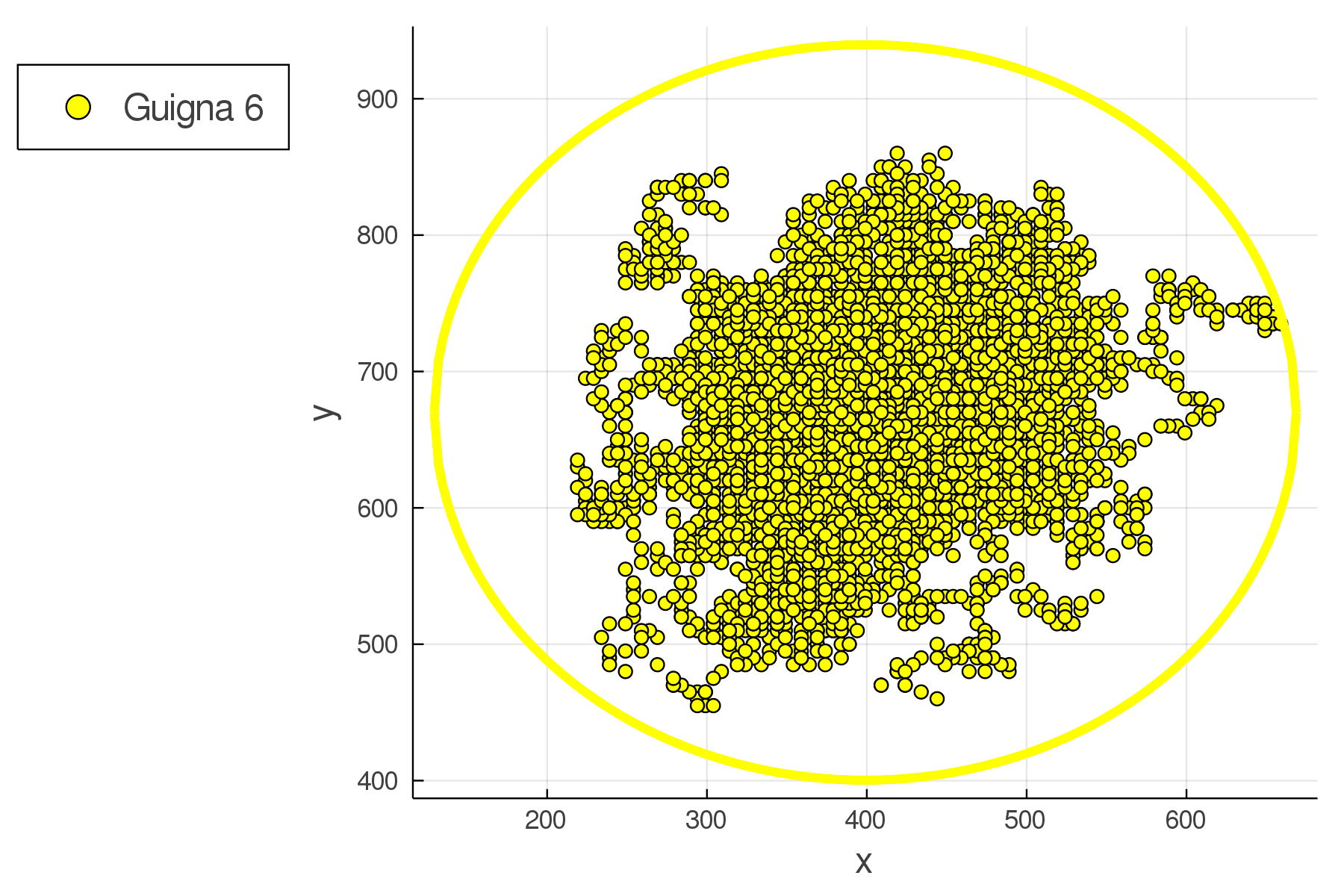}    
         \caption*{f) Positions of the guigna 6}       
    \end{subfigure} 
\centering
        \begin{subfigure}[c]{0.33\textwidth}
        \centering
 \includegraphics[width=\textwidth, height=0.85\textwidth]{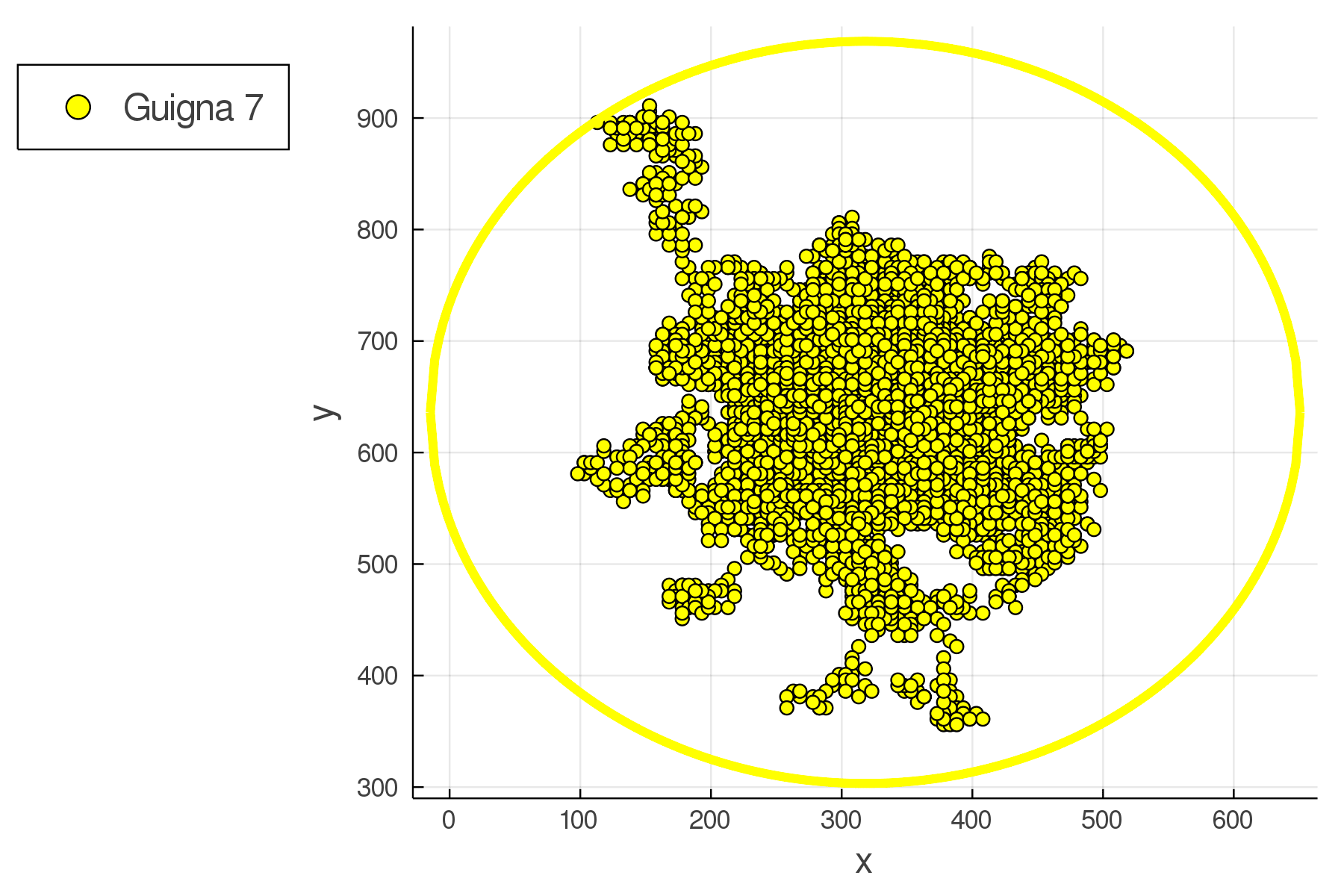}  
         \caption*{g) Positions of the guigna 7}    
    \end{subfigure}
        \begin{subfigure}[c]{0.33\textwidth}
        \centering
 \includegraphics[width=\textwidth, height=0.85\textwidth]{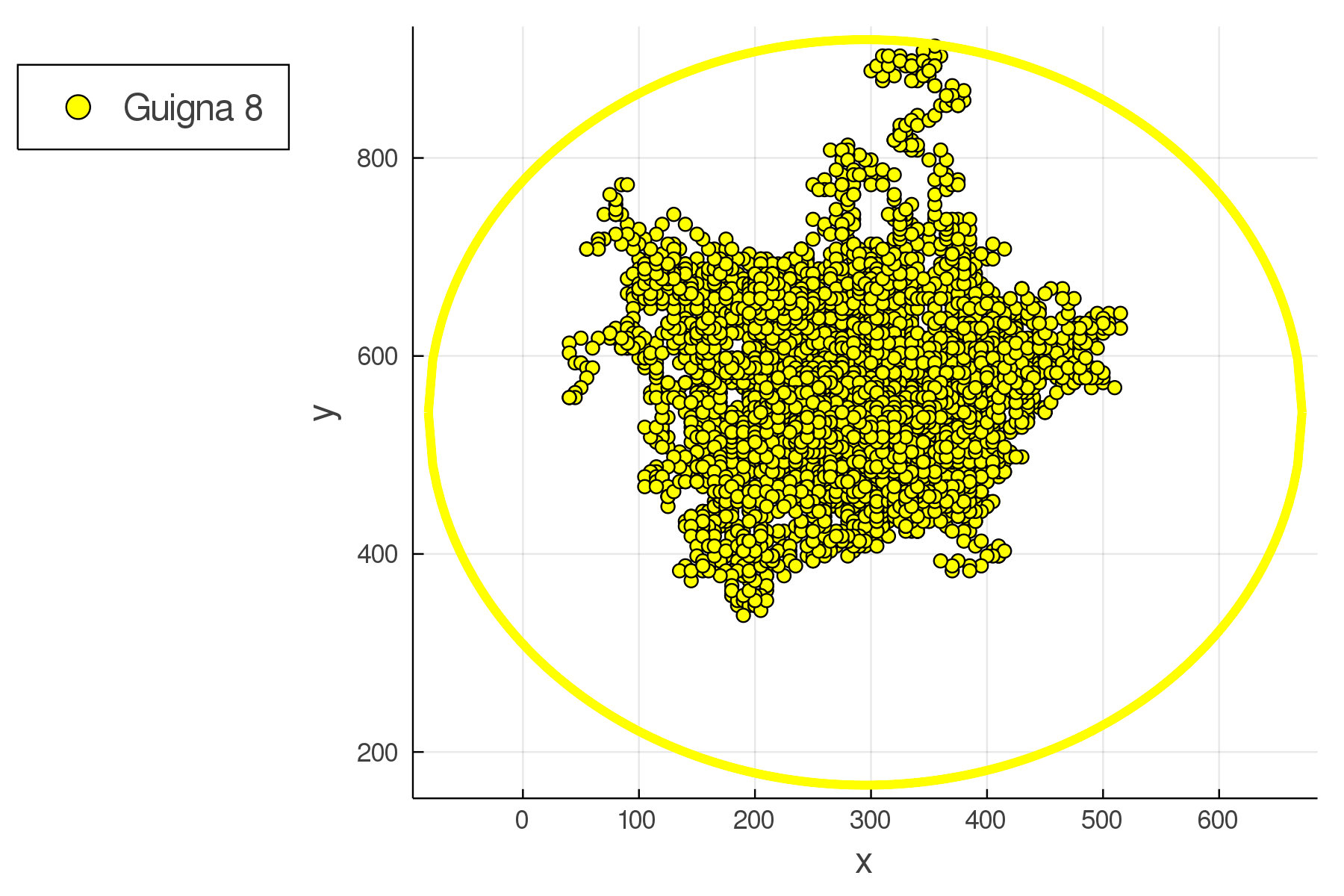}  
         \caption*{h) Positions of the guigna 8}      
    \end{subfigure} 
        \begin{subfigure}[c]{0.33\textwidth}
        \centering
 \includegraphics[width=\textwidth, height=0.85\textwidth]{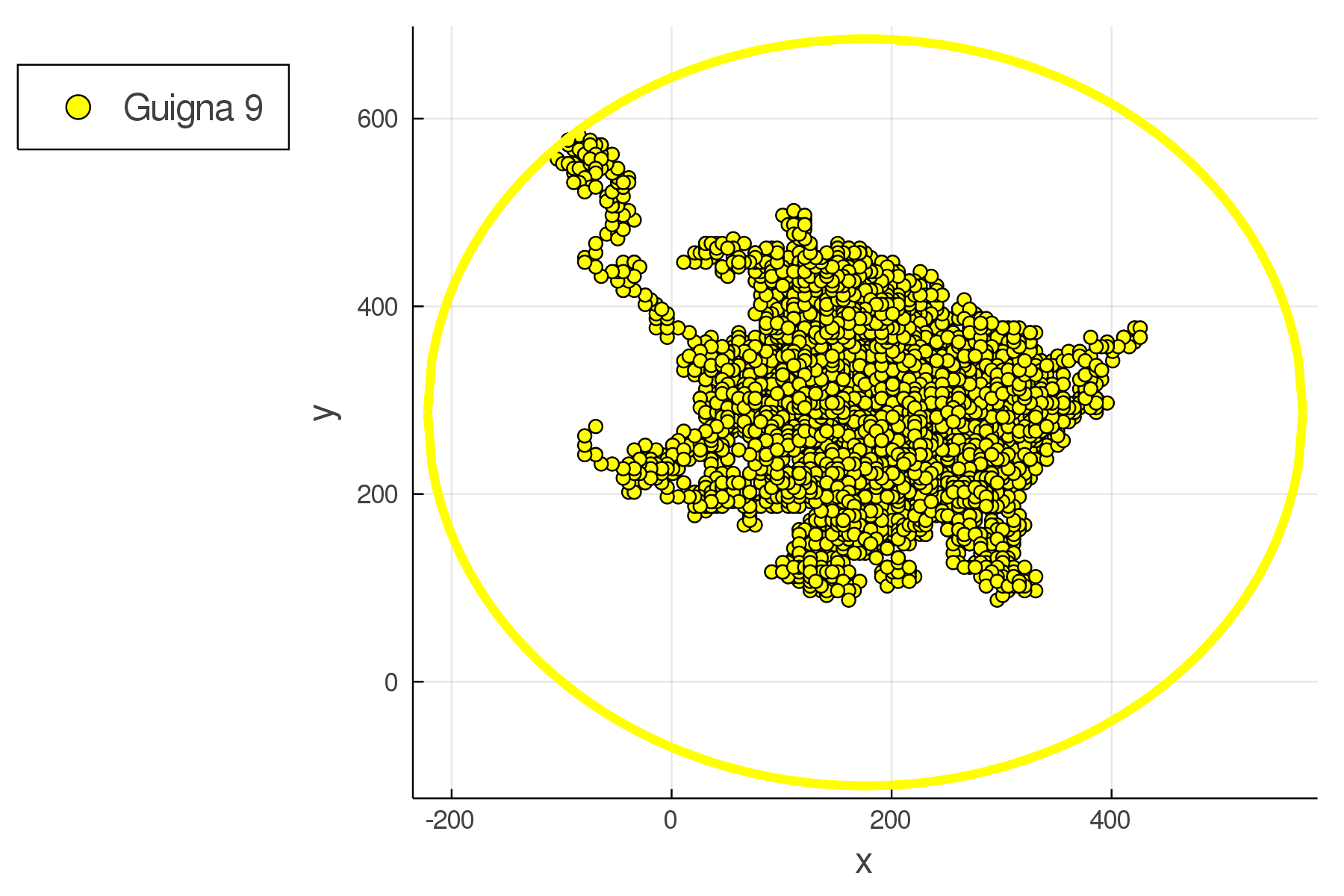}    
         \caption*{i) Positions of the guigna 9}       
    \end{subfigure}  
        \caption{Distributions of the positions of the different guignas.}\label{fig:07}
\end{figure}

So, through 50 sequences of random variables and using the previous process, the distributions of points shown in the Figure \ref{fig:07} were generated, with which it is possible to emulate distributions of positions analogous to those produced by animals with radiocollars when moving through a territory, which are analogous to the distributions of positions of domestic cats shown in the reference \cite{lopez2021free}.

Considering that in the simulation each guigna generates a distribution of positions given by the following succession of random variables

\begin{eqnarray}
\set{x_i}_{i=1}^{M} & \mbox{ with }&M=10^4,
\end{eqnarray}

it is possible to characterize the distances that each guigna moves away from its initial position $x_0$ by the following succession of distances in terms of the step size $r$

\begin{eqnarray}
\set{\frac{\norm{x_0-x_i}}{r}}_{i=1}^{M}.
\end{eqnarray}

\begin{figure}[!ht]
\centering
        \begin{subfigure}[c]{0.33\textwidth}
        \centering
 \includegraphics[width=\textwidth, height=0.85\textwidth]{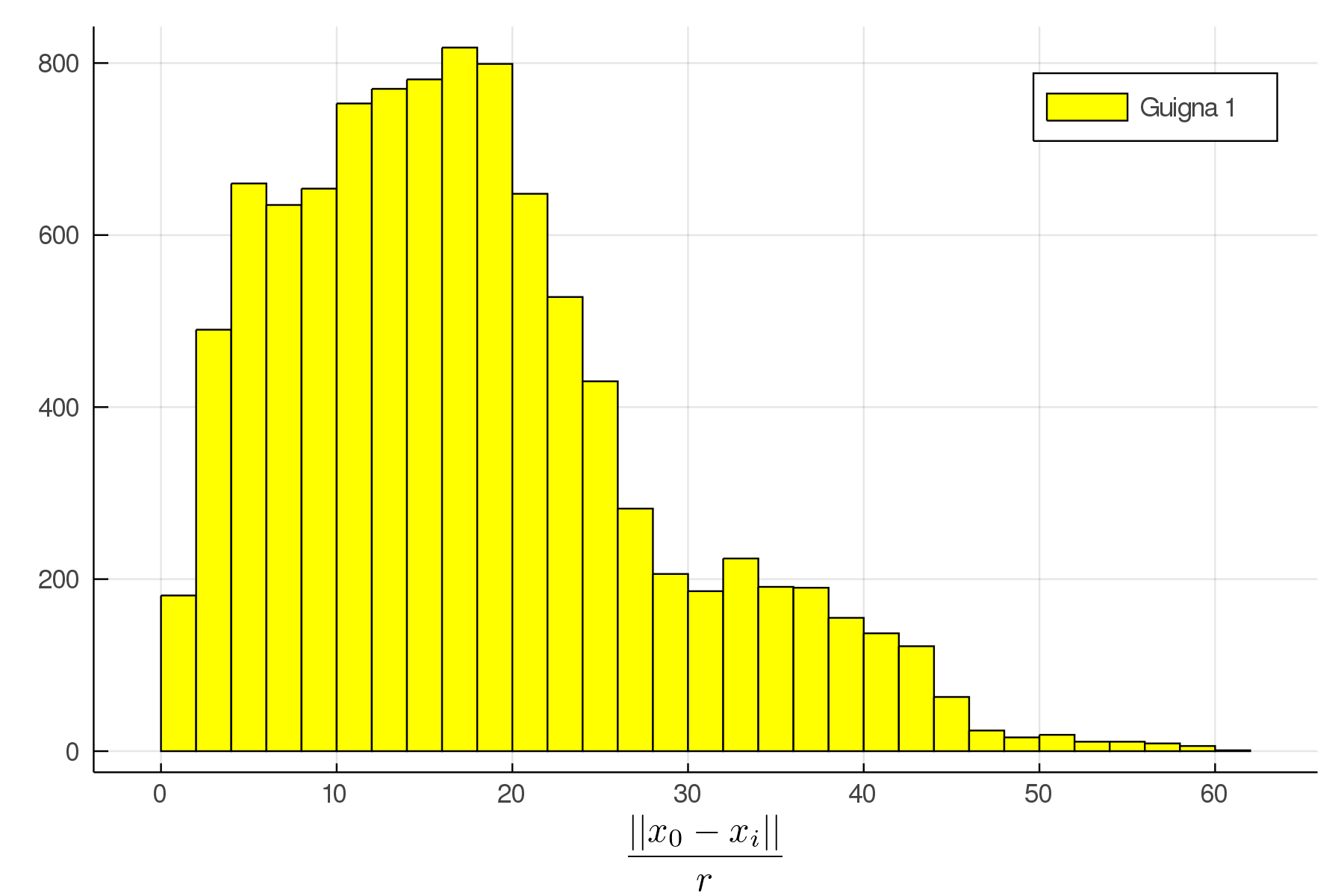}  
         \caption*{a) Histogram of distances of the guigna 1}    
    \end{subfigure}
        \begin{subfigure}[c]{0.33\textwidth}
        \centering
 \includegraphics[width=\textwidth, height=0.85\textwidth]{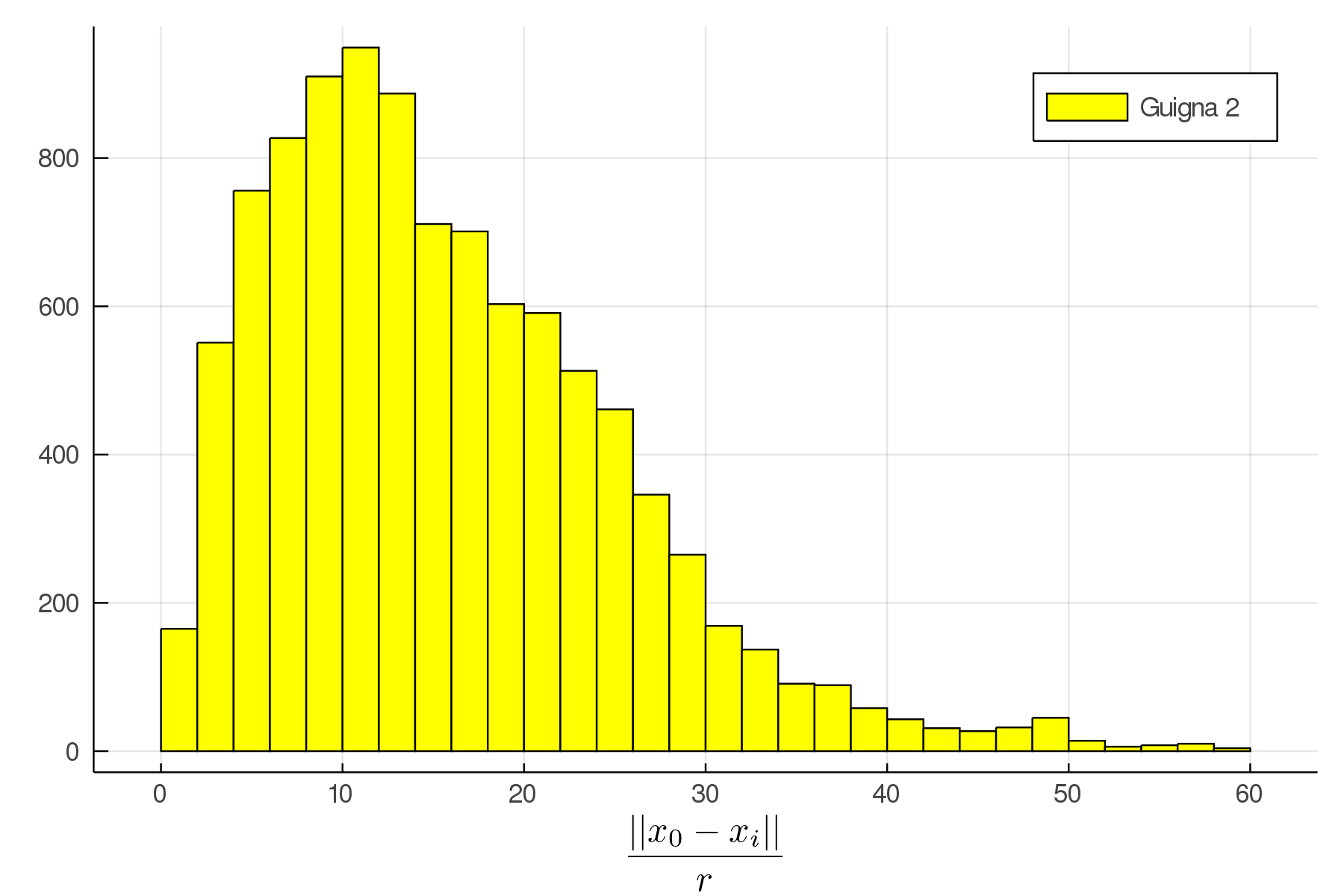}  
         \caption*{b) Histogram of distances of the guigna 2}      
    \end{subfigure} 
        \begin{subfigure}[c]{0.33\textwidth}
        \centering
 \includegraphics[width=\textwidth, height=0.85\textwidth]{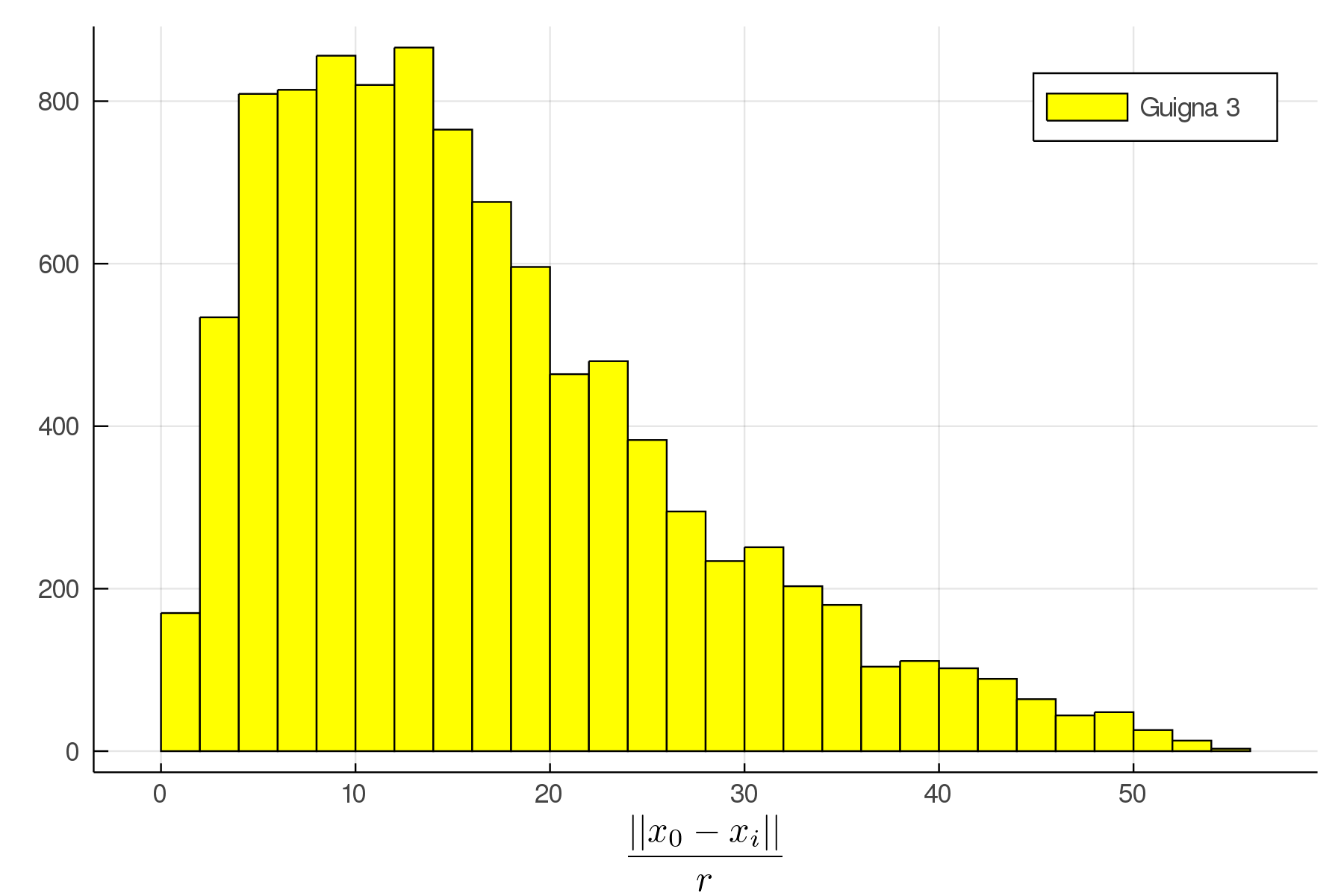}    
         \caption*{c) Histogram of distances of the guigna 3}       
    \end{subfigure} 
\centering
        \begin{subfigure}[c]{0.33\textwidth}
        \centering
 \includegraphics[width=\textwidth, height=0.85\textwidth]{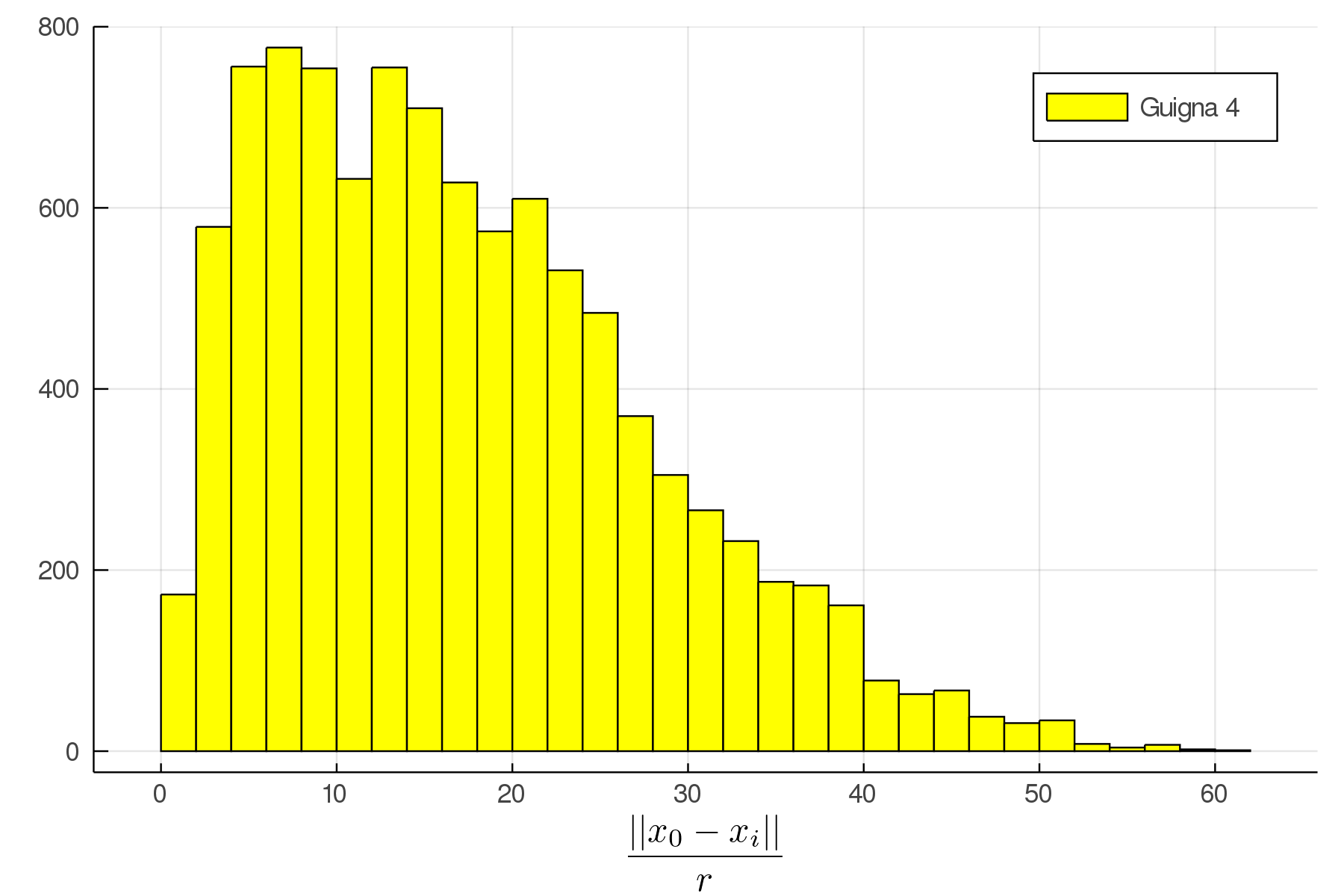}  
         \caption*{d) Histogram of distances of the guigna 4}    
    \end{subfigure}
        \begin{subfigure}[c]{0.33\textwidth}
        \centering
 \includegraphics[width=\textwidth, height=0.85\textwidth]{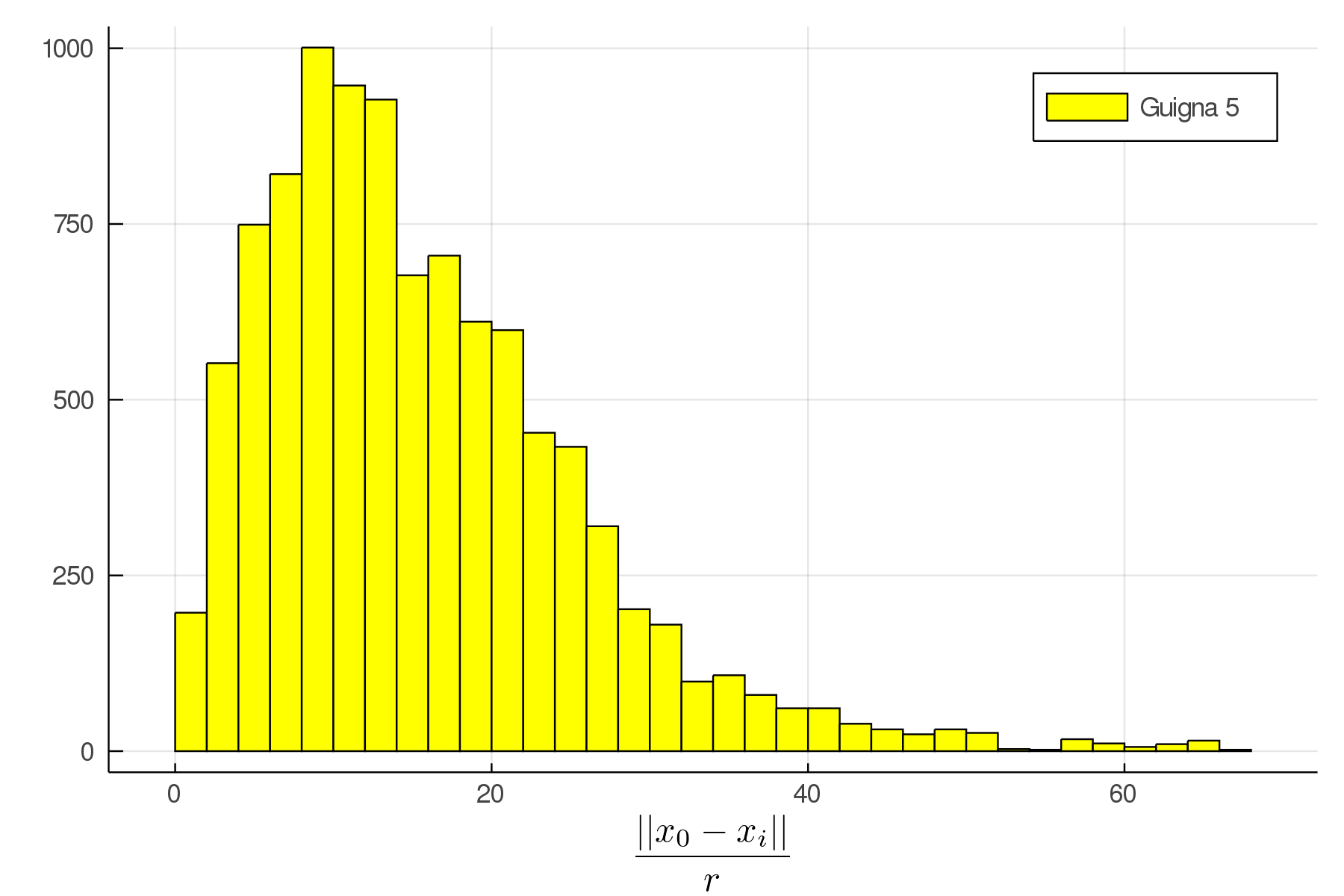}  
         \caption*{e) Histogram of distances of the guigna 5}      
    \end{subfigure} 
        \begin{subfigure}[c]{0.33\textwidth}
        \centering
 \includegraphics[width=\textwidth, height=0.85\textwidth]{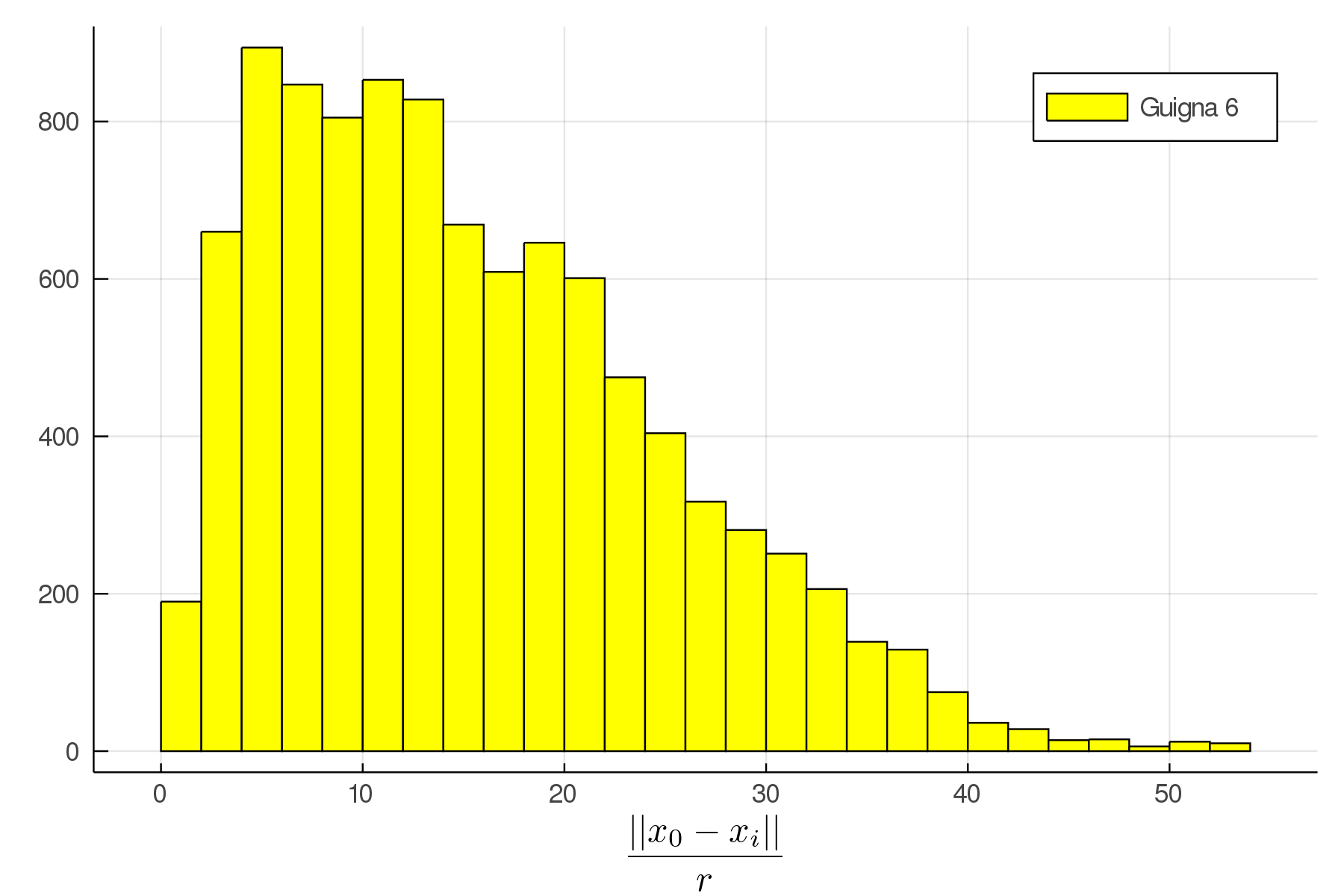}    
         \caption*{f) Histogram of distances of the guigna 6}       
    \end{subfigure} 
\centering
        \begin{subfigure}[c]{0.33\textwidth}
        \centering
 \includegraphics[width=\textwidth, height=0.85\textwidth]{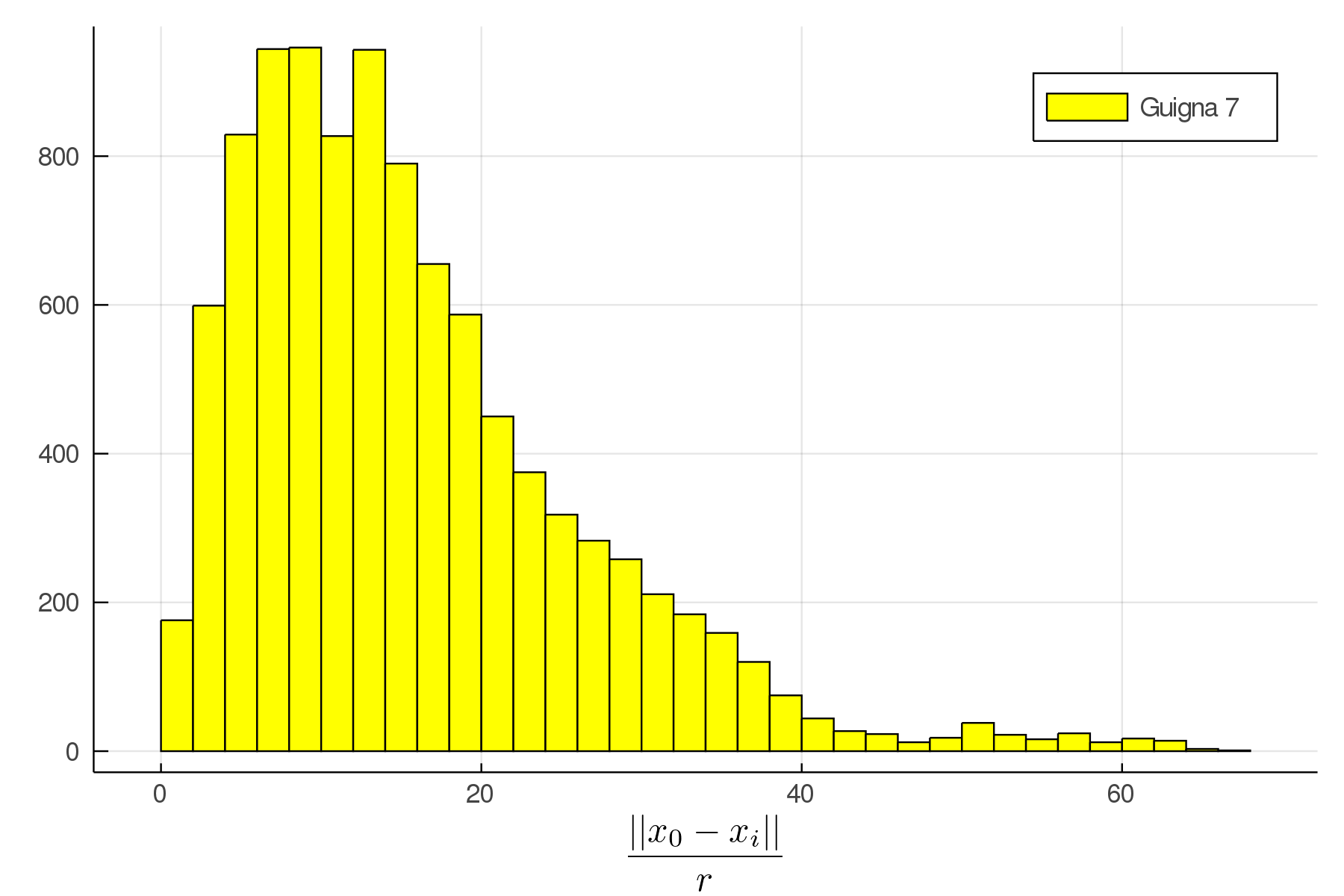}  
         \caption*{g) Histogram of distances of the guigna 7}    
    \end{subfigure}
        \begin{subfigure}[c]{0.33\textwidth}
        \centering
 \includegraphics[width=\textwidth, height=0.85\textwidth]{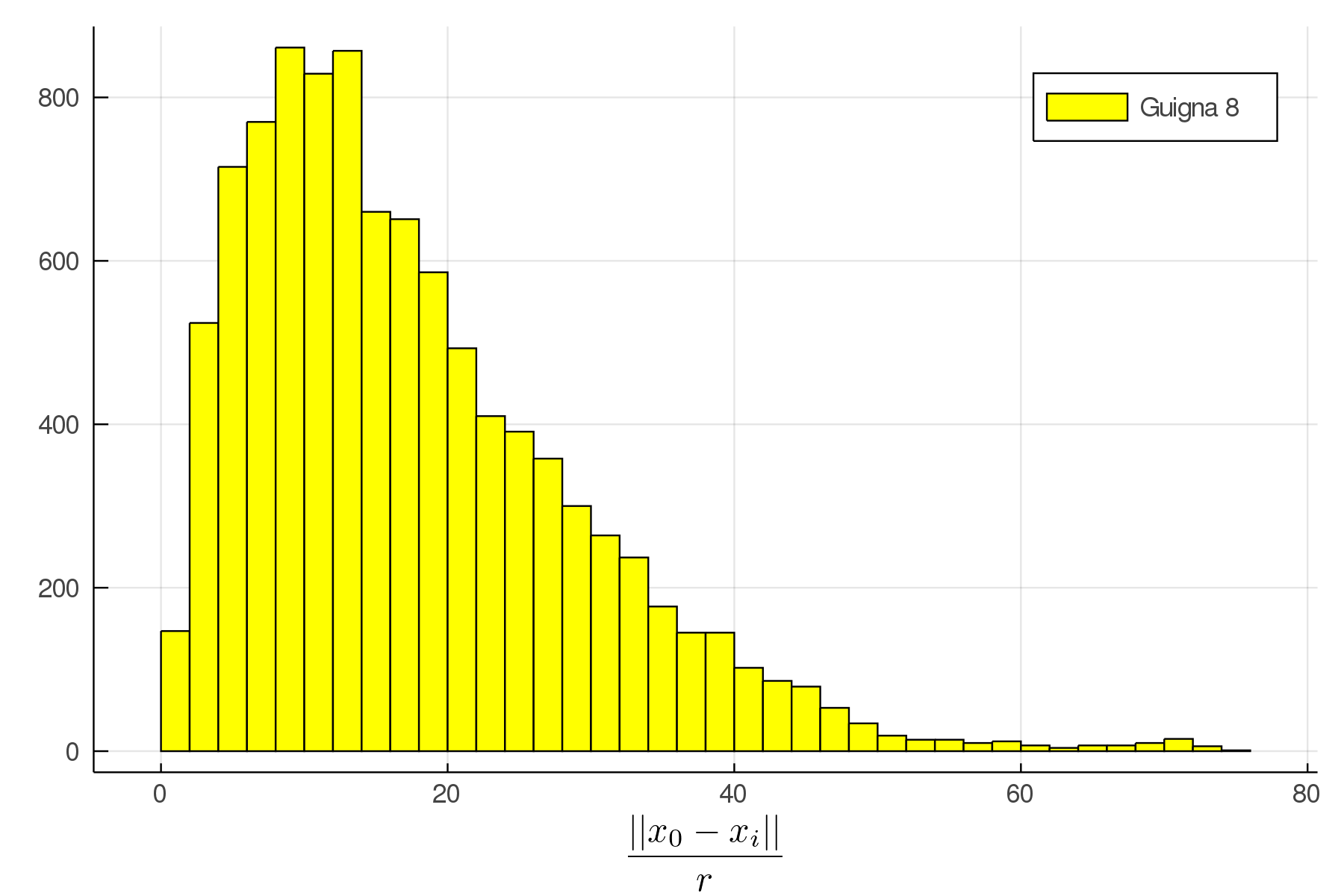}  
         \caption*{h) Histogram of distances of the guigna 8}      
    \end{subfigure} 
        \begin{subfigure}[c]{0.33\textwidth}
        \centering
 \includegraphics[width=\textwidth, height=0.85\textwidth]{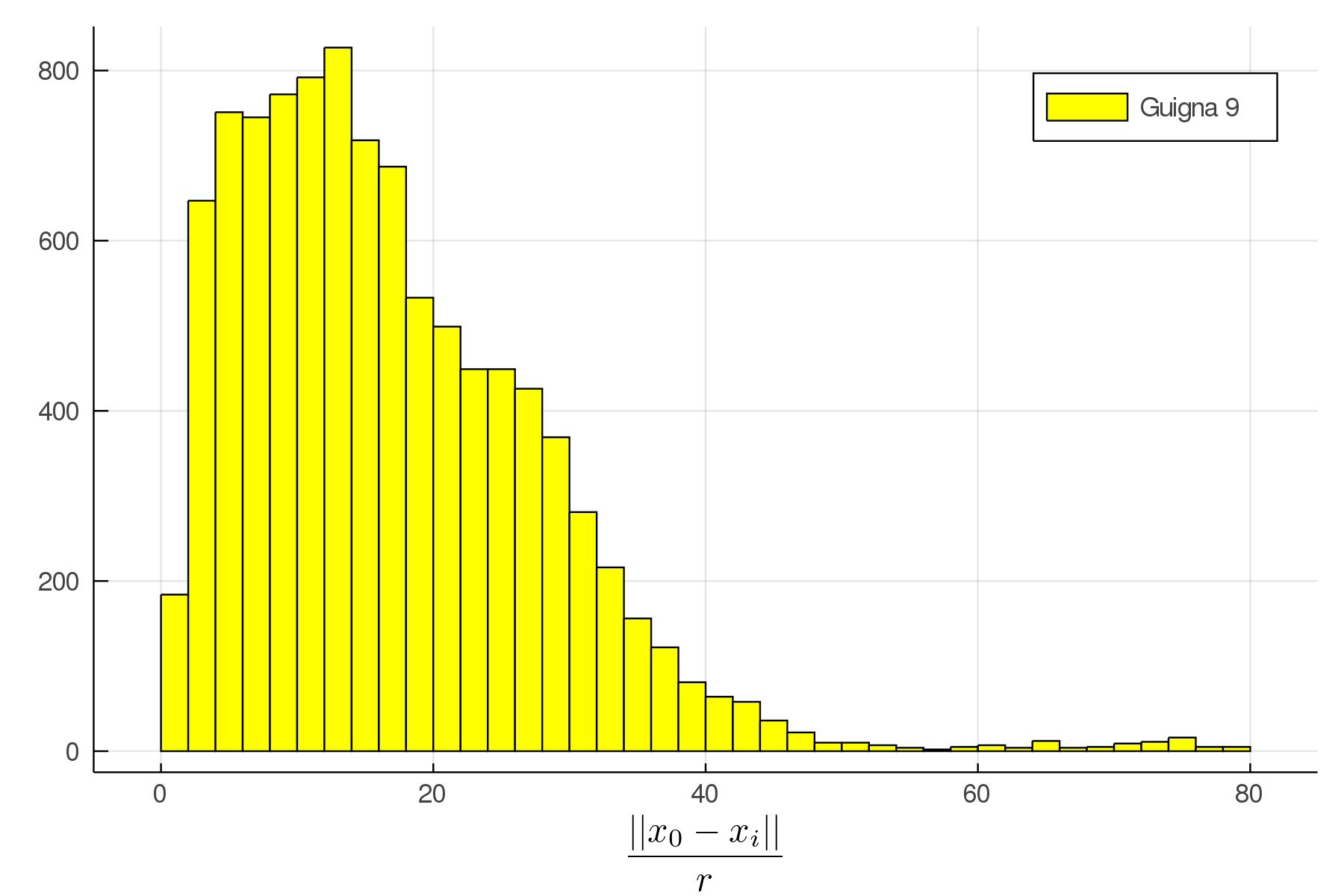}    
         \caption*{i) Histogram of distances of the guigna 9}       
    \end{subfigure}  
        \caption{Histograms of the sequences
of distances $\set{\frac{\norm{x_0-x_i}}{r}}_{i=1}^M$ of the different guignas.}\label{fig:08}
\end{figure}

\begin{figure}[!ht]
\centering
        \begin{subfigure}[c]{0.9\textwidth}
        \centering
 \includegraphics[width=\textwidth, height=0.6\textwidth]{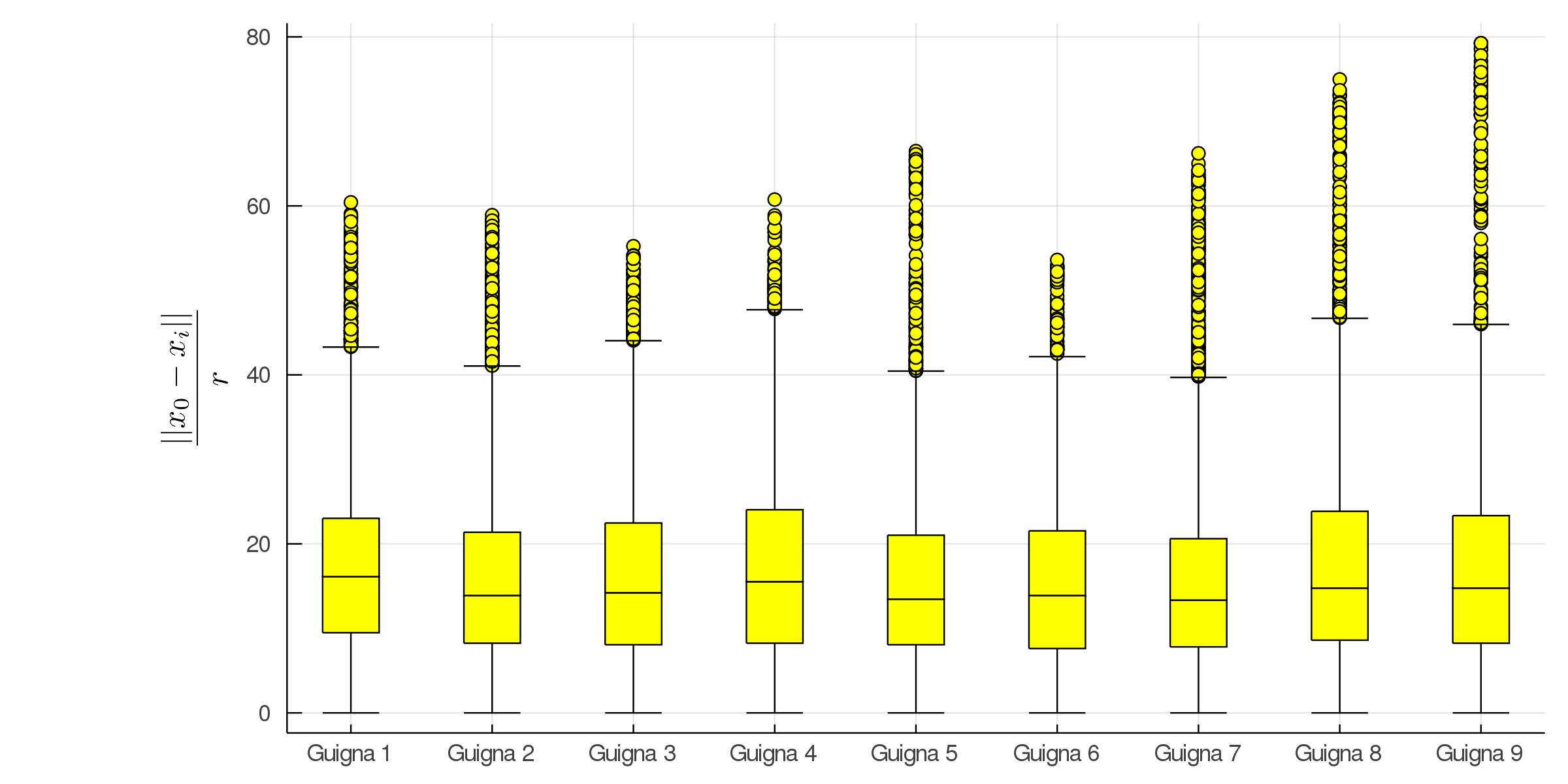}   
    \end{subfigure}  
        \caption{Box plots of the sequences
of distances $\set{\frac{\norm{x_0-x_i}}{r}}_{i=1}^M$ of the different guignas.}\label{fig:09}
\end{figure}

So, the distances at which each guigna tends to move away from its initial position $x_0$ may be analyzed using histograms and box plots as shown in Figure \ref{fig:08} and Figure \ref{fig:09}, respectively.

\newpage

It is necessary to mention that from the Figure \ref{fig:09} it may be concluded that the values of the distributions of the distances of the guignas in the simulation have a high level of dispersion, so that the mean distances of the distributions would not be representative measures of the data, since that mean values tend to be affected in the presence of anomalous data caused by dispersion. On the other hand, the median of the distributions are not considerably affected by the dispersion, so in the presence of data with a high level of dispersion it could be considered as an acceptable measure to characterize the behavior of the data. It is important to mention the above, because the behavior of the distances that could be obtained from an animal with radiocollar in its territory should not necessarily be uniform, since irregularities may occur due to interactions with external agents or irregularities of the terrain, so work with data with a considerable degree of dispersion is something that falls under the criteria of the simulation, and as a consequence, it may be considered that the most representative measures of the distributions of the data of the guignas generated in the simulation are those shown in the following table:

\begin{footnotesize}
\begin{longtable}{c|ccccc}
\toprule
    Guignas &  $\underset{i\geq 1}{max} \set{\frac{\norm{x_0-x_i}}{r}}$&$\underset{i\geq 1}{mean} \set{\frac{\norm{x_0-x_i}}{r}}$ & $\underset{i\geq 1}{median} \set{\frac{\norm{x_0-x_i}}{r}}$\\
\midrule
    Guigna 1 & 60.42 & 17.47 & 16.12 \\
    Guigna 2 & 58.90 & 15.57 & 13.89 \\
    Guigna 3 & 55.23 & 16.41 & 14.21 \\
    Guigna 4 & 60.75 & 17.23 & 15.52 \\
    Guigna 5 & 66.48 & 15.52 & 13.45 \\
    Guigna 6 & 53.60 & 15.42 & 13.89 \\
    Guigna 7 & 66.22 & 15.58 & 13.34 \\
    Guigna 8 & 74.97 & 17.38 & 14.76 \\
    Guigna 9 & 79.25 & 16.71 & 14.76 \\
    \bottomrule
\caption{Most representative distances of the sequences
of distances $\set{\frac{\norm{x_0-x_i}}{r}}_{i=1}^M$ of the different guignas.}
\end{longtable}
\end{footnotesize}

From the previous table it is possible to define the following values for each distribution of positions of the guignas

\begin{eqnarray}
R_M:=\underset{i\geq 1}{max} \set{\frac{\norm{x_0-x_i}}{r}},
\end{eqnarray}

\begin{eqnarray}
r_m:=\min \set{\underset{i\geq 1}{mean} \set{\frac{\norm{x_0-x_i}}{r}},\underset{i\geq 1}{median} \set{\frac{\norm{x_0-x_i}}{r}}},
\end{eqnarray}

with which it is possible to measure the interactions they have with predators through the following result:

\begin{eqnarray}
\mbox{If } \frac{\norm{x_0-x_p}}{r}\leq R_M \mbox{ and } \set{x_i}_{i=1}^M \cap B\left(x_p;\frac{r_m}{2}\right)\neq \emptyset \ \Rightarrow \ \exists \set{x_{i_p}}_{i_p=1}^m\in B\left(x_p;\frac{r_m}{2}\right),
\end{eqnarray}

where $x_p$ denotes the position of some predator, and the values $x_{i_p}$ represent positions of the guignas that could be considered dangerously close to some predator. On the other hand, through the values $x_{i_p}$ it is possible to characterize the number of interactions that each guigna has with some predators as shown in the following table:

\begin{footnotesize}
\begin{longtable}{c|ccccccccc}
\toprule
          & Guigna 1 & Guigna 2 & Guigna 3 & Guigna 4 & Guigna 5 & Guigna 6 & Guigna 7 & Guigna 8 & Guigna 9 \\ \midrule
    Predator 1 & 8     & 0     & 0     & 0     & 0     & 0     & 0     & 0     & 0 \\
    Predator 2 & 40    & 0     & 0     & 0     & 0     & 0     & 0     & 0     & 0 \\
    Predator 3 & 97    & 0     & 72    & 16    & 18    & 0     & 0     & 0     & 0 \\
    Predator 4 & 0     & 0     & 0     & 0     & 13    & 0     & 0     & 0     & 0 \\
    Predator 5 & 0     & 0     & 14    & 302   & 165   & 0     & 0     & 0     & 0 \\
    Predator 6 & 0     & 75    & 0     & 0     & 0     & 464   & 104   & 0     & 0 \\
    Predator 7 & 0     & 240   & 70    & 0     & 0     & 1232  & 190   & 53    & 0 \\
    Predator 8 & 0     & 39    & 72    & 43    & 0     & 411   & 274   & 272   & 0 \\
    Predator 9 & 0     & 0     & 37    & 32    & 0     & 47    & 54    & 595   & 0 \\
    Predator 10 & 0     & 0     & 0     & 321   & 36    & 0     & 20    & 3     & 27 \\
    Predator 11 & 0     & 0     & 0     & 333   & 0     & 0     & 0     & 0     & 653 \\
    Predator 12 & 0     & 0     & 0     & 95    & 161   & 0     & 0     & 0     & 209 \\
    Predator 13 & 0     & 0     & 0     & 22    & 201   & 0     & 0     & 0     & 141 \\
    Predator 14 & 0     & 0     & 0     & 0     & 61    & 0     & 0     & 0     & 17 \\
    \bottomrule
    \caption{Number of interactions of the guignas with the different predators.}
\end{longtable}
\end{footnotesize}

So, considering the following result:

\begin{eqnarray}
\mbox{If }\set{x_i}_{i=1}^m\in B\left(x_p;\dfrac{r_m}{2} \right) \ \Rightarrow \  \forall x_j \in \set{x_i}_{i=1}^m \hspace{0.2cm} \exists k_j= \dfrac{\norm{x_p-x_j}}{r},
\end{eqnarray}

it is possible to analyze the distribution of distances of each guigna with each of the predators with which they interacted as shown below:

\begin{figure}[!ht]
\begin{tabular}{cc}
\begin{minipage}[c]{0.37\textwidth}
\centering
\footnotesize
    \begin{tabular}{c|cc}
\cmidrule{1-1}    Guiga 1 &       &  \\ \toprule
    Predators & $\underset{i\geq 1}{mean} \set{\frac{\norm{x_p-x_i}}{r}}$  & $\underset{i\geq 1}{median} \set{\frac{\norm{x_p-x_i}}{r}}$ \\ \midrule
    Predator 1 & 7.28  & 7.32 \\
    Predator 2 & 6.86  & 6.84 \\
    Predator 3 & 5.06  & 5.40 \\
     \bottomrule
    \end{tabular}%
\end{minipage} &
\begin{minipage}[c]{0.60\textwidth}
\centering
 \includegraphics[width=\textwidth, height=0.5\textwidth]{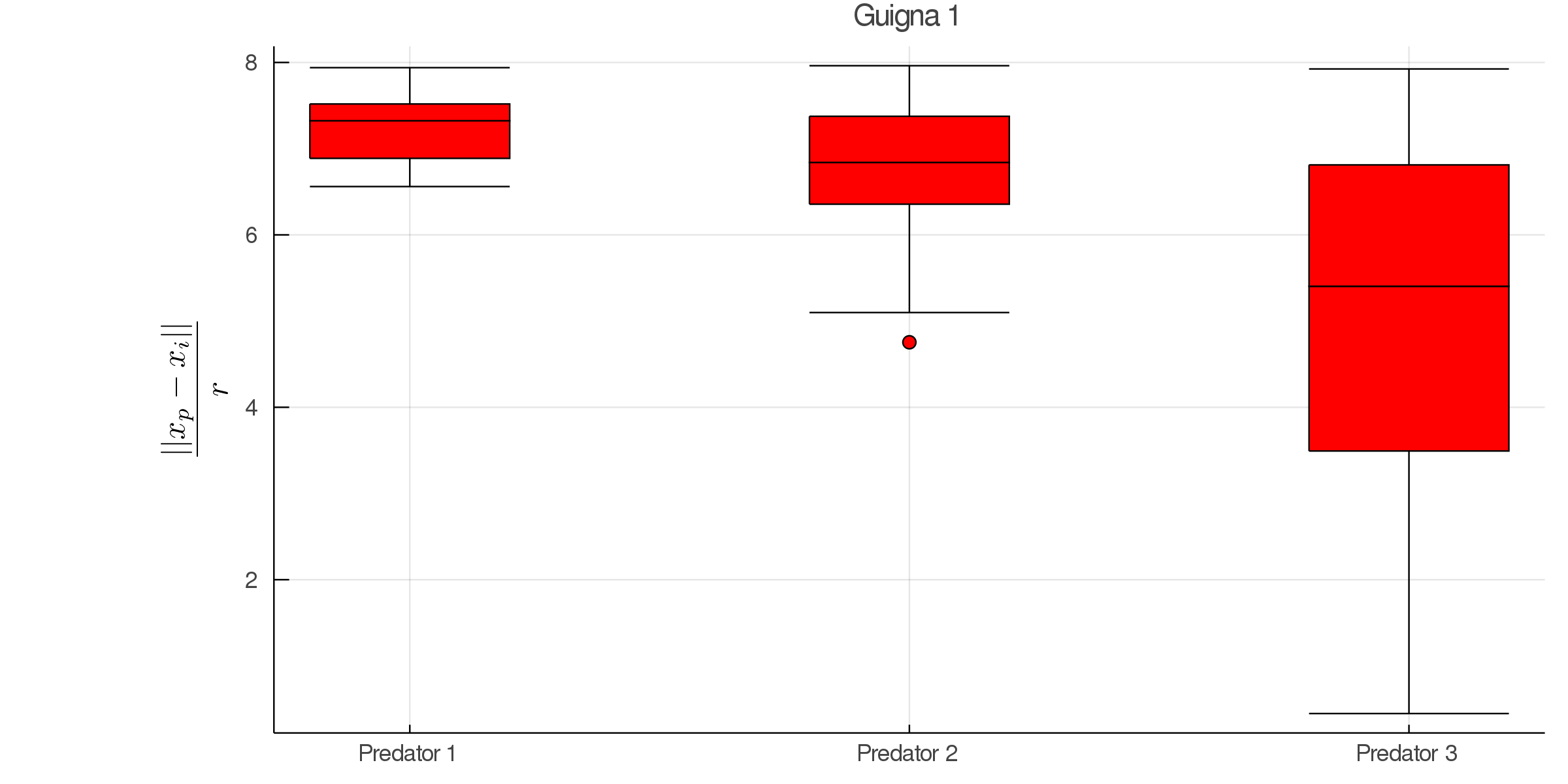} 
\end{minipage}
\end{tabular}
\caption{Most representative distances of the interactions of guigna 1 with some predators, and box plot of the distances $\set{\frac{\norm{x_p-x_i}}{r}}_{i=1}^m$ of the interactions of guigna 1 with some predators.}
\end{figure}

\newpage

\begin{figure}[!ht]
\begin{tabular}{cc}
\begin{minipage}[c]{0.37\textwidth}
\centering
\footnotesize
    \begin{tabular}{c|cc}
\cmidrule{1-1}    Guiga 2 &       &  \\ \toprule
    Predators & $\underset{i\geq 1}{mean} \set{\frac{\norm{x_p-x_i}}{r}}$  & $\underset{i\geq 1}{median} \set{\frac{\norm{x_p-x_i}}{r}}$ \\ \midrule
    Predator 6 & 5.12  & 5.80 \\
    Predator 7 & 4.86  & 5.20 \\
    Predator 8 & 5.64  & 5.60 \\
     \bottomrule
    \end{tabular}%
\end{minipage} &
\begin{minipage}[c]{0.60\textwidth}
\centering
 \includegraphics[width=\textwidth, height=0.5\textwidth]{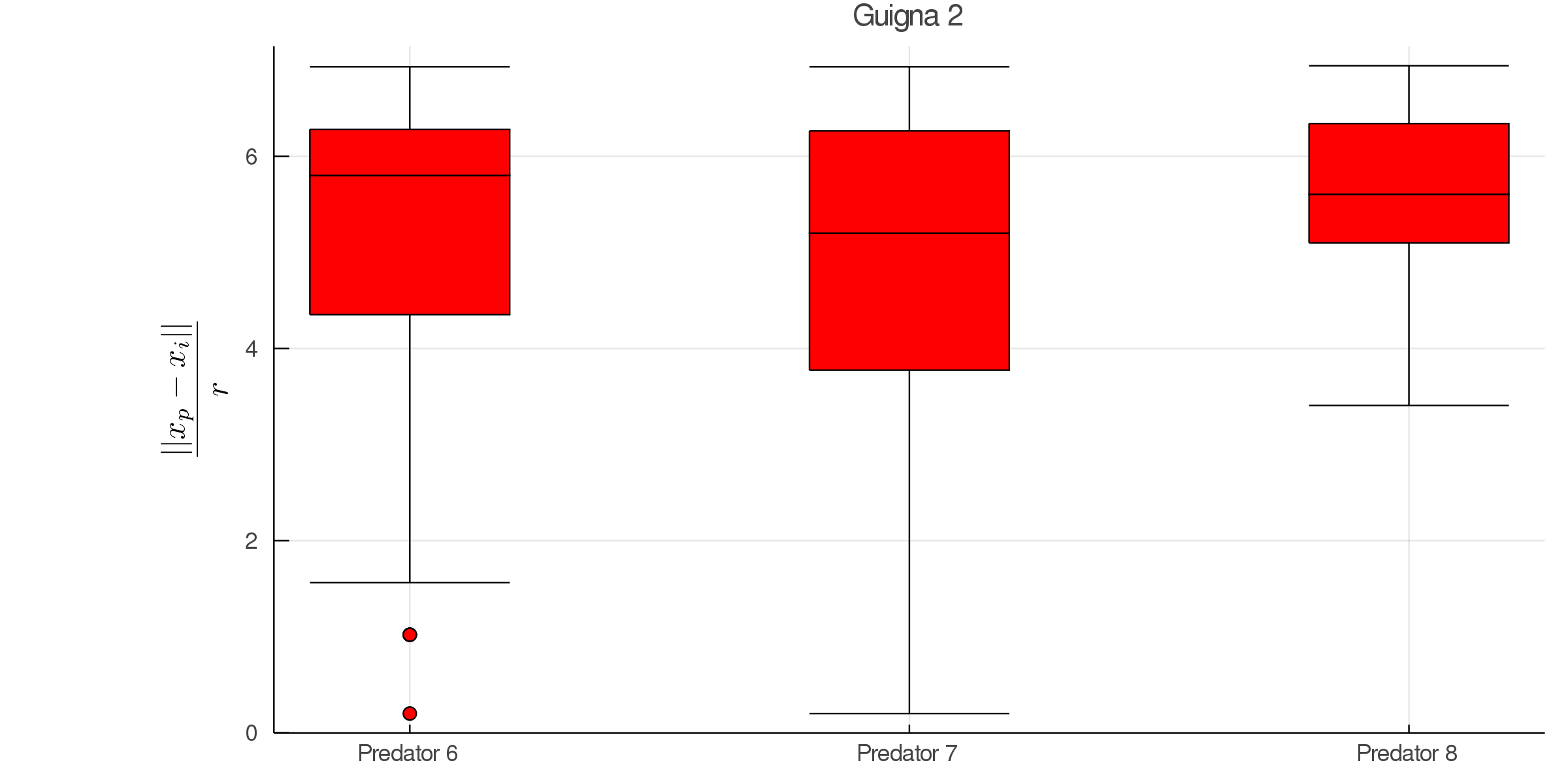} 
\end{minipage}
\end{tabular}
\caption{Most representative distances of the interactions of guigna 2 with some predators, and box plot of the distances $\set{\frac{\norm{x_p-x_i}}{r}}_{i=1}^m$ of the interactions of guigna 2 with some predators.}
\end{figure}

\begin{figure}[!ht]
\begin{tabular}{cc}
\begin{minipage}[c]{0.37\textwidth}
\centering
\footnotesize
    \begin{tabular}{c|cc}
\cmidrule{1-1}    Guiga 3 &       &  \\ \toprule
    Predators & $\underset{i\geq 1}{mean} \set{\frac{\norm{x_p-x_i}}{r}}$  & $\underset{i\geq 1}{median} \set{\frac{\norm{x_p-x_i}}{r}}$ \\ \midrule
    Predator 3 & 4.32  & 4.29 \\
    Predator 5 & 5.88  & 5.92 \\
    Predator 7 & 4.59  & 4.71 \\
    Predator 8 & 5.42  & 5.75 \\
    Predator 9 & 5.45  & 5.73 \\
     \bottomrule
    \end{tabular}%
\end{minipage} &
\begin{minipage}[c]{0.60\textwidth}
\centering
 \includegraphics[width=\textwidth, height=0.5\textwidth]{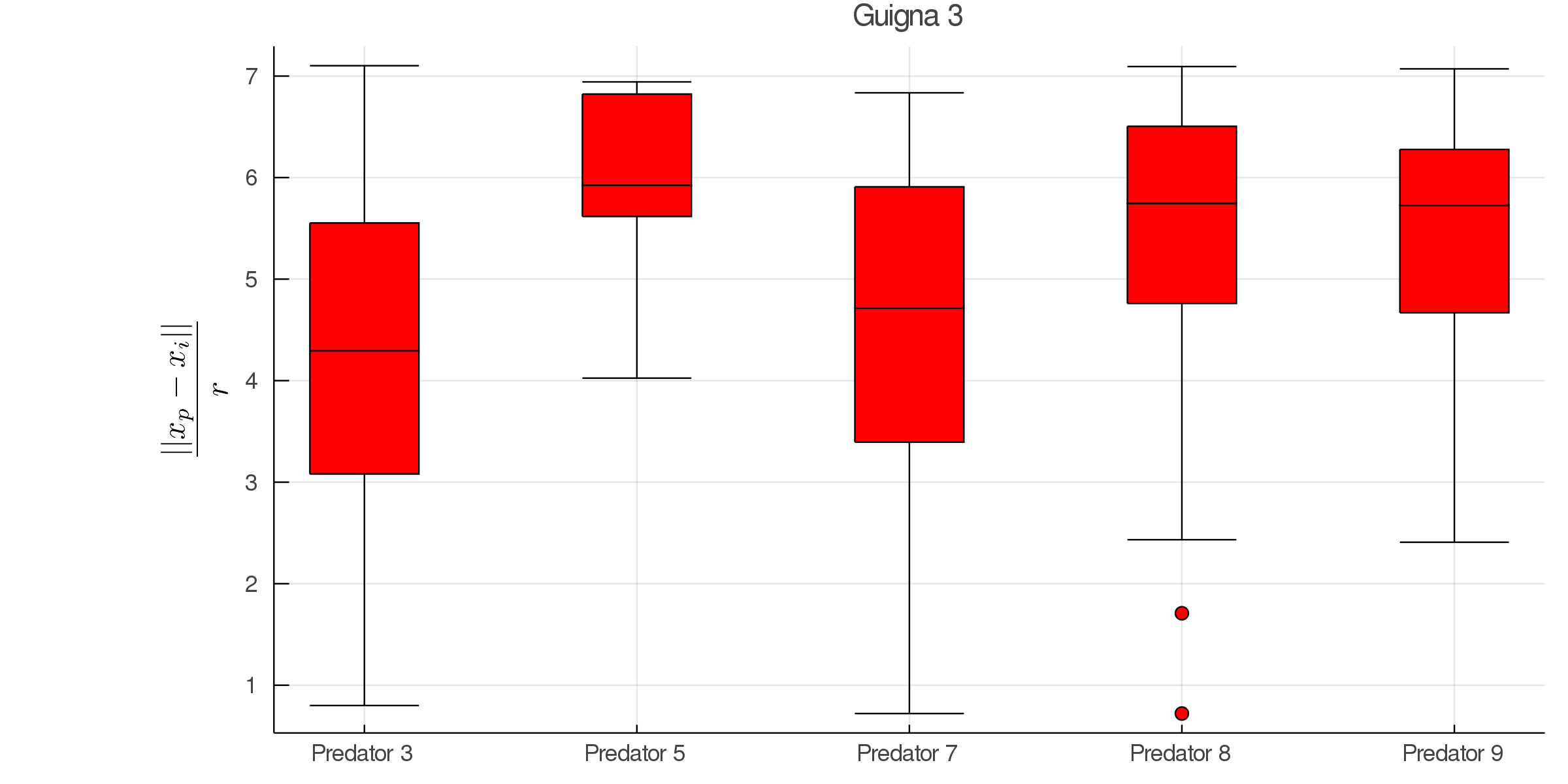} 
\end{minipage}
\end{tabular}
\caption{Most representative distances of the interactions of guigna 3 with some predators, and box plot of the distances $\set{\frac{\norm{x_p-x_i}}{r}}_{i=1}^m$ of the interactions of guigna 3 with some predators.}
\end{figure}

\begin{figure}[!ht]
\begin{tabular}{cc}
\begin{minipage}[c]{0.37\textwidth}
\centering
\footnotesize
    \begin{tabular}{c|cc}
\cmidrule{1-1}    Guiga 4 &       &  \\ \toprule
    Predators & $\underset{i\geq 1}{mean} \set{\frac{\norm{x_p-x_i}}{r}}$  & $\underset{i\geq 1}{median} \set{\frac{\norm{x_p-x_i}}{r}}$ \\ \midrule
    Predator 3 & 6.12  & 6.15 \\
    Predator 5 & 5.16  & 5.35 \\
    Predator 8 & 5.71  & 6.05 \\
    Predator 9 & 6.80  & 7.21 \\
    Predator 10 & 5.01  & 5.56 \\
    Predator 11 & 4.81  & 5.00 \\
    Predator 12 & 5.28  & 5.40 \\
    Predator 13 & 4.71  & 4.67 \\
     \bottomrule
    \end{tabular}%
\end{minipage} &
\begin{minipage}[c]{0.60\textwidth}
\centering
 \includegraphics[width=\textwidth, height=0.5\textwidth]{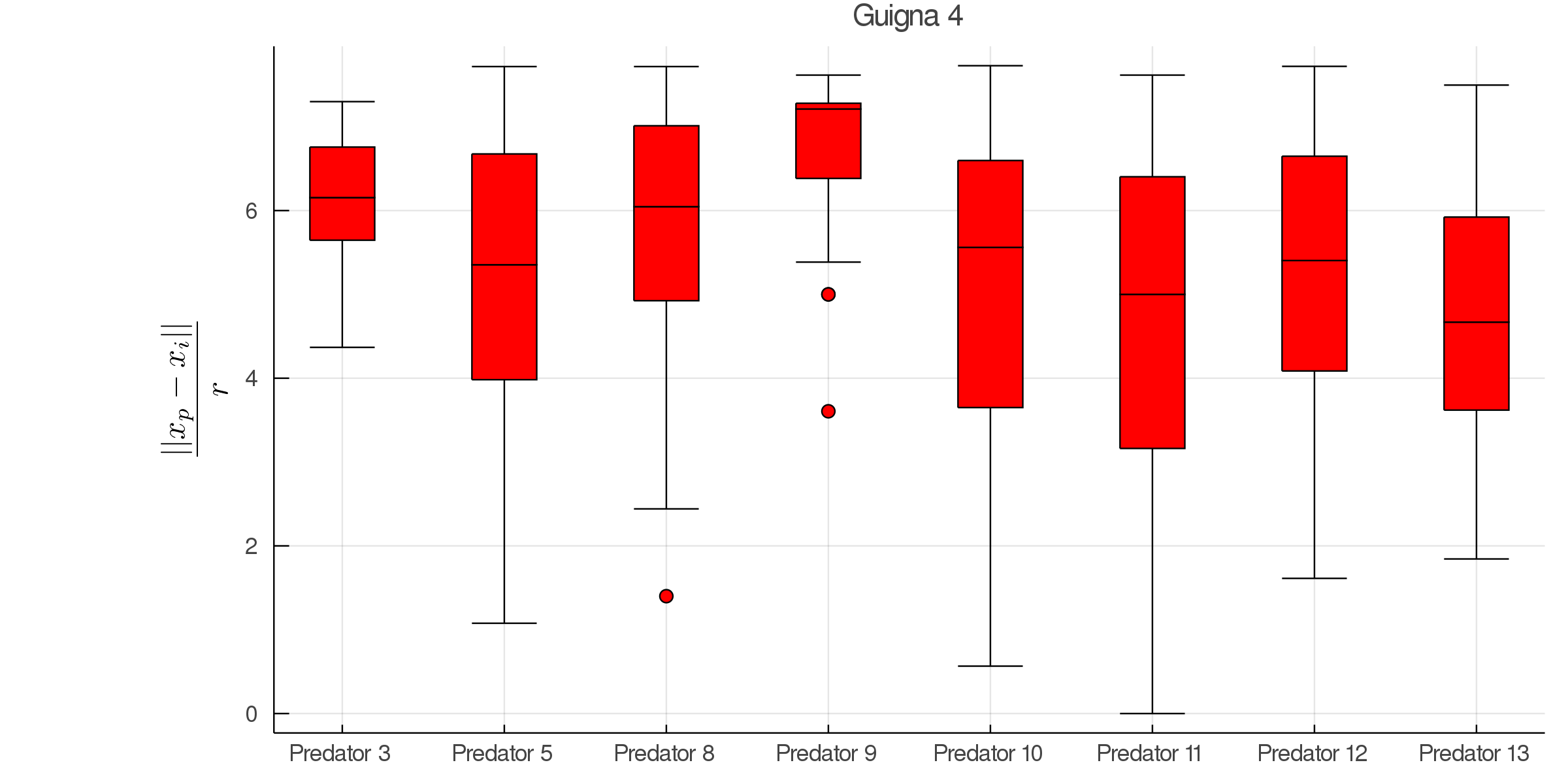} 
\end{minipage}
\end{tabular}
\caption{Most representative distances of the interactions of guigna 4 with some predators, and box plot of the distances $\set{\frac{\norm{x_p-x_i}}{r}}_{i=1}^m$ of the interactions of guigna 4 with some predators.}
\end{figure}

\newpage

\begin{figure}[!ht]
\begin{tabular}{cc}
\begin{minipage}[c]{0.37\textwidth}
\centering
\footnotesize
    \begin{tabular}{c|cc}
\cmidrule{1-1}    Guiga 5 &       &  \\ \toprule
    Predators & $\underset{i\geq 1}{mean} \set{\frac{\norm{x_p-x_i}}{r}}$  & $\underset{i\geq 1}{median} \set{\frac{\norm{x_p-x_i}}{r}}$ \\ \midrule
    Predator 3 & 3.67  & 3.51 \\
    Predator 4 & 4.84  & 5.20 \\
    Predator 5 & 4.43  & 4.62 \\
    Predator 10 & 5.36  & 5.30 \\
    Predator 12 & 4.19  & 4.40 \\
    Predator 13 & 4.32  & 4.53 \\
    Predator 14 & 5.13  & 5.30 \\
    \bottomrule
    \end{tabular}%
\end{minipage} &
\begin{minipage}[c]{0.60\textwidth}
\centering
 \includegraphics[width=\textwidth, height=0.5\textwidth]{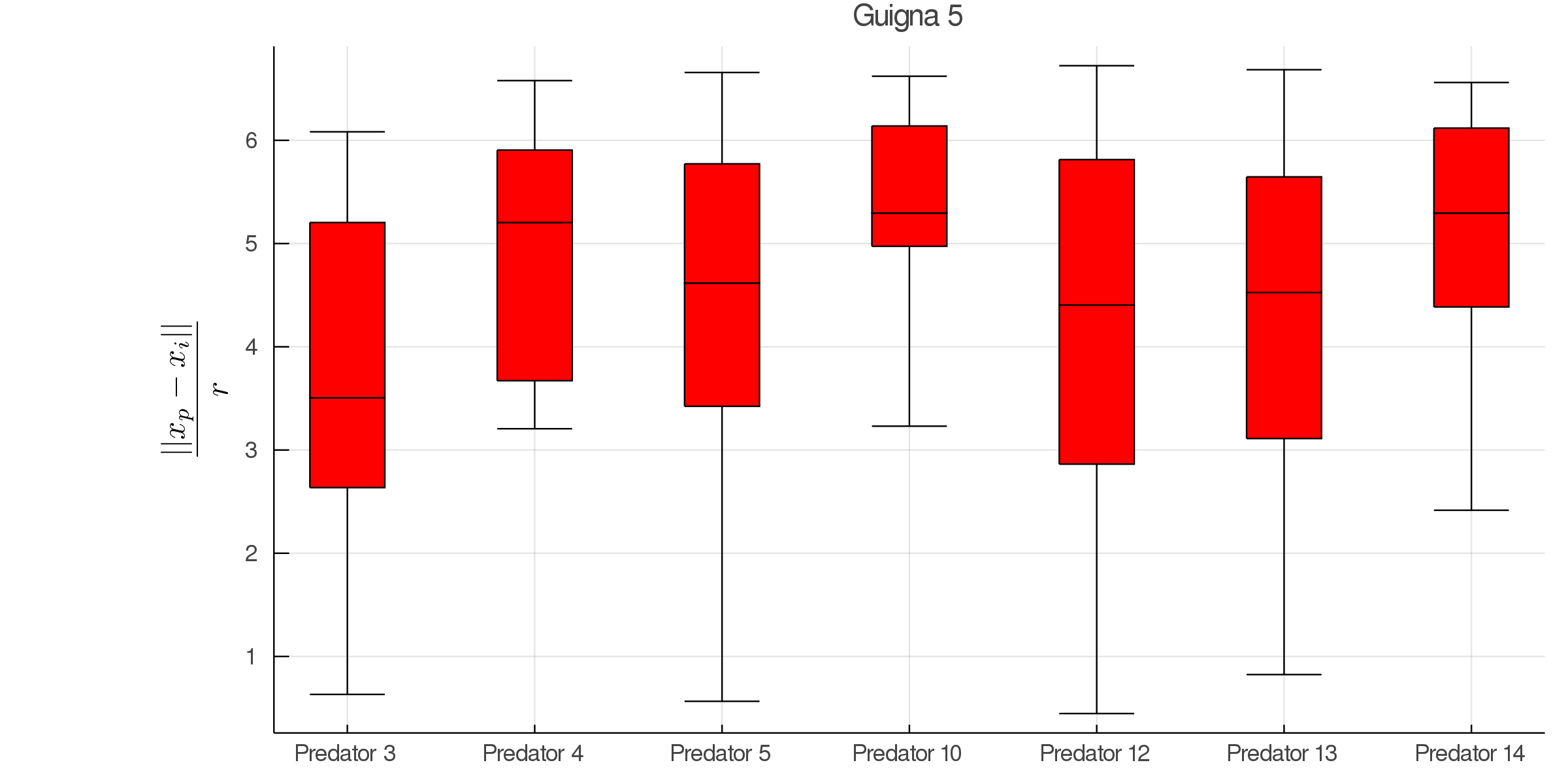} 
\end{minipage}
\end{tabular}
\caption{Most representative distances of the interactions of guigna 5 with some predators, and box plot of the distances $\set{\frac{\norm{x_p-x_i}}{r}}_{i=1}^m$ of the interactions of guigna 5 with some predators.}
\end{figure}

\begin{figure}[!ht]
\begin{tabular}{cc}
\begin{minipage}[c]{0.37\textwidth}
\centering
\footnotesize
    \begin{tabular}{c|cc}
\cmidrule{1-1}    Guiga 6 &       &  \\ \toprule
    Predators & $\underset{i\geq 1}{mean} \set{\frac{\norm{x_p-x_i}}{r}}$  & $\underset{i\geq 1}{median} \set{\frac{\norm{x_p-x_i}}{r}}$ \\ \midrule
    Predator 6 & 4.85  & 5.19 \\
    Predator 7 & 4.59  & 4.87 \\
    Predator 8 & 4.79  & 5.25 \\
    Predator 9 & 5.00  & 5.43 \\
    \bottomrule
    \end{tabular}%
\end{minipage} &
\begin{minipage}[c]{0.60\textwidth}
\centering
 \includegraphics[width=\textwidth, height=0.5\textwidth]{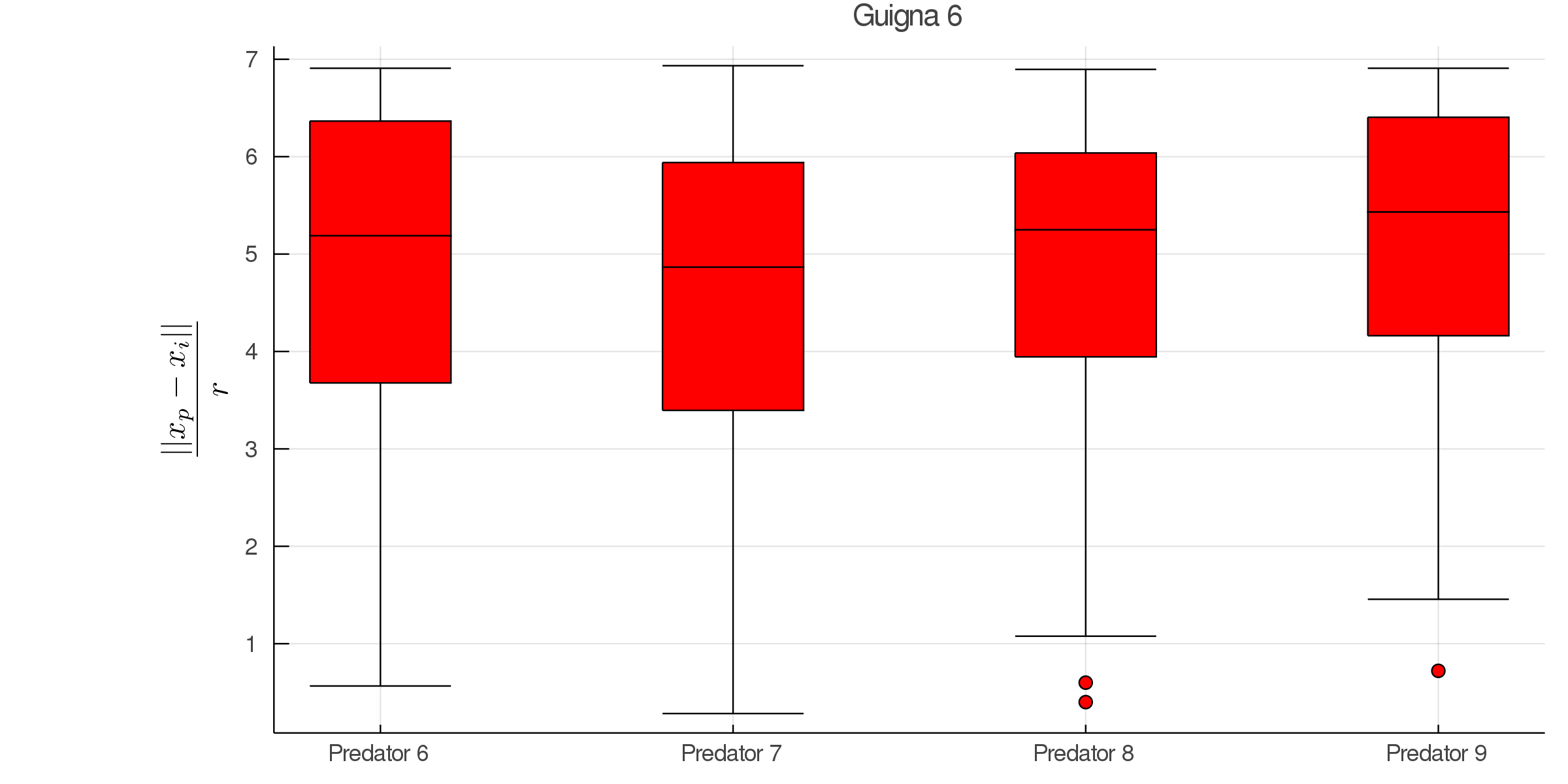} 
\end{minipage}
\end{tabular}
\caption{Most representative distances of the interactions of guigna 6 with some predators, and box plot of the distances $\set{\frac{\norm{x_p-x_i}}{r}}_{i=1}^m$ of the interactions of guigna 6 with some predators.}
\end{figure}

\begin{figure}[!ht]
\begin{tabular}{cc}
\begin{minipage}[c]{0.37\textwidth}
\centering
\footnotesize
    \begin{tabular}{c|cc}
\cmidrule{1-1}    Guiga 7 &       &  \\ \toprule
    Predators & $\underset{i\geq 1}{mean} \set{\frac{\norm{x_p-x_i}}{r}}$  & $\underset{i\geq 1}{median} \set{\frac{\norm{x_p-x_i}}{r}}$ \\ \midrule
    Predator 6 & 4.42  & 4.62 \\
    Predator 7 & 4.53  & 4.68 \\
    Predator 8 & 4.57  & 4.92 \\
    Predator 9 & 5.13  & 5.60 \\
    Predator 10 & 4.57  & 4.80 \\
    \bottomrule
    \end{tabular}%
\end{minipage} &
\begin{minipage}[c]{0.60\textwidth}
\centering
 \includegraphics[width=\textwidth, height=0.5\textwidth]{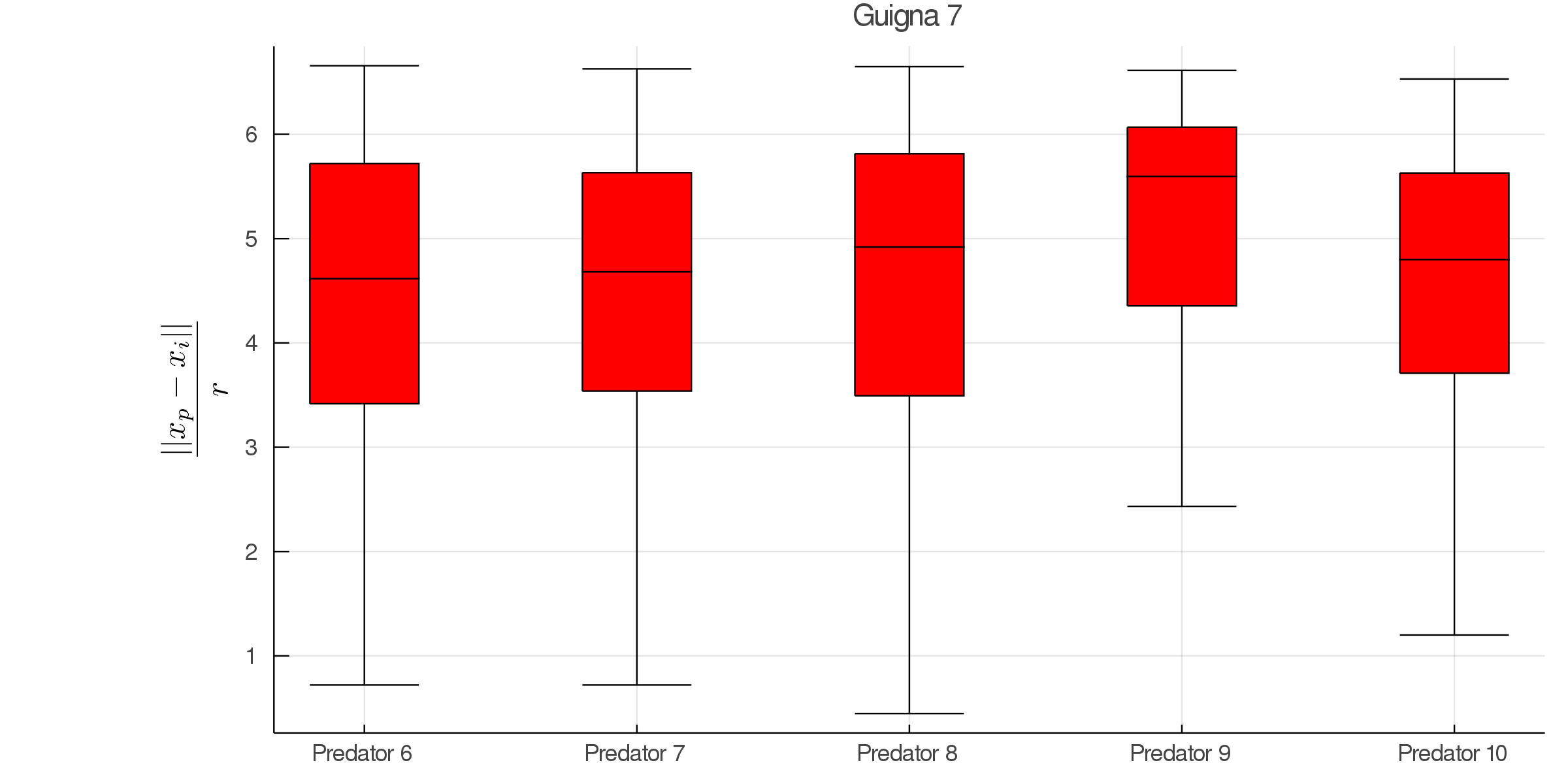} 
\end{minipage}
\end{tabular}
\caption{Most representative distances of the interactions of guigna 7 with some predators, and box plot of the distances $\set{\frac{\norm{x_p-x_i}}{r}}_{i=1}^m$ of the interactions of guigna 7 with some predators.}
\end{figure}

\newpage

\begin{figure}[!ht]
\begin{tabular}{cc}
\begin{minipage}[c]{0.37\textwidth}
\centering
\footnotesize
    \begin{tabular}{c|cc}
\cmidrule{1-1}    Guiga 8 &       &  \\ \toprule
    Predators & $\underset{i\geq 1}{mean} \set{\frac{\norm{x_p-x_i}}{r}}$  & $\underset{i\geq 1}{median} \set{\frac{\norm{x_p-x_i}}{r}}$ \\ \midrule
    Predator 7 & 5.10  & 5.20 \\
    Predator 8 & 4.91  & 5.20 \\
    Predator 9 & 5.12  & 5.31 \\
    Predator 10 & 6.62  & 6.51 \\
    \bottomrule
    \end{tabular}%
\end{minipage} &
\begin{minipage}[c]{0.60\textwidth}
\centering
 \includegraphics[width=\textwidth, height=0.5\textwidth]{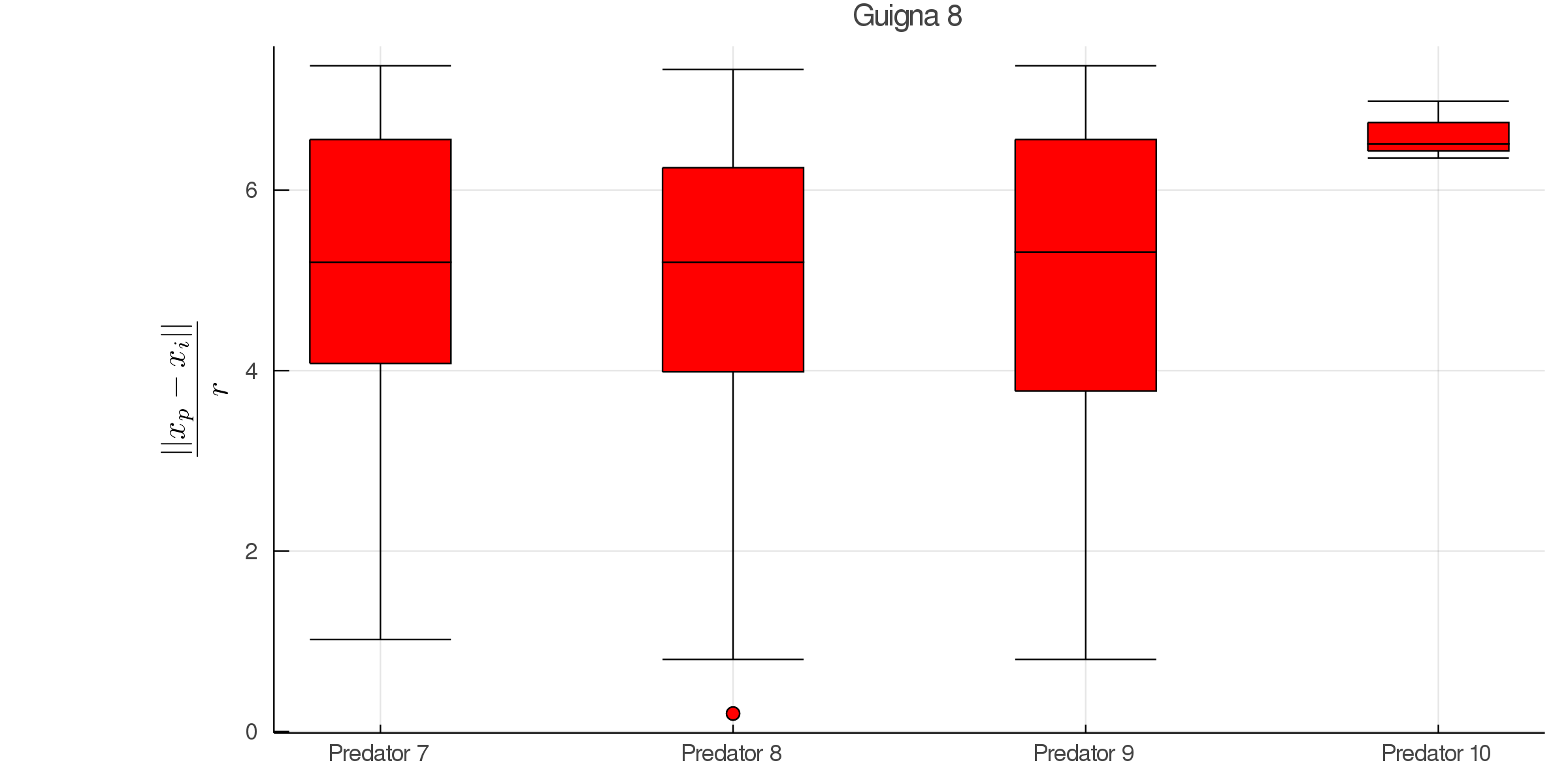} 
\end{minipage}
\end{tabular}
\caption{Most representative distances of the interactions of guigna 8 with some predators, and box plot of the distances $\set{\frac{\norm{x_p-x_i}}{r}}_{i=1}^m$ of the interactions of guigna 8 with some predators.}
\end{figure}

\begin{figure}[!ht]
\begin{tabular}{cc}
\begin{minipage}[c]{0.37\textwidth}
\centering
\footnotesize
    \begin{tabular}{c|cc}
\cmidrule{1-1}    Guiga 9 &       &  \\ \toprule
    Predators & $\underset{i\geq 1}{mean} \set{\frac{\norm{x_p-x_i}}{r}}$  & $\underset{i\geq 1}{median} \set{\frac{\norm{x_p-x_i}}{r}}$ \\ \midrule
    Predator 10 & 5.65  & 6.00 \\
    Predator 11 & 5.01  & 5.30 \\
    Predator 12 & 4.81  & 5.23 \\
    Predator 13 & 4.64  & 4.83 \\
    Predator 14 & 5.87  & 6.05 \\
    \bottomrule
    \end{tabular}%
\end{minipage} &
\begin{minipage}[c]{0.60\textwidth}
\centering
 \includegraphics[width=\textwidth, height=0.5\textwidth]{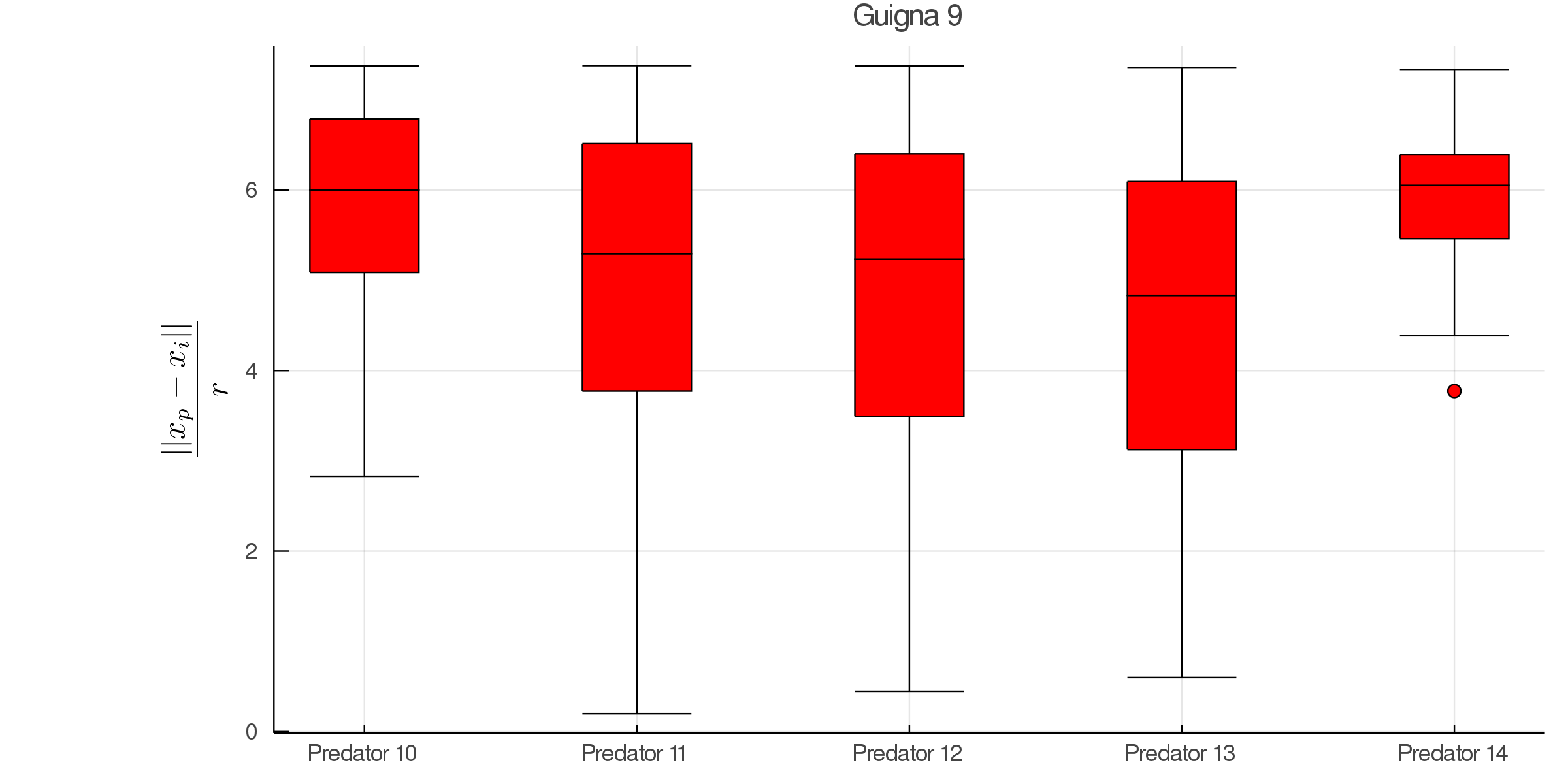} 
\end{minipage}
\end{tabular}
\caption{Most representative distances of the interactions of guigna 9 with some predators, and box plot of the distances $\set{\frac{\norm{x_p-x_i}}{r}}_{i=1}^m$ of the interactions of guigna 9 with some predators.}
\end{figure}

Finally, taking into account the possibility that the data have a high level of dispersion, the following value may be defined

\begin{eqnarray}
r_l:=\min \set{\underset{i\geq 1}{mean} \set{\frac{\norm{x_p-x_i}}{r}},\underset{i\geq 1}{median} \set{\frac{\norm{x_p-x_i}}{r}}},
\end{eqnarray}

with which it is possible to obtain the following result:

\begin{eqnarray}
\mbox{If } \set{x_{i}}_{i=1}^M \cap B\left(x_p;r_l\right)\neq \emptyset \ \Rightarrow \ \mbox{ there exists at least one value } x_j\in  \set{x_{i}}_{i=1}^M,
\end{eqnarray}

where $x_j$ represents a position in which some guigna is at a distance potentially lethal from the predator $x_p$.

\section{Conclusions}

In this paper, under the hypothesis that the guignas maintain a sedentary behavior in a specific area of a given territory, one way to simulate a distribution of points in a given territory was shown using random walkers to emulate the distribution of data that would be obtained by placing radiocollars in a population of guignas, with which it was possible to make estimates of the mean distances that move away from a certain fixed position, and the interactions that they may have with points of the territory that represent a high probability of lethality, such as farms, packs of dogs, roads, urban areas, etc. However, due to the few records obtained from guignas, a combination of test data with real data, obtained by camera traps, was used. It is necessary to mention that although the simulation fulfilled its objective, it would be necessary to consider the following points to improve the estimates of the behavior of the guignas:

\begin{itemize}
\item[$i$)] Establish a coordinate record of possible guignas territories in a region of interest using camera traps.
\item[$ii$)] Establish a coordinate record
of predator territories, as well as places potentially lethal to guignas.
\item[$iii$)] Establish a scale in a specific region of interest with the presence of guignas from a satellite image.
\item[$iv$)] Establish a step size $r$ considering the scale of the region of interest given in the previous point and the mean distance that a guigna usually travels in its territory, or a feline with similar habits, in a given interval of time such as one week, one month, etc.
\end{itemize}

Finally, it is necessary to mention that by estimating the possible interactions that a guignas population may have with possible predators in a territory with the help of a satellite image, it is possible to evaluate the points of a territory that represent a potentially lethal risk for the guignas, and thus generate relocation strategies that help preserve them. The latter would also be possible with populations of other animals that show sedentary behavior, so carrying out simulations with random walkers with an adequate step size $r$ could help the preservation of different species.

\textbf{Acknowledgments}: We would like to thank Francisca Javiera Miranda Orcaistegu, who obtained and provided the photographic records of L. guigna. Trap cameras have been installed in the Chiloé National Park, which is administered by the Chilean National Forest Corporation CONAF. Our thanks go out to every member of CONAF who dedicates their work to the conservation of the species.

\bibliography{Biblio}
\bibliographystyle{unsrt}

\end{document}